\documentclass[%
 aip,
 amsmath,amssymb,
 reprint,%
]{revtex4-1}

\usepackage{graphicx}
\usepackage{dcolumn}
\usepackage{bm}
\usepackage[mathlines]{lineno}
\usepackage[version=3]{mhchem}
\usepackage{soul}
\usepackage{float}
\usepackage{comment}
\usepackage{chemfig}
\usepackage{multirow}
\usepackage{makecell}
\usepackage{textcomp}
\usepackage{url}
\usepackage{comment}
\usepackage{color}
\usepackage[utf8]{inputenc}
\usepackage[T1]{fontenc}
\usepackage{mathptmx}

\begin{document}

\preprint{AIP/123-QED}

\title{Deep Potential generation scheme and simulation protocol for the \ce{Li10GeP2S12}-type superionic conductors}

\author{Jianxing Huang}
    \affiliation{State Key Laboratory of Physical Chemistry of Solid Surfaces, iChEM, College of Chemistry and Chemical Engineering, Xiamen University, Xiamen 361005, China}

\author{Linfeng Zhang}
    \affiliation{Program in Applied and Computational Mathematics, Princeton University, Princeton, NJ 08544, USA}

\author{Han Wang}
    \affiliation{Laboratory of Computational Physics, Institute of Applied Physics and Computational Mathematics, Fenghao East Road 2, Beijing 100094, P.R. China}

\author{Jinbao Zhao}     
    \email{jbzhao@xmu.edu.cn}
    \affiliation{State Key Laboratory of Physical Chemistry of Solid Surfaces, iChEM, College of Chemistry and Chemical Engineering, Xiamen University, Xiamen 361005, China}
 
\author{Jun Cheng}
    \email{chengjun@xmu.edu.cn}
    \affiliation{State Key Laboratory of Physical Chemistry of Solid Surfaces, iChEM, College of Chemistry and Chemical Engineering, Xiamen University, Xiamen 361005, China}

\author{Weinan E}
    \email{weinan@math.princeton.edu}
    \affiliation{Program in Applied and Computational Mathematics, Princeton University, Princeton, NJ 08544, USA}
    \affiliation{Department of Mathematics, 
    Princeton University, Princeton, NJ 08544, USA}

\date{\today}

\begin{abstract}
Solid-state electrolyte materials with superior lithium ionic conductivities are vital to the next-generation Li-ion batteries.
Molecular dynamics could provide atomic scale information to understand the diffusion process of Li-ion in these superionic conductor materials. 
Here we implement the Deep Potential Generator (DP-GEN) to set up an efficient protocol to automatically generate interatomic potentials for \ce{Li10GeP2S12}-type solid-state electrolyte materials (\ce{Li10GeP2S12}, \ce{Li10SiP2S12} and \ce{Li10SnP2S12}).
The reliability and accuracy of the fast interatomic potentials are validated. 
With the potentials, we extend the simulation of diffusion process to a wide temperature range (300K-1000K) and systems with large size (\textasciitilde 1000 atoms).
Important technical aspects like the statistical error and size effect are carefully investigated and benchmark tests including the effect of density functional, thermal expansion and configurational disorder are performed.
The computed data that considering these factors agrees well with the experimental results and we find that the 3 structures shows different behaviors with respect to configurational disorder.
Our work paves the way for further research on computation screening of solid-state electrolyte materials.
\end{abstract}
\maketitle

\section{Introduction}
All-solid-state Li-ion batteries are amongst the most promising candidates for the next-generation rechargeable batteries~\cite{goodenough2013li, tarascon2011issues, chu2012opportunities, hu2016batteries, zhang2018new}. 
Desired solid-state electrolyte (SSE) materials should have high \ce{Li+} conductivities and wide electrochemical windows. 
Several groups of promising candidates, with performance competitive to current commercial liquid electrolytes (e.g., \ce{Li10GeP2S12}~\cite{kamaya2011lithium}, \ce{Li7La3Zr2O2}~\cite{murugan2007fast}, and \ce{Li7P3S11}~\cite{seino2014sulphide}) have been reported. 
Due to their relevant highest ionic conductivities, the family of \ce{Li10GeP2S12}-type materials have attracted extensive studies~\cite{kamaya2011lithium, ong2013phase, ceder_ong_wang_2018, Mo2012}.

Improvement of SSE performance benefits from the fundamental understanding of the atomic-scale diffusion process. 
The \textit{ab initio} molecule dynamics (AIMD) calculation~\cite{car1985cpmd} has been utilized to investigate the microscopic details of the diffusion processes\cite{ong2013phase, ceder_ong_wang_2018, he2017origin, nolan2018computation, van2000lithium, Mo2012}. 
The diffusion coefficients of most superionic conductors ranges from $10^{-13} m^2/s$ to $10^{-10} m^2/s$.
Unfortunately, due to its high computational cost, AIMD is typically limited to a system size of hundreds of atoms in the time scale of tens of pico-seconds.
Accurate calculation of diffusion coefficients requires simulations in the time scale of nanosecond, which is unreachable for current AIMD methods.
This makes AIMD practically impossible to accurately estimate the diffusion coefficients of solid-state electrolyte materials at experimental conditions, i.e., at room temperature.
Therefore one often resorts to the extrapolation strategy: assuming that a single Arrhenius relationship applies to a wide temperature range (this implicitly assumes a temperature-independent diffusion mechanism), then one can predict the ionic conductivity under working conditions(-40 \textcelsius \textasciitilde 80 \text(celsius) by extrapolating  from the high-temperature regime  (\textgreater 500 K) to the low temperature regime(\textless 500 K).

\begin{figure*}[!htb]
\includegraphics[width=0.95\textwidth]{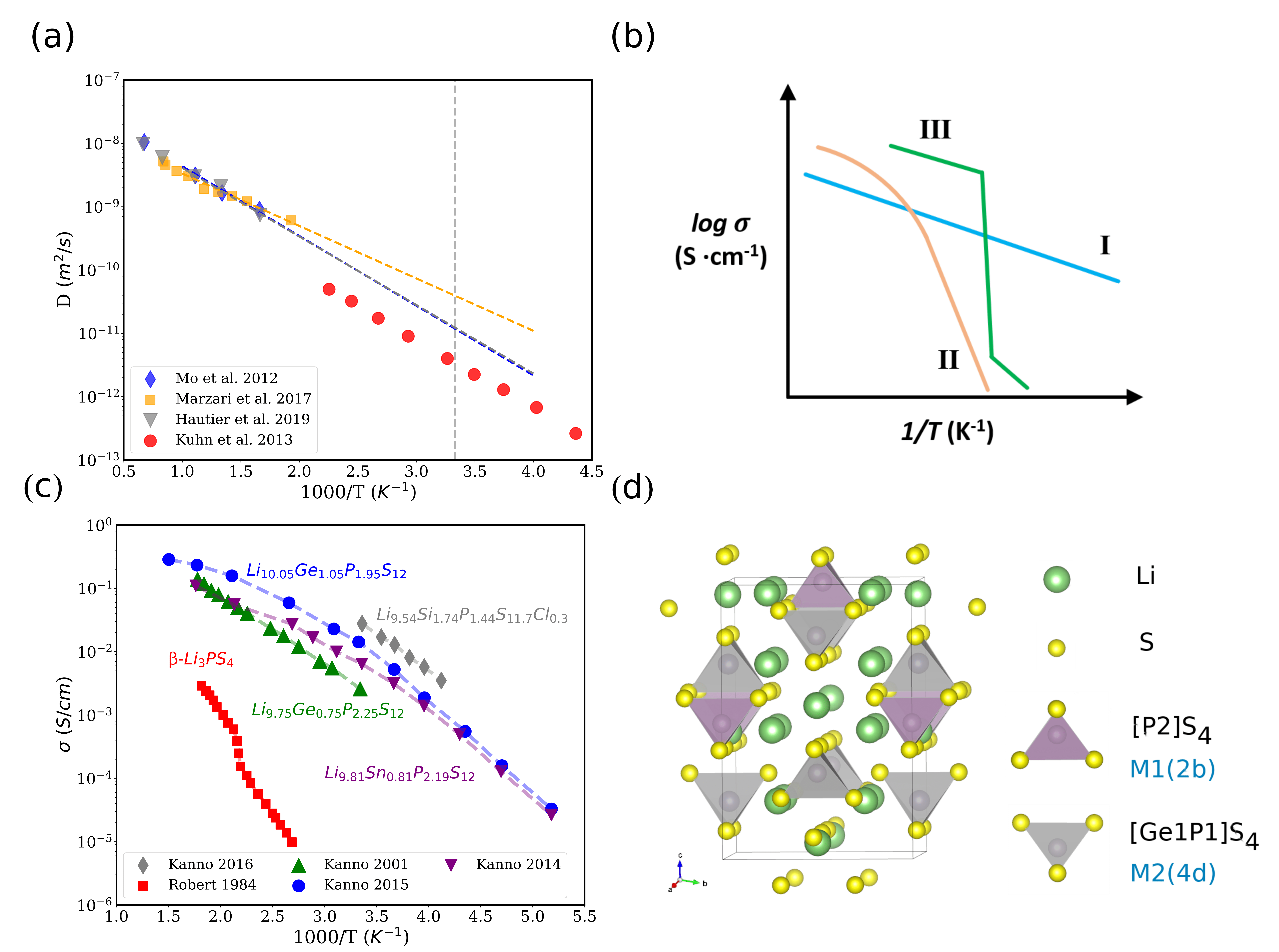}
\caption{\label{fig:background}
(a) Diffusion coefficients of \ce{Li10GeP2S12} by AIMD calculations~\cite{Mo2012, marcolongo2017ionic}(blue, orange and grey points), and solid state NMR measurement~\cite{Kuhn2013} (red points).
And the AIMD data are extrapolated to room temperature.
The vertical dashed line corresponds to the room temperature (300K).
(b) Schematic illustration of the three kinds of the temperature dependence of the conductivity, according to Ref.~\cite{boyce1979superionic}.
(c) Temperature dependence of the ionic conductivity of Li-ion superionic conductors from experiments. 
Data were taken from: $\beta$-\ce{Li_3PS_4}\cite{TACHEZ1984181}, \ce{Li_{10.05}Ge_{1.05}P_{1.95}S_{12}}\cite{Kwon2015} and \ce{Li_{9.75}Ge_{0.75}P_{2.25}S_{12}}\cite{Kanno2001}.
(d) Characteristics of \ce{Li10GeP2S12} structure. 
The occupation ratio of P in the M1(4d) site for all 3 systems are around 0.5~\cite{kamaya2011lithium, bron2013li10snp2s12, Kuhn2013}.
}
\end{figure*}
 
A commonly found protocol in literature based on this assumption is to collect high-temperatures diffusion coefficients (600 K \textasciitilde 1200 K) from 100~ps \textasciitilde 400~ps AIMD simulations~\cite{ong2013phase, ceder_ong_wang_2018, Mo2012} including 100 \textasciitilde 200 atoms and extrapolate these data to obtain room-temperature ionic conductivities.
However, this extrapolation strategy based on temperature will lead to deviations up to two orders of magnitude at room temperature.
As demonstrated in Fig.~\ref{fig:background} (a), AIMD calculation significantly overestimate diffusion coefficients at room temperature.
This issue, in particular when applied to SSE materials, has been comprehensively discussed by He \textit{et al.}~\cite{he2018statistical}.

Even more problematically, this extrapolation approach loses predictive power when the Arrhenius-type temperature dependence breaks down, which has been discussed in detail over 40 years ago~\cite{boyce1979superionic}. 
Fig.~\ref{fig:background} (b) shows three types of superionic conductors give rise to different transition behaviors of the ionic conductivity with respect to the inverse of the temperature.
The temperature dependence of ionic conductivities of several typical ionic conductors are depicted in Fig.~\ref{fig:background} (c). 
The examples of \ce{Li_{10.05}Ge_{1.05}P_{1.95}S_{12}} and $\beta$-\ce{Li_3PS_4} represent the failures of the extrapolation strategy in \ce{Li10GeP2S12}-like systems. 
The assumption behind the extrapolation strategy is only applicable to systems whose diffusion mechanisms are independent of temperature.

An explicit solution is to directly simulate the diffusion processes at room temperature, which requires larger systems and longer trajectories to ensure the convergence.
Thus, acceleration of simulations without losing the accuracy of DFT calculation is desired.
There have been ever-increasing efforts recently to develop empirical interatomic potential or model Hamiltonians involving simple analytical terms to speed up simulations, for SSE materials of interest~\cite{xiao2015candidate, kahle2018modeling, masanobu2020cuckoo, icf2020jiang}.
However, due to the relative large errors, the empirical models are only suitable for the crude screening of materials. 
More recent works have utilized machine learning (ML) tools~\cite{behler2007generalized, bartok2010gaussian, artrith2017atomic, NIPS2018_7696, zhang2018new} to represent the many-body and nonlinear dependence of the potential energy surface (PES) on atomic positions for modeling~\cite{artrith2018constructing, deringer2018towards, fujikake2018gaussian,lacivita2018structural, li2017study, deng2019electrostatic}.
In particular, applications to SSE materials, e.g., \ce{Li3PO4}\cite{li2017study}, \ce{LiPON}~\cite{lacivita2018structural}, \ce{Li10GeP2S12}\cite{Aris2019} and \ce{Li3N}~\cite{deng2019electrostatic}, have recently been explored. 

Despite these efforts, two major obstacles have remained.

First, a systematic and automatic procedure to generate uniformly accurate PES models, with a minimal set of training data, is still largely missing. 
The most straightforward approach is to perform extensive AIMD simulations at different temperatures and train models based on these AIMD trajectories. 
However, this procedure is computationally demanding, and the generated snapshots are highly correlated, reducing the quality of the training data.
For this reason, a great amount of trial-and-error process is involved in most of the ML-based models, and consequently, the reliability of these models is in doubt.
Besides, the performances of ML potentials depend on a few factors, including the reliability of locality assumption, quality of training data and hyperparameters of models. 
Specifically, though there have been some efforts to set up ML potential generation schemes for different materials~\cite{bernstein2019novo, Nyshadham2019, Aris2019, boron2018faraday, mortazavi2020-conductivity, podryabinkin2019-uspex, Gubaev2018}, there lacks a well-benchmarked, automatic, and efficient scheme for the generation and validation of ML potential for Li-ion superionic conductor materials.

Second, the diffusion process is determined by the structures, e.g., lattice volume~\cite{ong2013phase, Mo2012} and configurational disorder~\cite{stamminger2019ionic, ohno2019further, zhou2019entropically, hanghofer2019substitutional}.
The structure of \ce{Li10GeP2S12} is visualized in Fig.~\ref{fig:background} (d), in which \ce{MS4} (M = Ge, Si, Sn, and P) tetrahedrons form the solid-like backbone and the liquid-like Li-ions are free to flow across the channels. 
The tetrahedral units include two groups of sites: M1(2b) and M2(4d). The M2(4d) sites could be occupied by Ge/Si/Sn/P atoms while only P atom are found to occupy the M1(2b) sites. 
The fractional occupation results in the disordered arrangement of \ce{MS4} tetrahedral units. 
To the best of our knowledge, little attention has been paid to the effects of the thermal expansion of lattice volume and configurational disorder on diffusion processes.
To set up protocols to accurately compute diffusion coefficients of \ce{Li10GeP2S12}-type materials at different temperatures, technical issues of MD simulations and the effects of these factors shall be validated. 

In this work, we implement a concurrent learning scheme (DP-GEN) to generate uniformly accurate Deep Potential (DP) models for 3 \ce{Li10GeP2S12}-type superionic conductors (\ce{Li10GeP2S12}, \ce{Li10SiP2S12} and \ce{Li10SnP2S12}), respectively.
The reliability and performance of the DP models are examined, including locality test, model accuracy, and speed test. 
With the DP models, the effects of key simulation settings, thermal expansion, and configurational disorder have been investigated.
The validated protocol is applied to compute diffusion coefficients between 300 K and 1000 K.
And finally the simulation results are compared with experimental data and the differences are discussed.

\section{Methods}

\subsection{DP-GEN for \ce{Li10GeP2S12}-type structures}
\begin{figure}[!htb]
\includegraphics[width=0.45\textwidth]{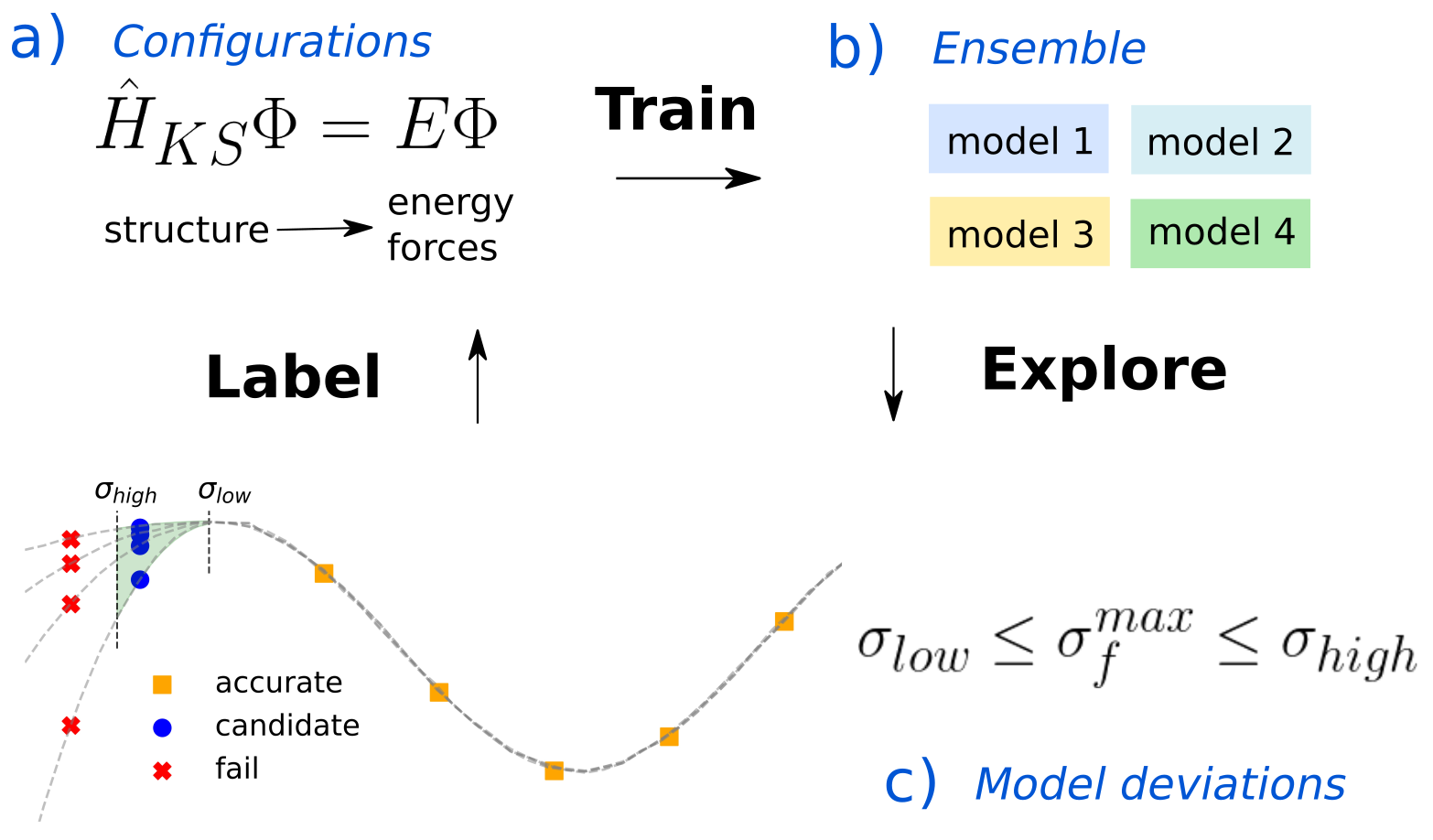}
\caption{\label{fig:dpgen}
Schematic illustration of DP-GEN process. 
Usually, a dataset with hundreds of configurations is required to start the iteration of DP-GEN.
Each DP-GEN iteration includes 3 stages: \textbf{labeling}, \textbf{training}, and \textbf{exploration}.
(a) First, configurations are \textbf{labeled} by high precision single-point DFT calculations.
(b) Then, an ensemble of DP models with the same architecture (i.e. number of neural network layers and nodes) but different random seeds are \textbf{trained} simultaneously using the whole dataset. 
(c) To \textbf{explore} larger configuration space, a few MD simulations at different thermodynamic conditions are driven by the ensemble of DP models from the previous stage.
The selected candidates are labeled by DFT calculation and then added to the dataset.
The exploration process is considered converged after a predetermined number of loops is reached or only a small percentage (e.g. 0.5\%) of candidates are found in the last exploration iteration.
}
\end{figure}

Using the Deep Potential Generator (DP-GEN)~\cite{zhang2019active,zhang2020dpgen}, a minimal set of training data is generated via an efficient and sufficient sampling process, thereby guaranteeing a reliable PES model produced by training. 
The flowchart of DP-GEN iteration is shown in Fig.~\ref{fig:dpgen}.
In the exploration step, model deviations are evaluated using the ensemble of trained models and new configurations are picked according to the maximum deviation of forces ($\sigma_f^{max}$), defined as:
\begin{equation}
\label{eqn:max_devi}
\sigma_f^{max} =  \mathop{max}\limits_{i} \sqrt{\langle||f_i-\langle f_i\rangle||^2\rangle},
\end{equation}
where $f_i$ is the force acting on atom $i$, and $\langle\cdots\rangle$ denotes the average of the DP model ensemble.
Configurations with small force deviations ($\sigma_{f}^{max} < \sigma_{low} $, yellow square in Fig.~\ref{fig:dpgen}(c)) are effectively covered by the training dataset with high probability. 
On the contrary, excessive force deviation ($\sigma_{f}^{max} > \sigma_{high} $, red cross in Fig.~\ref{fig:dpgen}(c)) implies that the configuration may diverge from the relevant physical trajectories.
Therefore none of them are picked. 
Only configurations whose $\sigma_f^{max}$ fall between a predetermined window are labeled as \textit{candidates} (blue circles in Fig.~\ref{fig:dpgen}(c)). 
In practice, after running several MD trajectories, the selection criterion usually produces hundreds or thousands of candidates.
A small fraction of them is representative enough to improve the model, and therefore a cutoff number ($N_{label}^{max}$) is set to restrict the number of candidates.
These candidates are labeled and added to the original dataset for the next training. 
The labeling and training stages are rather standard, while there is large flexibility for the sampling strategy on how to explore the relevant configuration space in each iteration.
According to Ref.~\cite{zhang2020dpgen}, a practical rule of thumb is to set $\sigma_{low}$ slightly larger than the training error achieved by the model,and set $\sigma_{high}$ 0.1-0.3 eV/\AA~higher than $\sigma_{low}$.
In this paper, $\sigma_{low}$ and $\sigma_{high}$ are set to 0.12 and 0.25 eV/\AA, respectively. 

\begin{table}[!htb]
\caption{Exploration settings of DP-GEN iterations.}
\label{tbl:dpgen}
\begin{tabular}{ccccccc}
\hline
Ensemble & Iteration & Temperature(K) & Structure & Supercell \\
\hline
NpT & 1-4 &  \makecell[c]{50, 100, 200,\\ 300, 500, 700,\\ 900, 1200} & \makecell[c]{ordered\\DFT-relaxed} & $1\times1\times1$ \\
NpT & 5-8 &  \makecell[c]{300, 700, 1200} & \makecell[c]{ordered\\DFT-relaxed} & $2\times2\times2$ \\
NpT & 9-12 &  \makecell[c]{50, 100, 200,\\ 300, 500, 700,\\ 900, 1200} & \makecell[c]{disordered\\DFT-relaxed} & $1\times1\times1$ \\
NpT & 13-16 & \makecell[c]{300, 700, 1200} & \makecell[c]{disordered\\DFT-relaxed} & $2\times2\times2$ \\
NVT & 19-21 & \makecell[c]{50, 100, 200,\\300, 500, 700,\\ 900, 1200} & \makecell[c]{ordered\\experiment} & $2\times2\times2$ \\
\hline
\end{tabular}
\end{table}

In this work, all crystal structures are fetched from the Materials Project\cite{jain2013commentary, ong2015materials} database as conventional cells. 
The material ID of them in the Materials Project are: mp-696138 (\ce{Li10Ge(PS6)2}), mp-696129 (\ce{Li10SiP2S12}) and mp-696123 (\ce{Li10SnP2S12}). 
Structure manipulation are dealt with \textit{pymatgen}~\cite{ong2013python}.

The DP-GEN is started with 590 structures that are generated via slightly perturbing DFT-relaxed structures. 
A smooth version of Deep Potential (v1.2)~\cite{wang2018deepmd, NIPS2018_7696}, which is end-to-end, i.e. capable of fitting many-component data of SSE materials with little human intervention, was used for the training step. 
The exploration is run on 5 systems step by step. 
Each system is composed of 3 or 4 iterations depending on its convergence.
The exploration time of each system is gradually lengthened from 1000~fs to 10000~fs.
The exploration is beginning with ordered structures relaxed by DFT (i.e. structures downloaded from the Materials Project database, in which the position of Ge/Si/Sn/P atoms are fixed). 
Then the exploration is changed to disordered structures whose 4d sites are randomly occupied by Ge/Si/Sn/P.
Exploration with the NpT ensemble means that configurations with different lattice parameters could be automatically sampled.
And finally the exploration is performed in the systems with experimental lattice parameters.
The exploration of each system is considered converged when the percentage of accurate configurations is larger than 99.5\%.
The detailed settings of the DP-GEN are listed in Table~\ref{tbl:dpgen}.
The production DP models are trained with 10 times longer steps, in which the number of batch and the step of learning rate decay are set to 4000000 and 20000, respectively. 

\subsection{Locality test}
A key assumption of most ML potentials is the \textit{locality}. 
In a nutshell, ML potentials assume that properties of each atom only depend on its \textit{local} neighboring atoms within a sphere, and the properties of the whole system could be calculated by summing up contributions from all atoms. 
Yet, this assumption may be violated in some scenarios where the long-range interactions are non-negligible. 
It is useful to verify the reliability of this assumption when applying DP models to SSE materials. 
Herein, we employed the locality test suggested by Bart\'ok \textit{et al.}~\cite{bartok2010gaussian}.
For each atom, the atom together with its neighbors within a predefined radius are fixed. Then a random perturbation is applied over other atoms outside the sphere.
The procedure is repeated several times to collect forces acting on the central atoms. 
The deviation of forces indicates the dependence of the atom's properties on its neighboring atoms, i.e., the locality of the system.
This test is performed with different cutoff radius (5.5\AA \textasciitilde 7.5\AA) and the procedure is run using a $2\times2\times2$ supercell to ensure all cells are at least twice as large as the cutoff distances. 

\subsection{DFT settings}
DFT calculations were performed using the projector augmented-wave (PAW) ~\cite{blochl1994projector} method applied in VASP 5.4.4~\cite{kresse1996efficient, kresse1999ultrasoft}. 
The convergence test of K-point sampling test showed that a dense reciprocal-space mesh (0.26 \AA$^{-1}$) are required to ensure that forces are converge to less than 1 meV/atom.
All single-point calculations were carried out with a 650 eV cutoff for plane-wave expansion and the criterion for electronic convergence was $10^{-6}$ eV. 

To select the functional for DP models, we optimized structures with different exchange-correlation functional settings (LDA~\cite{lda}, PBE~\cite{gga-pbe}, PBEsol~\cite{pbesol}, PBE with Van der Waals (vdW) correction, SCAN~\cite{sun2015scan}, and PBE0~\cite{pbe0-1, pbe0-2}). 

\begin{table}[hb]
\caption{Lattice parameters and unit cell volume of \ce{Li10GeP2S12}, \ce{Li10SiP2S12} and \ce{Li10SnP2S12}, relaxed with different functional settings (PBE, LDA, PBEsol, PBE with optB88-vdW).} 
\label{tbl:functional-lattice-parameter}
\begin{tabular}{crrrrr}
\hline
Structure & Method & a(\AA) & b(\AA) & c(\AA) & Volume(\AA$^3$) \\
\hline
\multirow{8}{*}{\ce{Li10GeP2S12}} & PBE &  8.591 & 8.879 & 12.977 & 989.1 \\
 & PBE+vdw & 8.525 & 8.822 & 12.878 & 967.9 \\
 & LDA & 8.312 & 8.656 & 12.482 & 897.0 \\
 & PBEsol & 8.503 & 8.811 & 12.760 & 955.1 \\ 
 & SCAN & 8.534 & 8.819 & 12.917 & 971.6 \\
 & PBE0 & 8.552 & 8.816 & 12.926 & 974.0 \\
 & Exp.\cite{kamaya2011lithium} & 8.6941 & 8.6941 & 12.5994 & 952.4 \\
 & Exp.\cite{C3EE41728J} & 8.7142 & 8.7142 & 12.6073 & 957.4 \\
\hline
\multirow{8}{*}{\ce{Li10SiP2S12}} & PBE & 8.774 & 8.774 & 12.599 & 970.0 \\
 & PBE+vdw & 8.700 & 8.700 & 12.490 & 945.5 \\
 & LDA & 8.534 & 8.534 & 12.144 & 884.3 \\
 & PBEsol & 8.696 & 8.696 & 12.368 & 935.3 \\
 & SCAN & 8.728 & 8.728 & 12.496 & 951.9\\
 & PBE0 & 8.722 & 8.722 & 12.518 & 952.3 \\
 & Exp.\cite{whiteley2014empowering}. & 8.6512 & 8.6512 & 12.5095 &  936.3 \\
\hline
\multirow{8}{*}{\ce{Li10SnP2S12}} & PBE &  8.835 & 8.835 & 12.882 & 1005.4 \\
 & PBE+vdw & 8.738 & 8.738 & 12.765 & 974.7 \\
 & LDA & 8.574 & 8.574 & 12.388 & 910.6 \\
 & PBEsol & 8.744 & 8.744 & 12.625 & 965.3 \\
 & SCAN & 8.774 & 8.774 & 12.759 & 982.3 \\
 & PBE0 & 8.766  & 8.766 & 12.785 & 982.5 \\
 & Exp.\cite{bron2013li10snp2s12} & 8.7057 & 8.7057 & 12.7389 & 965.5 \\
\hline
\end{tabular}
\end{table}

As shown in Table~\ref{tbl:functional-lattice-parameter}, all methods except LDA overestimate the lattice volume. 
The lattice volumes calculated by PBEsol agree well with the experimental data. 
By introducing vdW correction, the results of PBE were improved. 
Considering the widely adoption of the PBE functional in the investigation of SSE materials~\cite{ong2013phase} and the accuracy of PBEsol functional on the prediction of lattice parameters, we trained two sets of DP models (named as DP-PBE and DP-PBEsol) to benchmark the effect of the exchange-correlation functional.

\subsection{Molecular dynamics settings}

\begin{table}[!htb]
\caption{MD settings for the investigation of different factors, i.e. functional, simulation time and size, thermal expansion, configurational disorder are listed. The $N_{atom}$ means the number of atoms in the systems. }
\label{tbl:md-setting}
\begin{tabular}{ccccc}
\hline
Factor & Time (ns) & $N_{atom}$ & Temperature (K)\\
\hline
Density functional & 1 & 900 & \makecell[c]{400, 500, 666,\\ 800, 1000} \\
\hline
Simulation time and cell size & 10 & \makecell[c]{50, 200, 400,\\ 900, 1600} & \makecell[c]{300, 400, \\500, 600} \\
\hline
Thermal expansion & 1 & 900 & \makecell[c]{500, 666,\\ 800, 1000} \\
\hline
Configurational disorder & 1 & 900 & \makecell[c]{300, 400,\\ 500, 666} \\
\hline
\end{tabular}
\end{table}

LAMMPS\cite{plimpton1995fast} was employed to run all MD simulations. 
For each MD simulation, 4 DP models are used simultaneously to evaluate the model deviation ($\sigma_f^{max}$) of all snapshots in the trajectories. 
The tracer diffusion coefficient ($D_{tr}$) at each temperature is estimated by the time derivative of the mean-square displacement (MSD) of \ce{Li+} and the block-averaged method is adopted.
By default, calculations are run with experimental lattice parameters using NVT ensemble, and the timestep is 2~fs.
\textcolor{red}{The Nose-Hoover thermostat is applied and the relaxation time is set to 100 times of the timestep (200~fs). 
And simulations are run with in 3D cells including 900 atoms (\ce{Li360Ge36P72S192}).}
Detailed settings of the MD simulations are listed in Table~\ref{tbl:md-setting}.
As reported by the experimental work~\cite{Weber2016}, it would be reasonable to assume a constant thermal expansion coefficients ($3.4\times10^{-5}/K$) in the temperature range of 300K and 1000K. 
To the best of our knowledge, no relevant experimental data of the thermal expansion of \ce{Li10SiP2S12} and \ce{Li10SnP2S12} have been reported.
Here we assume the two systems have similar thermal expansion coefficients as \ce{Li10GeP2S12}. 
To study the finite-size effects, the unit cell of \ce{Li10GeP2S12}-type materials were scale to various sizes ($1\times1\times1$, $2\times2\times1$, $2\times2\times2$, $3\times3\times2$, $4\times4\times2$).
To study the effect of configurational disorder, 30 cation disordered structures are generated for the MD simulations. 
\textcolor{red}{For all MD simulations, the model deviations of snapshots are computed to ensure that energies and forces of all snapshots are reliable.}

\section{Results and Discussion}

\subsection{Locality test}
\begin{figure}[!htb]
\includegraphics[width=0.45\textwidth]{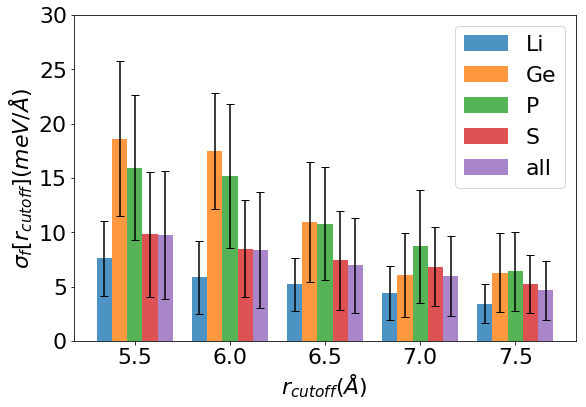}
\caption{\label{fig:locality} 
The locality test of \ce{Li10GeP2S12}. 
The averaged force deviation ($\sigma_f$) of each element, calculated with different cutoff radius are displayed.
The error bars correspond to the standard deviation of the force data.
}
\end{figure}

Results of locality test is illustrated in Fig.~\ref{fig:locality}.
The averaged deviations of forces are around 5 meV/\AA~to 10 meV/\AA, indicating that the assumption of the DP models is reliable for \ce{Li10GeP2S12}. 
The low deviations of forces acting on Li indicates the good ``locality'' of Li ions, and the motion of Li ions only depends on their closely neighboring atoms.
The relatively high deviations of \ce{Ge^{4+}} and \ce{P^{5+}} may due to their high charges. 
The long-range electrostatic forces may affect the forces on Ge and P atoms.
With the increasing of cutoff radius, the local environment of an atom is better described, leading to lower deviation of forces. 
Yet, the DP models need to trade off between accuracy and computational cost. 
In the \ce{Li10GeP2S12}-type materials, we found that 6\AA~would be sufficiently accurate to describe the diffusion process, which is adopted in this work as follows.
Thus, it would be useful to incorporate long-range effect in the machine learning potentials to improve the description of the interatomic interaction.
In conclusion, the locality test shows that the end-to-end Deep Potential models  would be accurate enough to investigate the diffusion of Li in \ce{Li10GeP2S12}-type materials.

\subsection{DP-GEN iteration}
\begin{table*}[!htb]
\caption{Percentage of accurate, candidate and failed configurations in each iteration. Data are labeled with PBE functional.}
\label{tbl:dpgen-iteration}
\begin{tabular}{ccrrrrrrrrrrrrrrrrrrr}
\hline
Type & 1 & 2 & 3 & 4 & 5 & 6 & 7 & 8 & 9 & 10 & 11 & 12 & 13 & 14 & 15 & 16 & 17 & 18 & 19 \\
\hline
candidate & 36.57 & 7.15 & 1.40 & 0.78 & 6.11 & 7.95 & 2.63 & 3.76 & 0.23 & 0.13 & 0.15 & 0.03 & 0.67 & 0.83 & 1.10 & 1.28 & 4.47 & 0.71 & 0.95 \\ 
accurate & 16.14 & 81.57 & 95.95 & 98.74 & 92.62 & 91.41 & 97.08 & 96.10 & 99.75 & 99.86 & 99.84 & 99.96 & 99.31 & 99.07 & 98.85 & 98.69 & 95.46 & 99.19 & 99.03 \\
failed & 47.28 & 11.26 &  2.64 &  0.47 &  1.26 &  0.63 &  0.27 &  0.12 &  0.00 &  0.00 &  0.00 &  0.00 &  0.01 &  0.08 &  0.03 &  0.02 & 0.08 & 0.11 & 0.02 \\
\hline
\end{tabular}
\end{table*}

\begin{figure}[!htb]
\includegraphics[width=0.45\textwidth]{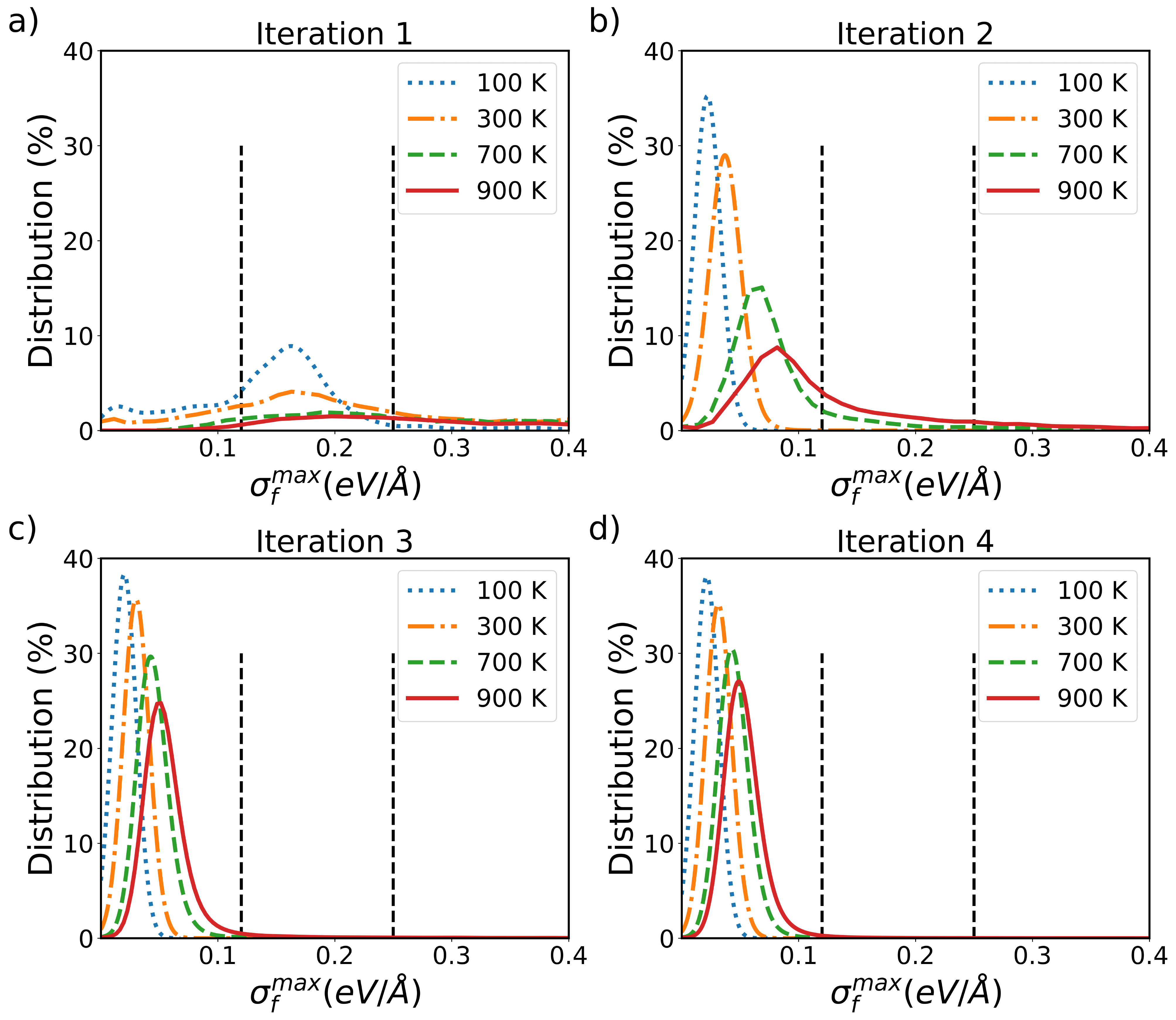}
\caption{\label{fig:deviation} 
For \ce{Li10GeP2S12}, distribution of maximum deviation of force ($\sigma_f^{max}$) from iteration 1 to 4. 
Distribution of deviation values at 4 temperatures are plotted and the two vertical lines (dashed) correspond to the lower and upper bound of the selection criteria (0.12 and 0.25 eV/\AA).
}
\end{figure}

To better illustrate the DP-GEN procedure, it is worth taking a thorough look at the exploration results of each iteration. 
Here, we study \ce{Li10GeP2S12} in depth as benchmark due to its importance and extensive previous work. 

Fig.~\ref{fig:deviation} shows the distribution of $\sigma_f^{max}$  at different temperatures in the first 4 iterations. 
And Table~\ref{tbl:dpgen-iteration} displays the percentage of the accurate, candidate and failed groups of configurations in each iteration. 
In the 1st iteration, it is not surprising that the trajectories given by the preliminary models include lots of unreasonable configurations and high-temperature simulations blow up very quickly. 
A large fraction of the snapshots sampled in this iteration have a $ \sigma_f^{max} $ larger than 0.4 eV/\AA~(Fig.~\ref{fig:deviation} (a)).
A large portion of the candidates with $ \sigma_f^{max} $ fallen in the selection range, 
selected for labelling are from low-temperature simulations.
This situation is drastically improved after just adding 300 labeled configurations to the training dataset. 
In the 2nd iteration, most low-temperature snapshots are labeled as ``accurate" and the majority of newly selected snapshots come from higher-temperature simulations.
Going from the 2nd iteration to the 3rd and the 4th, although the time duration of the simulation is extended (i.e. 1000~fs, 5000~fs, and 10000~fs, respectively), most snapshots have their $\sigma_f^{max}$ value at a satisfactory level, demonstrating a quick convergence of the DP-GEN process.
After 4 iterations, the models have converged in the original cell (50 atoms), i.e. the percentage of candidates being $ \sim $ 1~\%.

The 5th to 8th iterations are performed with $2\times 2 \times 2$ supercells (200 atoms) with the percentage of candidates gradually decreased to 0.6~\%.
Then the exploration moves to disorder structures from the 9th to the 16th iterations. 
Due to the similarity of the structures, only a small percentage of new configurations are labeled as ``candidate'' in these iterations. 
In the last 3 iterations, the exploration is run with NVT using the experimental lattice parameters, the same setting we use in the calculation of diffusion coefficients. 
Still a few candidates arise in these iterations, yet the DP models converge within 3 iterations. 
Similar trends are found in \ce{Li10SiP2S12} and \ce{Li10SnP2S12} systems.

Finally, around 4000 configurations in total are collected via DP-GEN automatically to train the DP models. 
The protocol could be further improved, e.g. by merging the exploration process of ordered and disordered structures, or train the universal DP models for ``Li-Si\/Ge\/Sn-PS" system.

\subsection{Accuracy and speed test of DP models}
\begin{table}[!h]
\caption{Root-mean square errors of the energies per atom (meV/atom) and forces (meV/\AA) of the DP-PBE and DP-PBEsol on the whole dataset generated from DP-GEN scheme. 
The standard deviations are evaluated using an ensemble of 4 models. 
}
\label{tbl:accuracy}
\begin{tabular}{ccccc}
\hline
Model & RMSEs &  \ce{Li10GeP2S12} & \ce{Li10SiP2S12} & \ce{Li10SnP2S12} \\
\hline
\multirow{2}{*}{DP-PBE} & $E$ & 1.65$\pm$0.03 & 1.82 $\pm$ 0.01 & 2.53 $\pm$ 0.02  \\
& $f$ & 82.4$\pm$0.91 & 82.7$\pm$0.19 & 92.5$\pm$0.39\\
\multirow{2}{*}{DP-PBEsol} & $E$ & 1.33$\pm$0.06 & 1.33 $\pm$ 0.01 & 1.27 $\pm$ 0.01  \\
& $f$ & 79.6$\pm$1.57 & 77.7$\pm$0.39 & 77.9 $\pm$ 0.29\\
\hline
\end{tabular}
\end{table}

\begin{figure*}[!htb]
\includegraphics[width=0.8\textwidth]{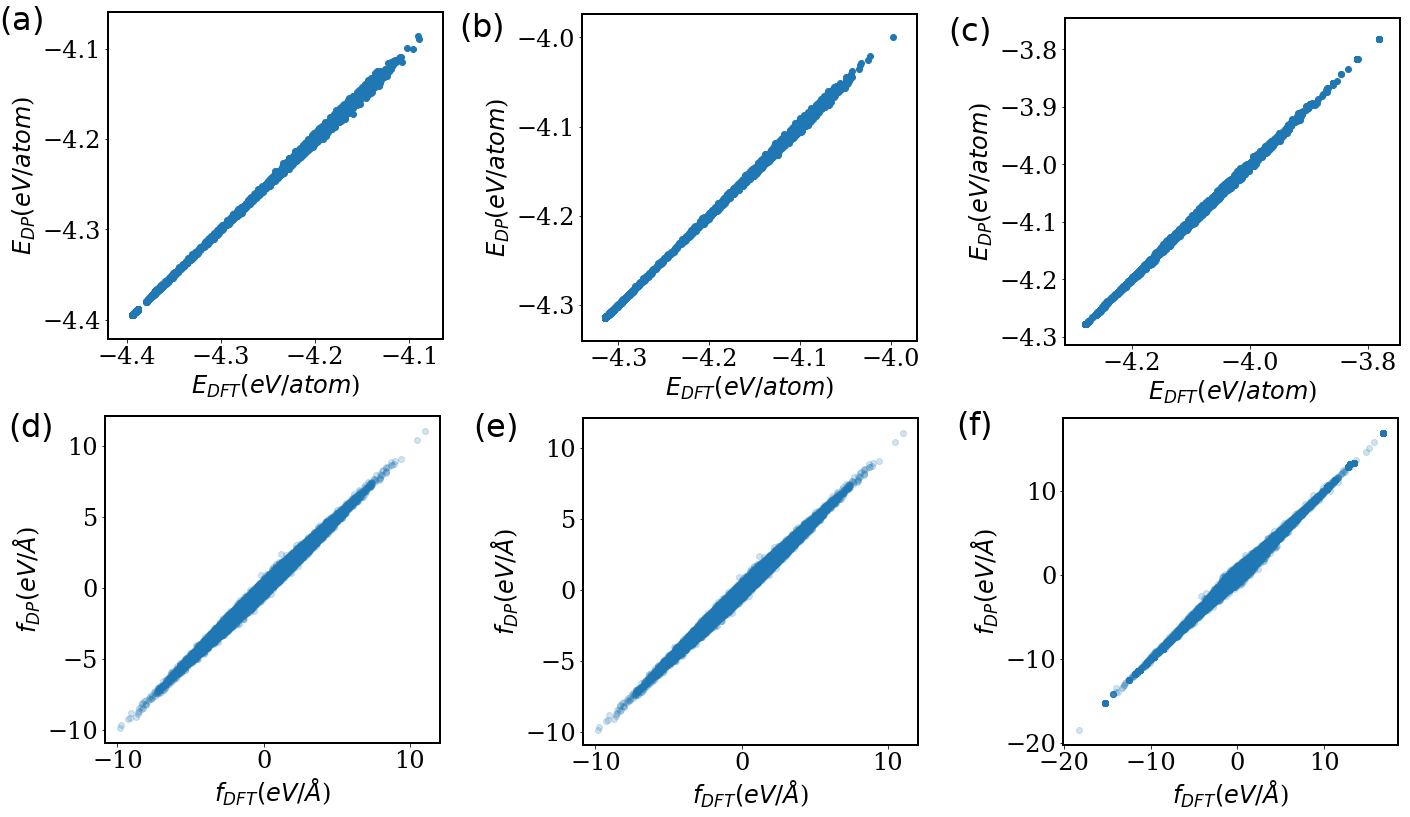}
\caption{\label{fig:accuracy} 
The energy and force computed by DP models and DFT of the 3 systems \ce{Li10SiP2S12} (a, d), \ce{Li10GeP2S12} (b, e) and \ce{Li10SnP2S12} (c, f), respectively.}
\end{figure*}

To ensure how much the DP models could speed up the simulation without losing DFT-level accuracy, the accuracy and speed test are performed on the whole DFT dataset.
As shown in Table~\ref{tbl:accuracy}, the root-mean-square errors (RMSEs) of energies per atom and forces are around 2 meV/atom and 80 meV/\AA~for two sets of models using PBE and PBEsol functional, and they have similar accuracy.
\textcolor{red}{The diagnostic plots between DFT and DP models are shown in Fig.~\ref{fig:accuracy}.}
The low standard deviations of the 4 models suggests that all potentials share similar accuracy and the DP-GEN scheme gives consistent errors for 3 different systems.

\begin{figure}[htb]
\includegraphics[width=0.4\textwidth]{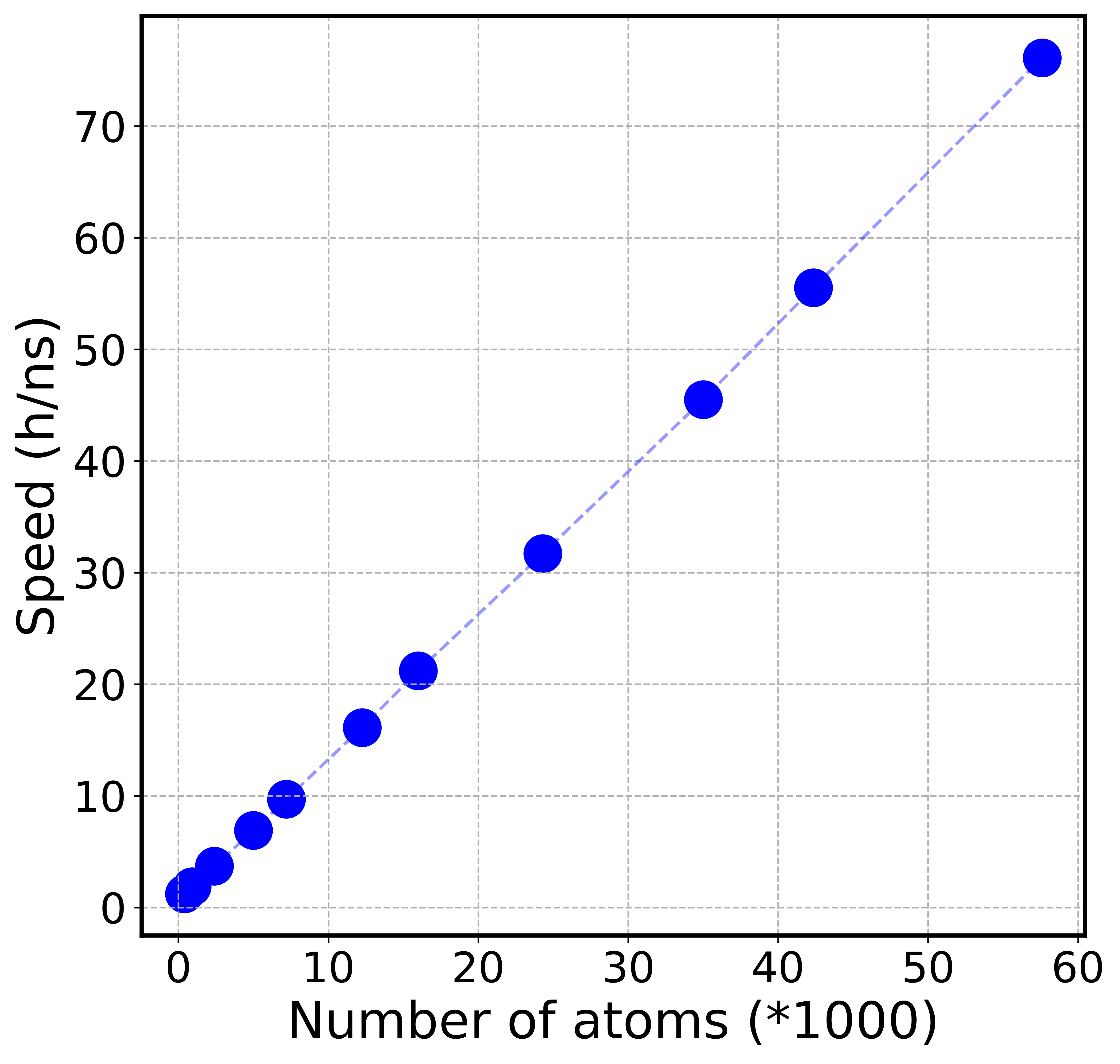}
\caption{\label{fig:speed} 
Speed test of DP models on a NVIDIA V100 GPU. 
The system (\ce{Li10GeP2S12}) are scaled to different sizes to test the required simulation time.}
\end{figure}

The speed test is run on one NVIDIA V100 GPU and the results are reported in Fig.~\ref{fig:speed}.
It only takes around 4 hours to simulate a 900-atom systems for 1~ns and the computational cost of DP scales nearly linearly with the system size.
The high accuracy and extraordinary speed make Deep Potential a powerful tool for large-scale atomic simulation.

\subsection{Simulation protocol for diffusion coefficients}
\subsubsection{Effect of finite-size and simulation time}
\begin{figure}[!htb]
\includegraphics[width=0.4\textwidth]{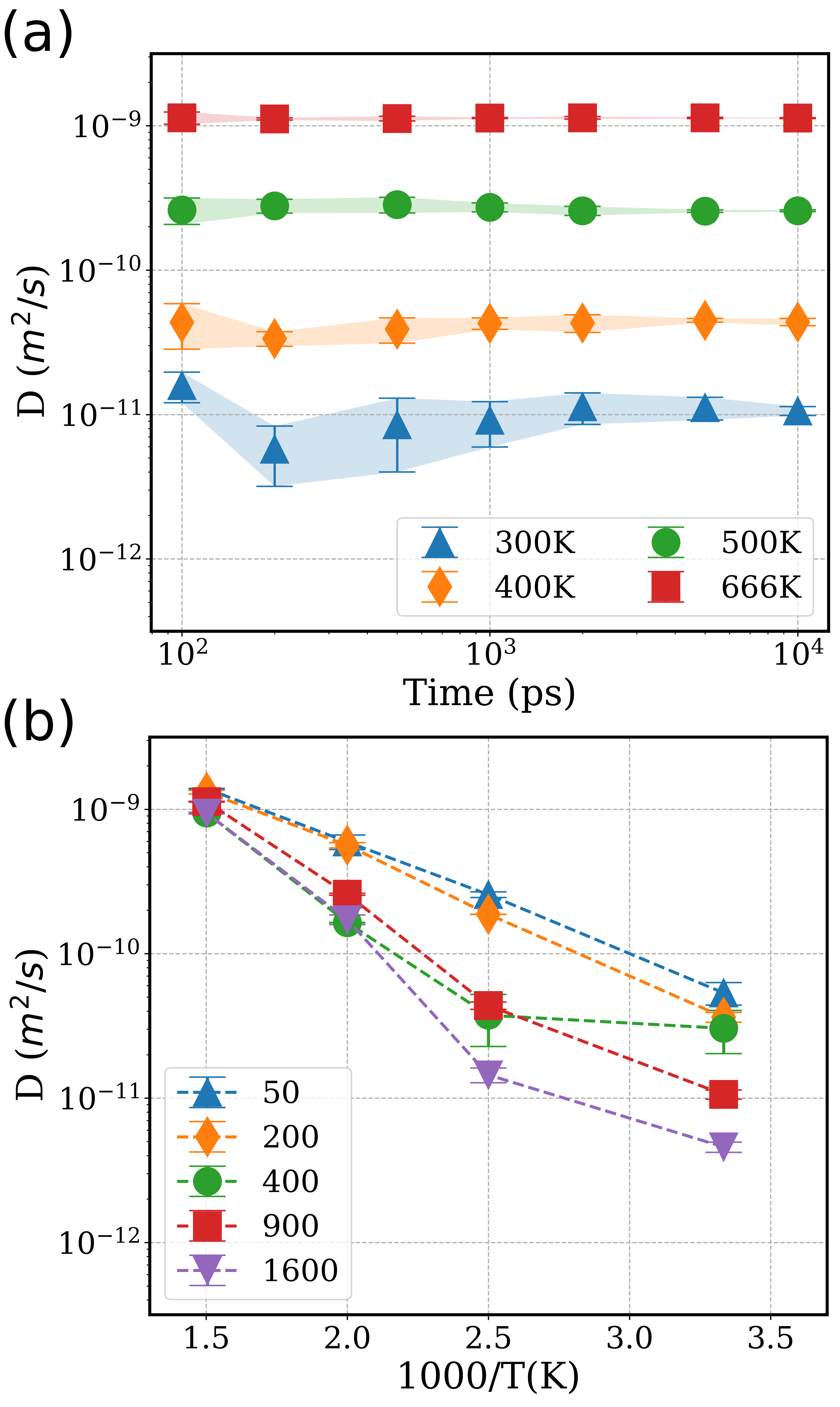}
\caption{\label{fig:size-time}
The convergence test of diffusion coefficients for \ce{Li10SiP2S12}. 
Data are computed with different (a) simulation time lengths (100~ps, 200~ps, 500~ps, 1~ns, 2~ns, 5~ns and 10~ns) and (b) system sizes (50, 200, 400, 900, and 1600 atoms).
}
\end{figure}
Previous studies~\cite{he2018statistical} based on AIMD have suggested that a 200~ps MD simulation would be sufficient to ensure the convergence of diffusivity at high temperatures (\textgreater 600K) and this is confirmed in Fig.~\ref{fig:size-time}(a).
Diffusion coefficients above the level of $10^{-10}$m${}^2$/s (400~K and 500~K) reach very small variances and converge within 1~ns.
Since the diffusivity decreases exponentially with temperature, the statistics of diffusion processes at room temperature require longer time to converge.
At 300~K, extending the simulation to 10~ns ensures convergence of all diffusivity data with an uncertainty of $10^{-12}$ m${}^2$/s.
Thus, the time scale of 10~ns is required for the simulation of room temperature diffusion processes.

The system-size dependence of the diffusion coefficient and viscosity from MD simulations with periodic boundary conditions is a classic topic and has been extensively discussed by, e.g., Yeh et. al~\cite{yeh2004system}.
Following the test of simulation time, analysis of the size effect is performed with 10~ns trajectories. 
Here, as shown in Fig.~\ref{fig:size-time}(b), a $2\times2\times1$ supercell size (200 atoms), which was used in most previous AIMD simulations of SSE materials~\cite{Mo2012,he2017origin}, overestimates diffusion coefficients by 10 to 100 times.
The diffusion coefficients start to converge when systems are enlarged to $2\times2\times2$ supercells (400 atoms).
Yet, this system still significantly overestimate diffusion coefficients at 300K.
By expanding the system to 900 atoms and 1600 atoms, we notice that the difference of diffusion coefficients between them is around $3\times10^{-12} m^2/s$.
Taking into account the results of the convergence test and speed test, we will run simulations with 900-atom systems for 1~ns for high temperatures (\textgreater 400K), and the simulation is lengthened to 10~ns when studying the diffusion process at room temperature.
This setting may still slightly overestimate diffusion coefficient at 300K but it should give correct magnitude of diffusion coefficients.

\subsubsection{Effect of density functional on diffusion coefficient}
\begin{figure}[!htb]
\includegraphics[width=0.4\textwidth]{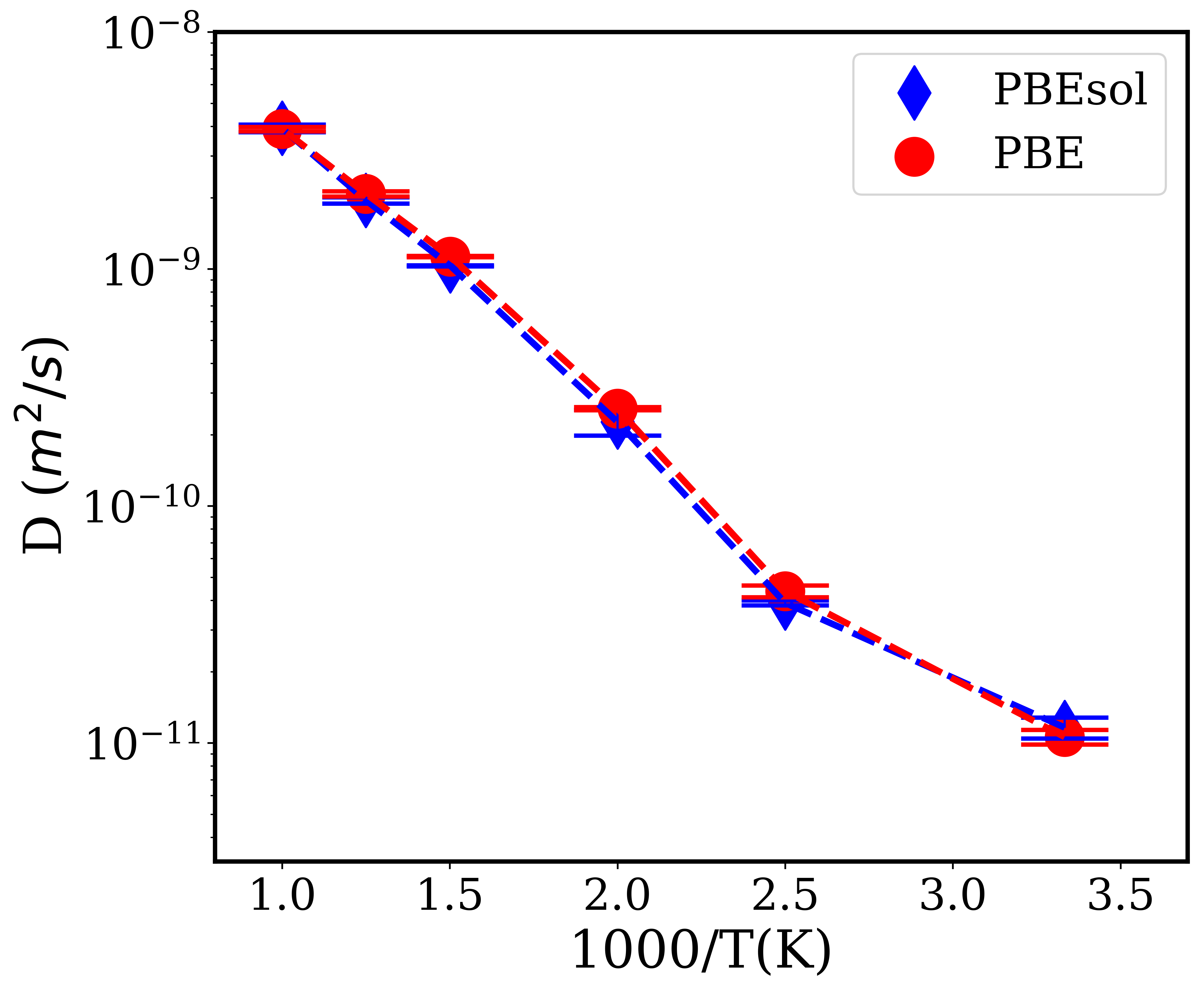}
\caption{\label{fig:functional} 
Diffusion coefficients of \ce{Li10SiP2S12}, simulated with DP-PBE and DP-PBEsol.}
\end{figure}
The quality of the interatomic potential depends on the DFT data. 
To the best of our knowledge, though it is suggested that the interatomic potentials labeled by different functionals may give different diffusion coefficients~\cite{Aris2019}, there has not been any relevant benchmark report. 
The diffusion coefficients calculated by DP-PBE and DP-PBEsol models, are displayed in Fig.~\ref{fig:functional}.
Simulated with the experimental lattice parameters, both sets of DP models give similar diffusion coefficients.
The consistency of the diffusion coefficients implies that though different methods give different lattice parameters (the lattice parameters computed by PBE are 3\%~larger than that by PBEsol), they do not significantly affect the calculated diffusion coefficients. 

\subsubsection{Effect of thermal expansion}

\begin{figure}[!htb]
\includegraphics[width=0.45\textwidth]{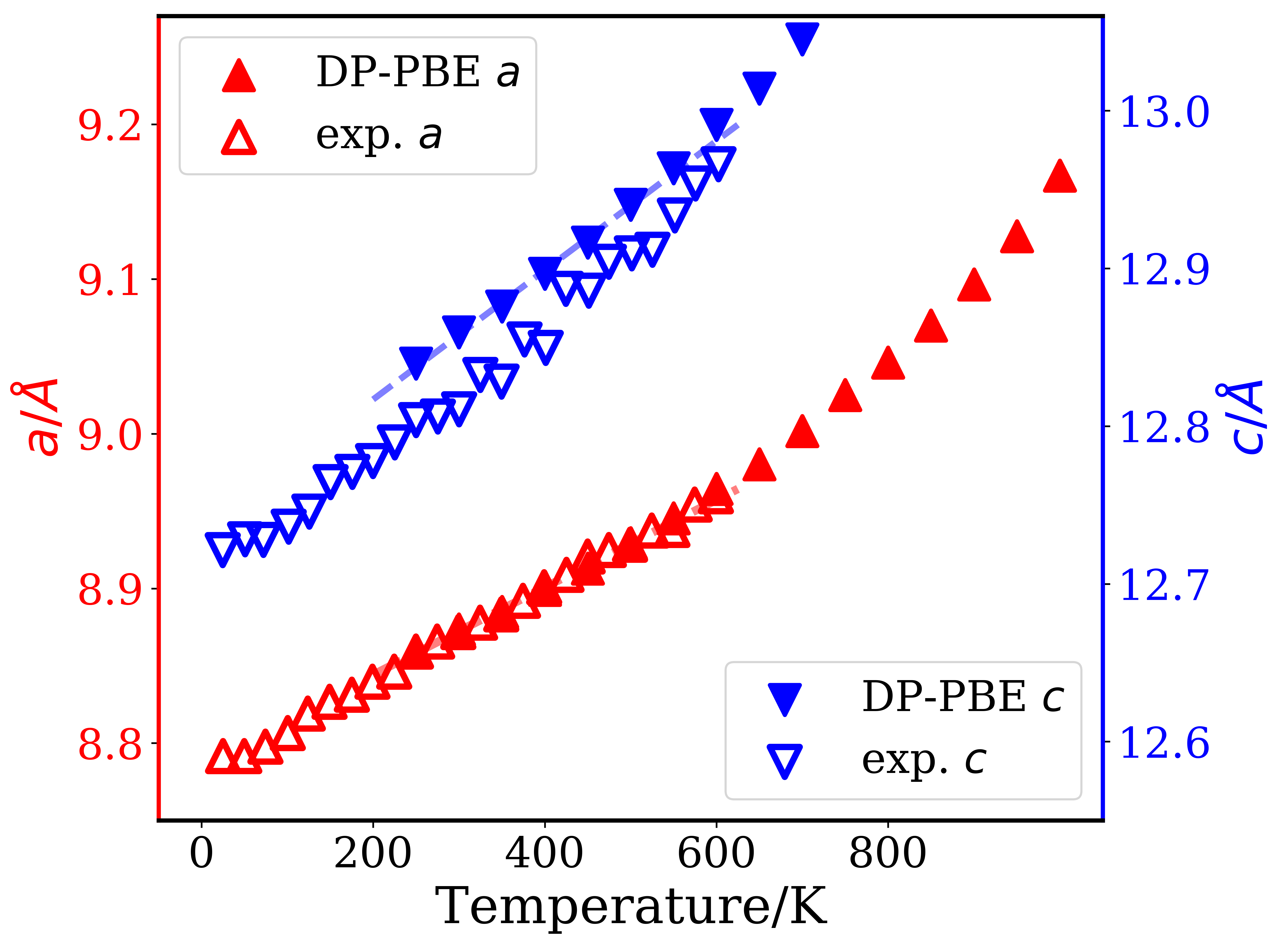}
\caption{\label{fig:NpT} 
Thermal expansion of \ce{Li10GeP2S12}. The lattice parameters \textit{a} and \textit{c} are shown.
The experimental data is extracted from Weiber \textit{et al.}~\cite{Weber2016}
The lattice parameters at different temperatures are estimated by the DP-PBE models.
The dashed lines is the fitting range of thermal expansion coefficients. 
}
\end{figure}

\begin{figure}[!htb]
\includegraphics[width=0.4\textwidth]{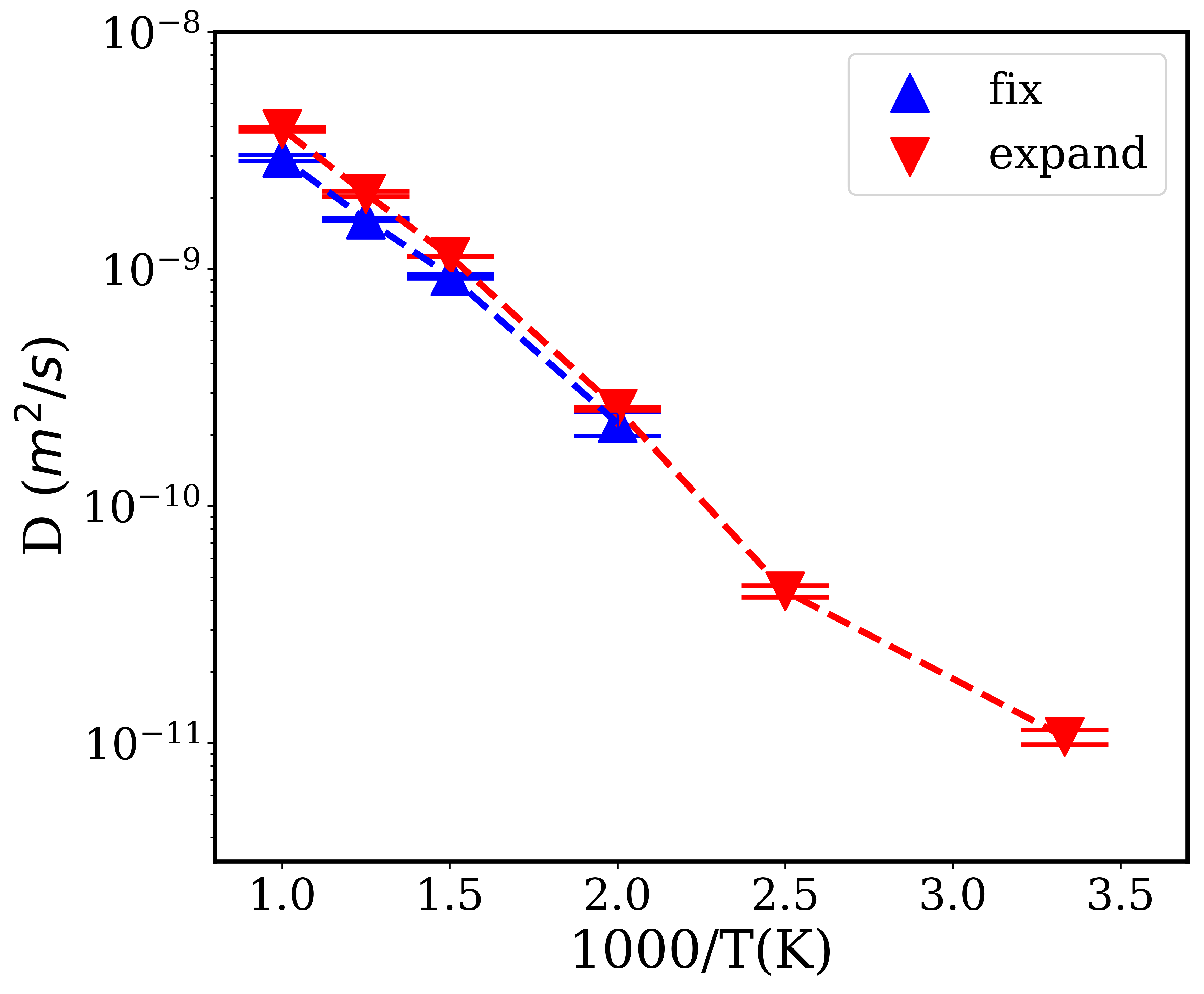}
\caption{\label{fig:thermal-expansion} 
Diffusion coefficients of \ce{Li10SiP2S12}, the `fix' and `expand' data correspond to simulations at different temperatures with fixed lattice parameters and lattice parameters assuming linear thermal expansion, respectively.}
\end{figure}

Lattice parameter is known to significantly affect the diffusion process~\cite{Ong2013,shao2016CR, C9TA05248H}. 
Ong \textit{et al.}~\cite{Ong2013} have reported the effect of expansion or contraction of lattice on calculated diffusion coefficients at the same temperature.
The thermal expansion is relatively small at low temperatures (less than 1\%), thus we only evaluate the effect at the high temperatures regime(\textgreater 700K).
It is shown in Fig.~\ref{fig:thermal-expansion} that at high temperatures, the thermal expansion already leads to noticeable differences in the computed diffusion coefficients.
Thus, the effect of thermal expansion should be considered at the whole temperature range.

To set up an automatic workflow, it is worth exploring whether DP models could predict the thermal expansion coefficients.
To obtain the lattice parameters at different temperatures, we equilibrated the simulation cell with N$p$T ensemble for 1~ns at different temperatures and average all configurations of trajectories to calculate the lattice parameters. 
The \ce{Li10GeP2S12}-type materials belong to tetragonal crystal lattice, of which the lattice parameter \textit{a} is equivalent to lattice parameter \textit{b}, and thus the thermal expansion of lattice parameter {a} and {c} are presented in Fig.~\ref{fig:NpT}.
Weber \textit{et al.}~\cite{Weber2016} found that lattice parameters \textit{a} and \textit{c} exhibit linear thermal expansion below 700~K and show slightly anisotropic expansion at higher temperature.
Here we focus on the linear expansion region that is of practical interest. 
The thermal expansion coefficient ($\alpha^L_{300K}$) calculated by DP-PBE and DP-PBEsol are both $3.2 \times 10^{-5}$ K${}^{-1}$, consistent with the value $3.5 \times 10^{-5}$ K${}^{-1}$ from experiment~\cite{Weber2016}. 
Thus we conclude that the N$p$T simulations could be adopted in the protocol to estimate thermal expansion coefficients of SSE materials for the accurate simulation of diffusion processes.

\subsubsection{Effect of configurational disorder}
\begin{figure}[!htb]
\includegraphics[width=0.4\textwidth]{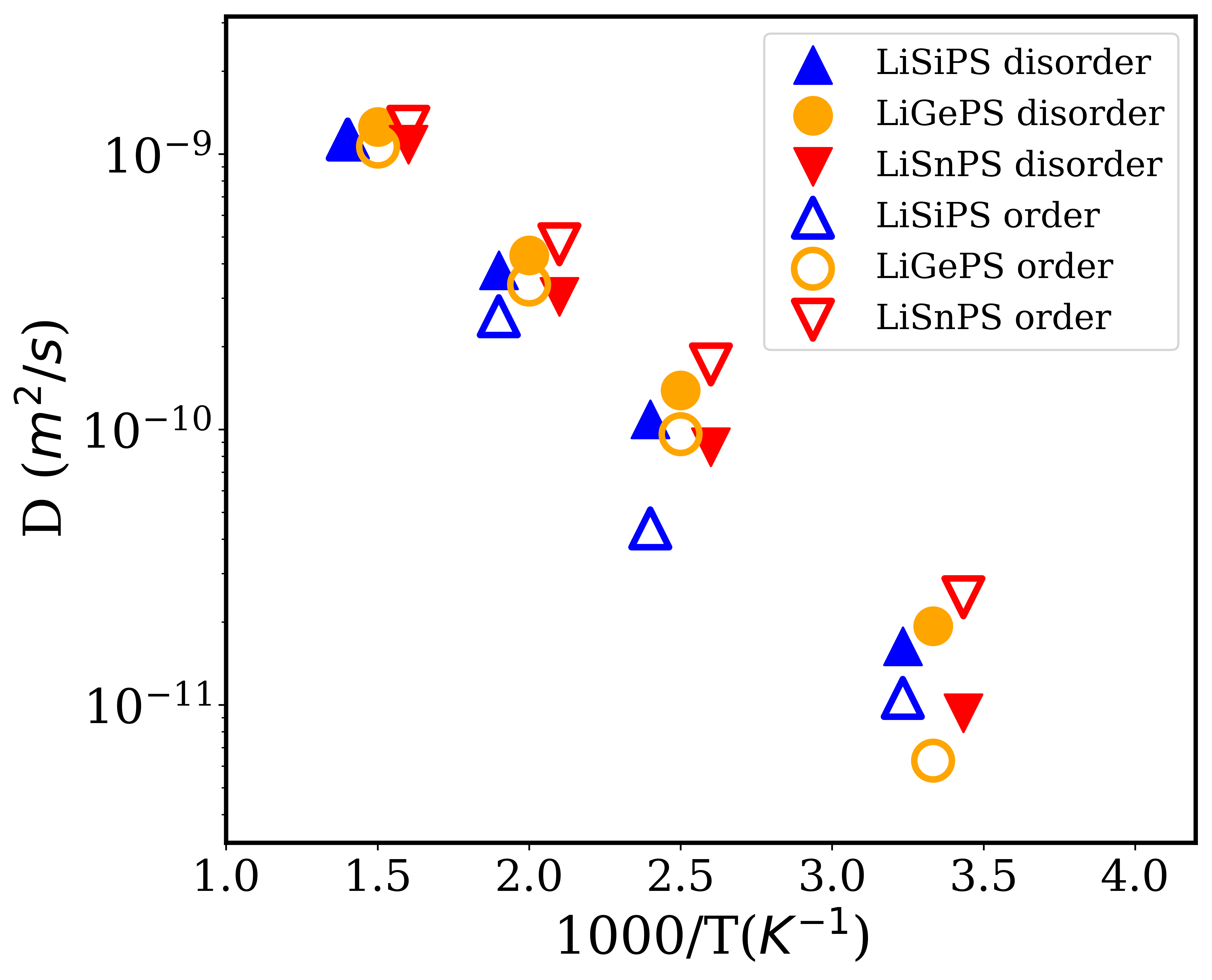}
\caption{\label{fig:order-disorder} 
Diffusion coefficients of \ce{Li10GeP2S12}, \ce{Li10SiP2S12} and \ce{Li10SnP2S12}, averaged from cation site ordered/disordered configurations.
Note that the set of temperatures studied are the same for the three materials and the data of \ce{Li10SiP2S12} and \ce{Li10SnP2S12} are slightly shifted to left and right for better visualization, respectively.
}
\end{figure}

\begin{figure*}[!htb]
\includegraphics[width=0.95\textwidth]{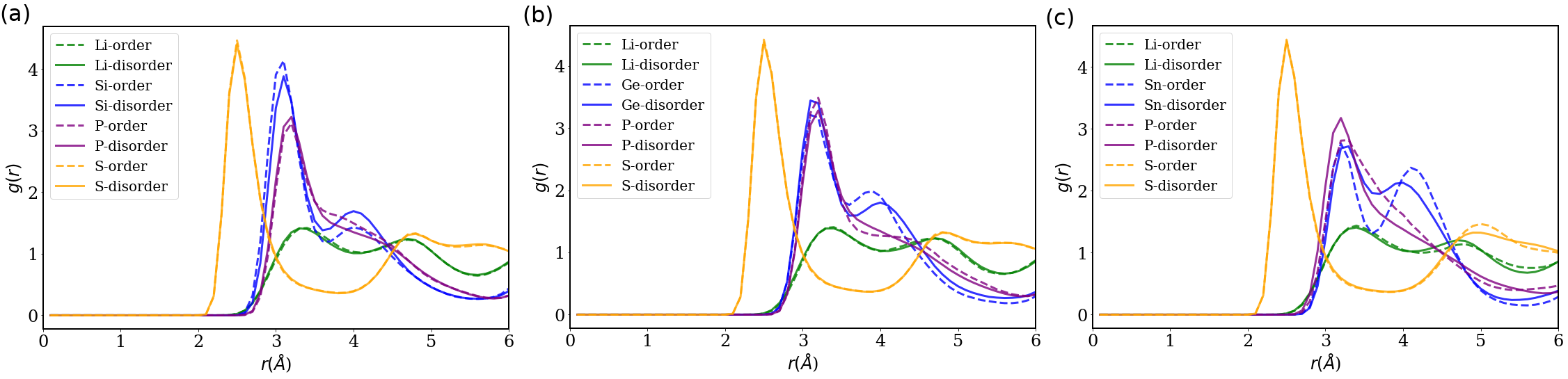}
\caption{\label{fig:rdf} 
Radial distribution function of Li and other ions in \ce{Li10SiP2S12} (a), \ce{Li10GeP2S12}(b) and \ce{Li10SnP2S12} (c) are shown.
RDF data of ordered and disordered configurations are averaged from 5 trajectories.
The dashed and solid lines correspond to ordered and disordered structures, respectively.
}
\end{figure*}

It has been suggested that the disordered arrangement in SSE materials may improve diffusion coefficients~\cite{stamminger2019ionic, ohno2019further, zhou2019entropically, hanghofer2019substitutional}.
Yet, an atomic insight based on sufficient MD simulations is still lacking. 
To sample cation disordered configurations, a large simulation system is required, which is beyond the ability of AIMD. 
By including the disordered configurations in DP-GEN iterations, our DP models are able to accurately simulate disordered configurations. 
In this test, 30 disordered configurations with 900 atoms (\ce{Li360M36P72S432}, M=Si, Ge, Sn) are randomly generated to compute diffusion coefficients.

In Fig.~\ref{fig:order-disorder}, we plot diffusion coefficients calculated with cation site disordered structures. 
At low temperatures, the disorder of cations (\ce{Ge^{4+}} and \ce{P^{5+}}) increases the diffusion coefficients by 2$\sim$4 times. 
The diffusion process of Li in the \ce{Li10GeP2S12}-type materials could be described as the jumping events of Li between several connected sites, each with unique local environment.
A simplified model to explain the effect of disorder is that the disordered arrangement `flatten' the potential energy surface between these sites~\cite{ohno2019further}, i.e. decrease the possible maximum energy barriers. 
Due to the speed-up of the rate-determining step, the jumping process is enhanced, leading to the improvement of diffusion coefficients.
The effect is not very significant at high temperatures because the benefit is relatively small in the systems with high diffusion coefficients.

Surprisingly, \ce{Li10GeP2S12} and \ce{Li10SiP2S12} benefit from the disorder while configurational disorder decreases diffusion coefficients of \ce{Li10SnP2S12}. 
These data may explain the fact that though \ce{Li10SnP2S12} has the largest volume among the three materials, the configurational disorder of this material decreases its diffusion coefficients.  

Besides, it is interesting that diffusion coefficients of the disordered systems seem to follow the Arrhenius relationship at low temperatures while the ordered systems do not.
According to the classification of the 3 kinds of diffusion behavior mentioned in Fig.~\ref{fig:background}, one possible reason is that the ordered structures may undergo phase transition when they are cooled down.
Another possible reason is that the diffusion mechanism may change, e.g. the diffusion path across \textit{ab} plane is shut.

\textcolor{red}{To investigate the difference between ordered and disordered structures, we analyze the radial distribution function (RDF) of the 3 systems and the results are plotted in Fig~\ref{fig:rdf}.
For all 3 systems, configurational disorder does not change the RDF of Li-Li and Li-S. 
And the RDF of Li-P (M=Si/Ge) is slightly changed.
Compared to \ce{Li10SiP2S12} and \ce{Li10GeP2S12}, the RDF of Li-M (M=Si/Ge/Sn) of \ce{Li10SnP2S12} significantly changed due to the disorder. 
The rearrangement of Sn and P sites may change the local environment of Li sites in \ce{Li10SnP2S12}, leading to significant energy differences among Li sites, which results in the suppression of diffusion processes.
Besides, for ordered and disordered systems, the RDF of Li-P and Li-Ge in \ce{Li10GeP12S12} are quite similar, meaning that the Li sites may have smaller energy differences than the other two systems, which is benefit to the diffusion process.
}
The physics of \ce{Li10SnP2S12} and the possible different transition behaviors between ordered and disordered structures require future analysis of the diffusion process, including the statistics of jumping events between local environments.

\subsection{Comparison to experiment}
\begin{figure*}[!htb]
\includegraphics[width=0.95\textwidth]{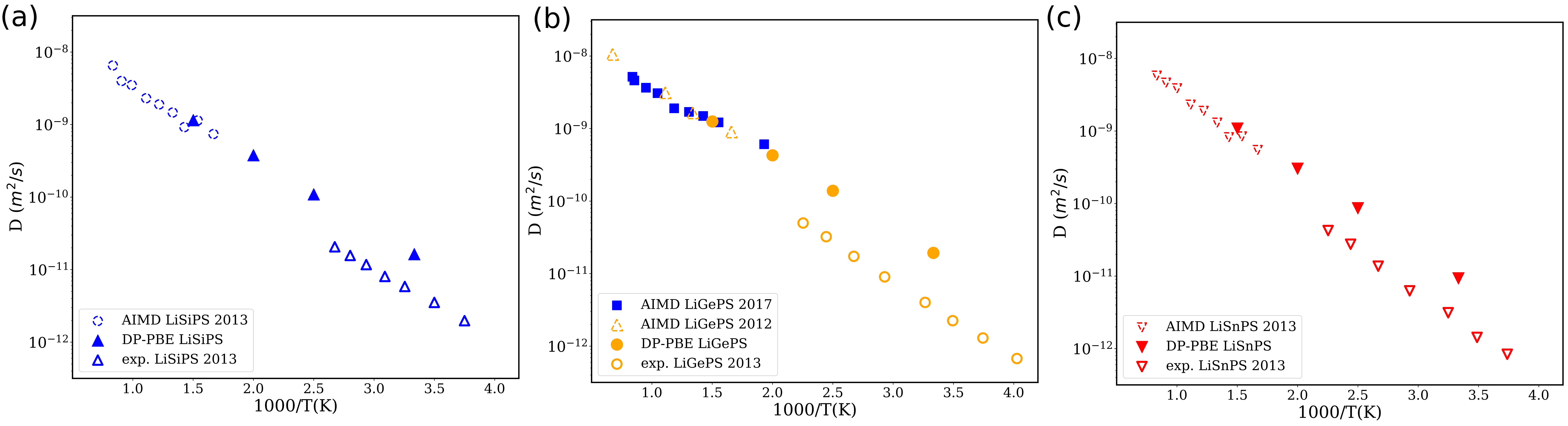}
\caption{\label{fig:dp-exp-aimd} 
Diffusion coefficients of the \ce{Li10SiP2S12}(a), \ce{Li10GeP2S12}(b), and \ce{Li10SnP2S12}(c), obtained from our DP-PBE models, previous AIMD simulations~\cite{Mo2012, marcolongo2017ionic}, and experimental solid-state NMR results~\cite{Kuhn2013}.
}
\end{figure*}
Based on the above investigation, we concluded that the effect of thermal expansion and configurational disorder shall be considered to obtain accurate simulation results.
Our results, data of previous AIMD simulations, and experiments are plotted in Fig.~\ref{fig:dp-exp-aimd} 
The computed diffusion coefficients of the 3 systems (\ce{Li10SiP2S12}, \ce{Li10GeP2S12} and \ce{Li10SnP2S12}) are $16\pm2 \times 10^{-12} m^2/s$, $19\pm2 \times 10^{-12}$, $9\pm2 \times 10^{-12}$.
And the experimental data~\cite{Kuhn2013} are $5.8\pm2 \times 10^{-12} m^2/s$, $4.1\pm2 \times 10^{-12} m^2/s$, and $3.1\pm2 \times 10^{-12} m^2/s$ respectively.
The DP models overestimate the diffusion coefficients at room temperature for around 5\textasciitilde15$\times 10^{-12} m^2/s$.

The difference between experiments and simulation data computed by DP models may due to the following reasons:
First, according to the \textit{interlaboratory reproducibility} study by Zeier and the collaborators~\cite{zeier2020centainty}, the experimental reported ionic conductivities have a uncertainty up to 4.5 mS/cm (corresponding to an uncertainty of 1\textasciitilde5$\times 10^{-12} m^2/s$ for \ce{Li10GeP2S12}-type materials)), which is similar to the overestimation given by the DP models.
The uncertainty of experimental measurement may due to the synthesize settings and the impurity of materials~\cite{bron2013li10snp2s12, whiteley2014empowering}, which usually decrease the diffusion coefficients.
Second, as shown by the test of finite-size effect, we would expect an slightly overestimation of our diffusion coefficients which are sampled by 900-atom systems.
Third, the models trained in this work could still be improved in several aspects, e.g. by including the long-range interactions, trading off between accuracy and computational cost. 
Despite these, our DP models presented in this work achieve state-of-the-art performance on the simulation of the diffusion processes of \ce{Li10GeP2S12}-type materials.

\section{Conclusion}
In this work, we present a systematic benchmark study for the generation, validation and application of DP-GEN for \ce{Li10GeP2S12}-type materials. 
We provide an efficient automated protocol to generate DP models and the key properties (accuracy, locality, thermal expansion and diffusion coefficient) of the DP models are examined.
With the DP models, we establish a reliable protocol to compute the diffusion coefficients in a wide temperature range (300 K \textasciitilde 1000 K).
The results show that current protocols based on AIMD simulation significantly overestimate diffusion coefficients due to finite-size effect and the ignorance of configurational disorder.
Though, our computed diffusion coefficients still slightly overestimate the diffusion coefficients, the errors are within experimental uncertainty.
The protocol should be sufficiently accurate to be applied to run simulations to understand the diffusion mechanisms of SSE materials.
With the verified protocols, we notice the effect of configurational disorder may depend on the materials, i.e. \ce{Li10SnP2S12} show different behavior from \ce{Li10GeP2S12} and \ce{Li10SiP2S12}.
The data generated in this work could be a starting point for research of doping and replacement of \ce{Li10GeP2S12}-tpye materials in the future.

\begin{acknowledgments}
J.-X. H. and J.C. are grateful for funding support from the National Natural Science Foundation of China (Grants No, 21861132015, 21991150, 21991151, 91745103, 22021001).
J.-X. H. and J.-B. Z. are supported by the National Key Research and Development Program of China (Grant No. 2017YFB0102000).  
The work of L. Z. and W. E was supported in part by a gift from iFlytek to Princeton University, the ONR grant N00014-13-1-0338, and the Center Chemistry in Solution and at Interfaces (CSI) funded by the DOE Award DE-SC0019394.
The work of H. W. was supported by the National Science Foundation of China under Grant No. 11871110, the National Key Research and Development Program of China under Grants No. 2016YFB0201200 and No. 2016YFB0201203, and Beijing Academy of Artificial Intelligence (BAAI). 
\end{acknowledgments}

\section{Data availability}
All DP models and the labeled DFT data could be downloaded from \url{http://dplibrary.deepmd.net/}.

\nocite{*}
\bibliography{aipsamp}

\begin{thebibliography}{97}%
\makeatletter
\providecommand \@ifxundefined [1]{%
 \@ifx{#1\undefined}
}%
\providecommand \@ifnum [1]{%
 \ifnum #1\expandafter \@firstoftwo
 \else \expandafter \@secondoftwo
 \fi
}%
\providecommand \@ifx [1]{%
 \ifx #1\expandafter \@firstoftwo
 \else \expandafter \@secondoftwo
 \fi
}%
\providecommand \natexlab [1]{#1}%
\providecommand \enquote  [1]{``#1''}%
\providecommand \bibnamefont  [1]{#1}%
\providecommand \bibfnamefont [1]{#1}%
\providecommand \citenamefont [1]{#1}%
\providecommand \href@noop [0]{\@secondoftwo}%
\providecommand \href [0]{\begingroup \@sanitize@url \@href}%
\providecommand \@href[1]{\@@startlink{#1}\@@href}%
\providecommand \@@href[1]{\endgroup#1\@@endlink}%
\providecommand \@sanitize@url [0]{\catcode `\\12\catcode `\$12\catcode
  `\&12\catcode `\#12\catcode `\^12\catcode `\_12\catcode `\%12\relax}%
\providecommand \@@startlink[1]{}%
\providecommand \@@endlink[0]{}%
\providecommand \url  [0]{\begingroup\@sanitize@url \@url }%
\providecommand \@url [1]{\endgroup\@href {#1}{\urlprefix }}%
\providecommand \urlprefix  [0]{URL }%
\providecommand \Eprint [0]{\href }%
\providecommand \doibase [0]{http://dx.doi.org/}%
\providecommand \selectlanguage [0]{\@gobble}%
\providecommand \bibinfo  [0]{\@secondoftwo}%
\providecommand \bibfield  [0]{\@secondoftwo}%
\providecommand \translation [1]{[#1]}%
\providecommand \BibitemOpen [0]{}%
\providecommand \bibitemStop [0]{}%
\providecommand \bibitemNoStop [0]{.\EOS\space}%
\providecommand \EOS [0]{\spacefactor3000\relax}%
\providecommand \BibitemShut  [1]{\csname bibitem#1\endcsname}%
\let\auto@bib@innerbib\@empty
\bibitem [{\citenamefont {Goodenough}\ and\ \citenamefont
  {Park}(2013)}]{goodenough2013li}%
  \BibitemOpen
  \bibfield  {author} {\bibinfo {author} {\bibfnamefont {J.~B.}\ \bibnamefont
  {Goodenough}}\ and\ \bibinfo {author} {\bibfnamefont {K.-S.}\ \bibnamefont
  {Park}},\ }\bibfield  {title} {\enquote {\bibinfo {title} {The li-ion
  rechargeable battery: a perspective},}\ }\href@noop {} {\bibfield  {journal}
  {\bibinfo  {journal} {Journal of the American Chemical Society}\ }\textbf
  {\bibinfo {volume} {135}},\ \bibinfo {pages} {1167--1176} (\bibinfo {year}
  {2013})}\BibitemShut {NoStop}%
\bibitem [{\citenamefont {Tarascon}\ and\ \citenamefont
  {Armand}(2011)}]{tarascon2011issues}%
  \BibitemOpen
  \bibfield  {author} {\bibinfo {author} {\bibfnamefont {J.-M.}\ \bibnamefont
  {Tarascon}}\ and\ \bibinfo {author} {\bibfnamefont {M.}~\bibnamefont
  {Armand}},\ }\bibfield  {title} {\enquote {\bibinfo {title} {Issues and
  challenges facing rechargeable lithium batteries},}\ }in\ \href@noop {}
  {\emph {\bibinfo {booktitle} {Materials for Sustainable Energy: A Collection
  of Peer-Reviewed Research and Review Articles from Nature Publishing
  Group}}}\ (\bibinfo  {publisher} {World Scientific},\ \bibinfo {year}
  {2011})\ pp.\ \bibinfo {pages} {171--179}\BibitemShut {NoStop}%
\bibitem [{\citenamefont {Chu}\ and\ \citenamefont
  {Majumdar}(2012)}]{chu2012opportunities}%
  \BibitemOpen
  \bibfield  {author} {\bibinfo {author} {\bibfnamefont {S.}~\bibnamefont
  {Chu}}\ and\ \bibinfo {author} {\bibfnamefont {A.}~\bibnamefont {Majumdar}},\
  }\bibfield  {title} {\enquote {\bibinfo {title} {Opportunities and challenges
  for a sustainable energy future},}\ }\href@noop {} {\bibfield  {journal}
  {\bibinfo  {journal} {nature}\ }\textbf {\bibinfo {volume} {488}},\ \bibinfo
  {pages} {294} (\bibinfo {year} {2012})}\BibitemShut {NoStop}%
\bibitem [{\citenamefont {Hu}(2016)}]{hu2016batteries}%
  \BibitemOpen
  \bibfield  {author} {\bibinfo {author} {\bibfnamefont {Y.-S.}\ \bibnamefont
  {Hu}},\ }\bibfield  {title} {\enquote {\bibinfo {title} {Batteries: getting
  solid},}\ }\href@noop {} {\bibfield  {journal} {\bibinfo  {journal} {Nature
  Energy}\ }\textbf {\bibinfo {volume} {1}},\ \bibinfo {pages} {16042}
  (\bibinfo {year} {2016})}\BibitemShut {NoStop}%
\bibitem [{\citenamefont {Zhang}\ \emph
  {et~al.}(2018{\natexlab{a}})\citenamefont {Zhang}, \citenamefont {Shao},
  \citenamefont {Lotsch}, \citenamefont {Hu}, \citenamefont {Li}, \citenamefont
  {Janek}, \citenamefont {Nazar}, \citenamefont {Nan}, \citenamefont {Maier},
  \citenamefont {Armand} \emph {et~al.}}]{zhang2018new}%
  \BibitemOpen
  \bibfield  {author} {\bibinfo {author} {\bibfnamefont {Z.}~\bibnamefont
  {Zhang}}, \bibinfo {author} {\bibfnamefont {Y.}~\bibnamefont {Shao}},
  \bibinfo {author} {\bibfnamefont {B.}~\bibnamefont {Lotsch}}, \bibinfo
  {author} {\bibfnamefont {Y.-S.}\ \bibnamefont {Hu}}, \bibinfo {author}
  {\bibfnamefont {H.}~\bibnamefont {Li}}, \bibinfo {author} {\bibfnamefont
  {J.}~\bibnamefont {Janek}}, \bibinfo {author} {\bibfnamefont {L.~F.}\
  \bibnamefont {Nazar}}, \bibinfo {author} {\bibfnamefont {C.-W.}\ \bibnamefont
  {Nan}}, \bibinfo {author} {\bibfnamefont {J.}~\bibnamefont {Maier}}, \bibinfo
  {author} {\bibfnamefont {M.}~\bibnamefont {Armand}},  \emph {et~al.},\
  }\bibfield  {title} {\enquote {\bibinfo {title} {New horizons for inorganic
  solid state ion conductors},}\ }\href@noop {} {\bibfield  {journal} {\bibinfo
   {journal} {Energy \& Environmental Science}\ }\textbf {\bibinfo {volume}
  {11}},\ \bibinfo {pages} {1945--1976} (\bibinfo {year}
  {2018}{\natexlab{a}})}\BibitemShut {NoStop}%
\bibitem [{\citenamefont {Kamaya}\ \emph {et~al.}(2011)\citenamefont {Kamaya},
  \citenamefont {Homma}, \citenamefont {Yamakawa}, \citenamefont {Hirayama},
  \citenamefont {Kanno}, \citenamefont {Yonemura}, \citenamefont {Kamiyama},
  \citenamefont {Kato}, \citenamefont {Hama}, \citenamefont {Kawamoto} \emph
  {et~al.}}]{kamaya2011lithium}%
  \BibitemOpen
  \bibfield  {author} {\bibinfo {author} {\bibfnamefont {N.}~\bibnamefont
  {Kamaya}}, \bibinfo {author} {\bibfnamefont {K.}~\bibnamefont {Homma}},
  \bibinfo {author} {\bibfnamefont {Y.}~\bibnamefont {Yamakawa}}, \bibinfo
  {author} {\bibfnamefont {M.}~\bibnamefont {Hirayama}}, \bibinfo {author}
  {\bibfnamefont {R.}~\bibnamefont {Kanno}}, \bibinfo {author} {\bibfnamefont
  {M.}~\bibnamefont {Yonemura}}, \bibinfo {author} {\bibfnamefont
  {T.}~\bibnamefont {Kamiyama}}, \bibinfo {author} {\bibfnamefont
  {Y.}~\bibnamefont {Kato}}, \bibinfo {author} {\bibfnamefont {S.}~\bibnamefont
  {Hama}}, \bibinfo {author} {\bibfnamefont {K.}~\bibnamefont {Kawamoto}},
  \emph {et~al.},\ }\bibfield  {title} {\enquote {\bibinfo {title} {A lithium
  superionic conductor},}\ }\href@noop {} {\bibfield  {journal} {\bibinfo
  {journal} {Nature materials}\ }\textbf {\bibinfo {volume} {10}},\ \bibinfo
  {pages} {682} (\bibinfo {year} {2011})}\BibitemShut {NoStop}%
\bibitem [{\citenamefont {Murugan}, \citenamefont {Thangadurai},\ and\
  \citenamefont {Weppner}(2007)}]{murugan2007fast}%
  \BibitemOpen
  \bibfield  {author} {\bibinfo {author} {\bibfnamefont {R.}~\bibnamefont
  {Murugan}}, \bibinfo {author} {\bibfnamefont {V.}~\bibnamefont
  {Thangadurai}}, \ and\ \bibinfo {author} {\bibfnamefont {W.}~\bibnamefont
  {Weppner}},\ }\bibfield  {title} {\enquote {\bibinfo {title} {Fast lithium
  ion conduction in garnet-type li7la3zr2o12},}\ }\href@noop {} {\bibfield
  {journal} {\bibinfo  {journal} {Angewandte Chemie International Edition}\
  }\textbf {\bibinfo {volume} {46}},\ \bibinfo {pages} {7778--7781} (\bibinfo
  {year} {2007})}\BibitemShut {NoStop}%
\bibitem [{\citenamefont {Seino}\ \emph {et~al.}(2014)\citenamefont {Seino},
  \citenamefont {Ota}, \citenamefont {Takada}, \citenamefont {Hayashi},\ and\
  \citenamefont {Tatsumisago}}]{seino2014sulphide}%
  \BibitemOpen
  \bibfield  {author} {\bibinfo {author} {\bibfnamefont {Y.}~\bibnamefont
  {Seino}}, \bibinfo {author} {\bibfnamefont {T.}~\bibnamefont {Ota}}, \bibinfo
  {author} {\bibfnamefont {K.}~\bibnamefont {Takada}}, \bibinfo {author}
  {\bibfnamefont {A.}~\bibnamefont {Hayashi}}, \ and\ \bibinfo {author}
  {\bibfnamefont {M.}~\bibnamefont {Tatsumisago}},\ }\bibfield  {title}
  {\enquote {\bibinfo {title} {A sulphide lithium super ion conductor is
  superior to liquid ion conductors for use in rechargeable batteries},}\
  }\href@noop {} {\bibfield  {journal} {\bibinfo  {journal} {Energy \&
  Environmental Science}\ }\textbf {\bibinfo {volume} {7}},\ \bibinfo {pages}
  {627--631} (\bibinfo {year} {2014})}\BibitemShut {NoStop}%
\bibitem [{\citenamefont {Ong}\ \emph {et~al.}(2013{\natexlab{a}})\citenamefont
  {Ong}, \citenamefont {Mo}, \citenamefont {Richards}, \citenamefont {Miara},
  \citenamefont {Lee},\ and\ \citenamefont {Ceder}}]{ong2013phase}%
  \BibitemOpen
  \bibfield  {author} {\bibinfo {author} {\bibfnamefont {S.~P.}\ \bibnamefont
  {Ong}}, \bibinfo {author} {\bibfnamefont {Y.}~\bibnamefont {Mo}}, \bibinfo
  {author} {\bibfnamefont {W.~D.}\ \bibnamefont {Richards}}, \bibinfo {author}
  {\bibfnamefont {L.}~\bibnamefont {Miara}}, \bibinfo {author} {\bibfnamefont
  {H.~S.}\ \bibnamefont {Lee}}, \ and\ \bibinfo {author} {\bibfnamefont
  {G.}~\bibnamefont {Ceder}},\ }\bibfield  {title} {\enquote {\bibinfo {title}
  {Phase stability, electrochemical stability and ionic conductivity of the li
  10$\pm$1 mp 2 x 12 (m= ge, si, sn, al or p, and x= o, s or se) family of
  superionic conductors},}\ }\href@noop {} {\bibfield  {journal} {\bibinfo
  {journal} {Energy \& Environmental Science}\ }\textbf {\bibinfo {volume}
  {6}},\ \bibinfo {pages} {148--156} (\bibinfo {year}
  {2013}{\natexlab{a}})}\BibitemShut {NoStop}%
\bibitem [{\citenamefont {Ceder}, \citenamefont {Ong},\ and\ \citenamefont
  {Wang}(2018)}]{ceder_ong_wang_2018}%
  \BibitemOpen
  \bibfield  {author} {\bibinfo {author} {\bibfnamefont {G.}~\bibnamefont
  {Ceder}}, \bibinfo {author} {\bibfnamefont {S.~P.}\ \bibnamefont {Ong}}, \
  and\ \bibinfo {author} {\bibfnamefont {Y.}~\bibnamefont {Wang}},\ }\bibfield
  {title} {\enquote {\bibinfo {title} {Predictive modeling and design rules for
  solid electrolytes},}\ }\href {\doibase 10.1557/mrs.2018.210} {\bibfield
  {journal} {\bibinfo  {journal} {MRS Bulletin}\ }\textbf {\bibinfo {volume}
  {43}},\ \bibinfo {pages} {746–751} (\bibinfo {year} {2018})}\BibitemShut
  {NoStop}%
\bibitem [{\citenamefont {Mo}, \citenamefont {Ong},\ and\ \citenamefont
  {Ceder}(2012)}]{Mo2012}%
  \BibitemOpen
  \bibfield  {author} {\bibinfo {author} {\bibfnamefont {Y.}~\bibnamefont
  {Mo}}, \bibinfo {author} {\bibfnamefont {S.~P.}\ \bibnamefont {Ong}}, \ and\
  \bibinfo {author} {\bibfnamefont {G.}~\bibnamefont {Ceder}},\ }\bibfield
  {title} {\enquote {\bibinfo {title} {{First Principles Study of the Li 10 GeP
  2 S 12 Lithium Super Ionic Conductor Material}},}\ }\href {\doibase
  10.1021/cm203303y} {\bibfield  {journal} {\bibinfo  {journal} {Chemistry of
  Materials}\ }\textbf {\bibinfo {volume} {24}},\ \bibinfo {pages} {15--17}
  (\bibinfo {year} {2012})}\BibitemShut {NoStop}%
\bibitem [{\citenamefont {Car}\ and\ \citenamefont
  {Parrinello}(1985)}]{car1985cpmd}%
  \BibitemOpen
  \bibfield  {author} {\bibinfo {author} {\bibfnamefont {R.}~\bibnamefont
  {Car}}\ and\ \bibinfo {author} {\bibfnamefont {M.}~\bibnamefont
  {Parrinello}},\ }\bibfield  {title} {\enquote {\bibinfo {title} {Unified
  approach for molecular dynamics and density-functional theory},}\ }\href@noop
  {} {\bibfield  {journal} {\bibinfo  {journal} {Physical Review Letters}\
  }\textbf {\bibinfo {volume} {55}},\ \bibinfo {pages} {2471} (\bibinfo {year}
  {1985})}\BibitemShut {NoStop}%
\bibitem [{\citenamefont {He}, \citenamefont {Zhu},\ and\ \citenamefont
  {Mo}(2017)}]{he2017origin}%
  \BibitemOpen
  \bibfield  {author} {\bibinfo {author} {\bibfnamefont {X.}~\bibnamefont
  {He}}, \bibinfo {author} {\bibfnamefont {Y.}~\bibnamefont {Zhu}}, \ and\
  \bibinfo {author} {\bibfnamefont {Y.}~\bibnamefont {Mo}},\ }\bibfield
  {title} {\enquote {\bibinfo {title} {Origin of fast ion diffusion in
  super-ionic conductors},}\ }\href@noop {} {\bibfield  {journal} {\bibinfo
  {journal} {Nature communications}\ }\textbf {\bibinfo {volume} {8}},\
  \bibinfo {pages} {15893} (\bibinfo {year} {2017})}\BibitemShut {NoStop}%
\bibitem [{\citenamefont {Nolan}\ \emph {et~al.}(2018)\citenamefont {Nolan},
  \citenamefont {Zhu}, \citenamefont {He}, \citenamefont {Bai},\ and\
  \citenamefont {Mo}}]{nolan2018computation}%
  \BibitemOpen
  \bibfield  {author} {\bibinfo {author} {\bibfnamefont {A.~M.}\ \bibnamefont
  {Nolan}}, \bibinfo {author} {\bibfnamefont {Y.}~\bibnamefont {Zhu}}, \bibinfo
  {author} {\bibfnamefont {X.}~\bibnamefont {He}}, \bibinfo {author}
  {\bibfnamefont {Q.}~\bibnamefont {Bai}}, \ and\ \bibinfo {author}
  {\bibfnamefont {Y.}~\bibnamefont {Mo}},\ }\bibfield  {title} {\enquote
  {\bibinfo {title} {Computation-accelerated design of materials and interfaces
  for all-solid-state lithium-ion batteries},}\ }\href@noop {} {\bibfield
  {journal} {\bibinfo  {journal} {Joule}\ } (\bibinfo {year}
  {2018})}\BibitemShut {NoStop}%
\bibitem [{\citenamefont {Van~der Ven}\ and\ \citenamefont
  {Ceder}(2000)}]{van2000lithium}%
  \BibitemOpen
  \bibfield  {author} {\bibinfo {author} {\bibfnamefont {A.}~\bibnamefont
  {Van~der Ven}}\ and\ \bibinfo {author} {\bibfnamefont {G.}~\bibnamefont
  {Ceder}},\ }\bibfield  {title} {\enquote {\bibinfo {title} {Lithium diffusion
  in layered li x coo2},}\ }\href@noop {} {\bibfield  {journal} {\bibinfo
  {journal} {Electrochemical and Solid-State Letters}\ }\textbf {\bibinfo
  {volume} {3}},\ \bibinfo {pages} {301--304} (\bibinfo {year}
  {2000})}\BibitemShut {NoStop}%
\bibitem [{\citenamefont {Marcolongo}\ and\ \citenamefont
  {Marzari}(2017)}]{marcolongo2017ionic}%
  \BibitemOpen
  \bibfield  {author} {\bibinfo {author} {\bibfnamefont {A.}~\bibnamefont
  {Marcolongo}}\ and\ \bibinfo {author} {\bibfnamefont {N.}~\bibnamefont
  {Marzari}},\ }\bibfield  {title} {\enquote {\bibinfo {title} {Ionic
  correlations and failure of nernst-einstein relation in solid-state
  electrolytes},}\ }\href@noop {} {\bibfield  {journal} {\bibinfo  {journal}
  {Physical Review Materials}\ }\textbf {\bibinfo {volume} {1}},\ \bibinfo
  {pages} {025402} (\bibinfo {year} {2017})}\BibitemShut {NoStop}%
\bibitem [{\citenamefont {Kuhn}, \citenamefont {K{\"{o}}hler},\ and\
  \citenamefont {Lotsch}(2013)}]{Kuhn2013}%
  \BibitemOpen
  \bibfield  {author} {\bibinfo {author} {\bibfnamefont {A.}~\bibnamefont
  {Kuhn}}, \bibinfo {author} {\bibfnamefont {J.}~\bibnamefont {K{\"{o}}hler}},
  \ and\ \bibinfo {author} {\bibfnamefont {B.~V.}\ \bibnamefont {Lotsch}},\
  }\bibfield  {title} {\enquote {\bibinfo {title} {{Single-crystal X-ray
  structure analysis of the superionic conductor Li10GeP2S12}},}\ }\href
  {\doibase 10.1039/c3cp51985f} {\bibfield  {journal} {\bibinfo  {journal}
  {Physical Chemistry Chemical Physics}\ }\textbf {\bibinfo {volume} {15}},\
  \bibinfo {pages} {11620} (\bibinfo {year} {2013})}\BibitemShut {NoStop}%
\bibitem [{\citenamefont {Boyce}\ and\ \citenamefont
  {Huberman}(1979)}]{boyce1979superionic}%
  \BibitemOpen
  \bibfield  {author} {\bibinfo {author} {\bibfnamefont {J.~B.}\ \bibnamefont
  {Boyce}}\ and\ \bibinfo {author} {\bibfnamefont {B.~A.}\ \bibnamefont
  {Huberman}},\ }\bibfield  {title} {\enquote {\bibinfo {title} {Superionic
  conductors: Transitions, structures, dynamics},}\ }\href@noop {} {\bibfield
  {journal} {\bibinfo  {journal} {Physics Reports}\ }\textbf {\bibinfo {volume}
  {51}},\ \bibinfo {pages} {189--265} (\bibinfo {year} {1979})}\BibitemShut
  {NoStop}%
\bibitem [{\citenamefont {Tachez}\ \emph {et~al.}(1984)\citenamefont {Tachez},
  \citenamefont {Malugani}, \citenamefont {Mercier},\ and\ \citenamefont
  {Robert}}]{TACHEZ1984181}%
  \BibitemOpen
  \bibfield  {author} {\bibinfo {author} {\bibfnamefont {M.}~\bibnamefont
  {Tachez}}, \bibinfo {author} {\bibfnamefont {J.-P.}\ \bibnamefont
  {Malugani}}, \bibinfo {author} {\bibfnamefont {R.}~\bibnamefont {Mercier}}, \
  and\ \bibinfo {author} {\bibfnamefont {G.}~\bibnamefont {Robert}},\
  }\bibfield  {title} {\enquote {\bibinfo {title} {Ionic conductivity of and
  phase transition in lithium thiophosphate li3ps4},}\ }\href {\doibase
  https://doi.org/10.1016/0167-2738(84)90097-3} {\bibfield  {journal} {\bibinfo
   {journal} {Solid State Ionics}\ }\textbf {\bibinfo {volume} {14}},\ \bibinfo
  {pages} {181 -- 185} (\bibinfo {year} {1984})}\BibitemShut {NoStop}%
\bibitem [{\citenamefont {Kwon}\ \emph {et~al.}(2015)\citenamefont {Kwon},
  \citenamefont {Hirayama}, \citenamefont {Suzuki}, \citenamefont {Kato},
  \citenamefont {Saito}, \citenamefont {Yonemura}, \citenamefont {Kamiyama},\
  and\ \citenamefont {Kanno}}]{Kwon2015}%
  \BibitemOpen
  \bibfield  {author} {\bibinfo {author} {\bibfnamefont {O.}~\bibnamefont
  {Kwon}}, \bibinfo {author} {\bibfnamefont {M.}~\bibnamefont {Hirayama}},
  \bibinfo {author} {\bibfnamefont {K.}~\bibnamefont {Suzuki}}, \bibinfo
  {author} {\bibfnamefont {Y.}~\bibnamefont {Kato}}, \bibinfo {author}
  {\bibfnamefont {T.}~\bibnamefont {Saito}}, \bibinfo {author} {\bibfnamefont
  {M.}~\bibnamefont {Yonemura}}, \bibinfo {author} {\bibfnamefont
  {T.}~\bibnamefont {Kamiyama}}, \ and\ \bibinfo {author} {\bibfnamefont
  {R.}~\bibnamefont {Kanno}},\ }\bibfield  {title} {\enquote {\bibinfo {title}
  {{Synthesis, structure, and conduction mechanism of the lithium superionic
  conductor Li10+$\delta$Ge1+$\delta$P2-$\delta$S12}},}\ }\href {\doibase
  10.1039/c4ta05231e} {\bibfield  {journal} {\bibinfo  {journal} {Journal of
  Materials Chemistry A}\ }\textbf {\bibinfo {volume} {3}},\ \bibinfo {pages}
  {438--446} (\bibinfo {year} {2015})}\BibitemShut {NoStop}%
\bibitem [{\citenamefont {Kanno}\ and\ \citenamefont
  {Murayama}(2001)}]{Kanno2001}%
  \BibitemOpen
  \bibfield  {author} {\bibinfo {author} {\bibfnamefont {R.}~\bibnamefont
  {Kanno}}\ and\ \bibinfo {author} {\bibfnamefont {M.}~\bibnamefont
  {Murayama}},\ }\bibfield  {title} {\enquote {\bibinfo {title} {{Lithium Ionic
  Conductor Thio-LISICON: The Li 2S-GeS 2-P 2S 5 System}},}\ }\href {\doibase
  10.1149/1.1379028} {\bibfield  {journal} {\bibinfo  {journal} {Journal of the
  Electrochemical Society}\ }\textbf {\bibinfo {volume} {148}},\ \bibinfo
  {pages} {742--746} (\bibinfo {year} {2001})}\BibitemShut {NoStop}%
\bibitem [{\citenamefont {Bron}\ \emph {et~al.}(2013)\citenamefont {Bron},
  \citenamefont {Johansson}, \citenamefont {Zick}, \citenamefont {Schmedt
  auf~der Günne}, \citenamefont {Dehnen},\ and\ \citenamefont
  {Roling}}]{bron2013li10snp2s12}%
  \BibitemOpen
  \bibfield  {author} {\bibinfo {author} {\bibfnamefont {P.}~\bibnamefont
  {Bron}}, \bibinfo {author} {\bibfnamefont {S.}~\bibnamefont {Johansson}},
  \bibinfo {author} {\bibfnamefont {K.}~\bibnamefont {Zick}}, \bibinfo {author}
  {\bibfnamefont {J.}~\bibnamefont {Schmedt auf~der Günne}}, \bibinfo {author}
  {\bibfnamefont {S.}~\bibnamefont {Dehnen}}, \ and\ \bibinfo {author}
  {\bibfnamefont {B.}~\bibnamefont {Roling}},\ }\bibfield  {title} {\enquote
  {\bibinfo {title} {Li10snp2s12: an affordable lithium superionic
  conductor},}\ }\href@noop {} {\bibfield  {journal} {\bibinfo  {journal}
  {Journal of the American Chemical Society}\ }\textbf {\bibinfo {volume}
  {135}},\ \bibinfo {pages} {15694--15697} (\bibinfo {year}
  {2013})}\BibitemShut {NoStop}%
\bibitem [{\citenamefont {He}\ \emph {et~al.}(2018)\citenamefont {He},
  \citenamefont {Zhu}, \citenamefont {Epstein},\ and\ \citenamefont
  {Mo}}]{he2018statistical}%
  \BibitemOpen
  \bibfield  {author} {\bibinfo {author} {\bibfnamefont {X.}~\bibnamefont
  {He}}, \bibinfo {author} {\bibfnamefont {Y.}~\bibnamefont {Zhu}}, \bibinfo
  {author} {\bibfnamefont {A.}~\bibnamefont {Epstein}}, \ and\ \bibinfo
  {author} {\bibfnamefont {Y.}~\bibnamefont {Mo}},\ }\bibfield  {title}
  {\enquote {\bibinfo {title} {Statistical variances of diffusional properties
  from ab initio molecular dynamics simulations},}\ }\href@noop {} {\bibfield
  {journal} {\bibinfo  {journal} {npj Computational Materials}\ }\textbf
  {\bibinfo {volume} {4}},\ \bibinfo {pages} {18} (\bibinfo {year}
  {2018})}\BibitemShut {NoStop}%
\bibitem [{\citenamefont {Xiao}, \citenamefont {Li},\ and\ \citenamefont
  {Chen}(2015)}]{xiao2015candidate}%
  \BibitemOpen
  \bibfield  {author} {\bibinfo {author} {\bibfnamefont {R.}~\bibnamefont
  {Xiao}}, \bibinfo {author} {\bibfnamefont {H.}~\bibnamefont {Li}}, \ and\
  \bibinfo {author} {\bibfnamefont {L.}~\bibnamefont {Chen}},\ }\bibfield
  {title} {\enquote {\bibinfo {title} {Candidate structures for inorganic
  lithium solid-state electrolytes identified by high-throughput bond-valence
  calculations},}\ }\href@noop {} {\bibfield  {journal} {\bibinfo  {journal}
  {Journal of Materiomics}\ }\textbf {\bibinfo {volume} {1}},\ \bibinfo {pages}
  {325--332} (\bibinfo {year} {2015})}\BibitemShut {NoStop}%
\bibitem [{\citenamefont {Kahle}, \citenamefont {Marcolongo},\ and\
  \citenamefont {Marzari}(2018)}]{kahle2018modeling}%
  \BibitemOpen
  \bibfield  {author} {\bibinfo {author} {\bibfnamefont {L.}~\bibnamefont
  {Kahle}}, \bibinfo {author} {\bibfnamefont {A.}~\bibnamefont {Marcolongo}}, \
  and\ \bibinfo {author} {\bibfnamefont {N.}~\bibnamefont {Marzari}},\
  }\bibfield  {title} {\enquote {\bibinfo {title} {Modeling lithium-ion
  solid-state electrolytes with a pinball model},}\ }\href@noop {} {\bibfield
  {journal} {\bibinfo  {journal} {Physical Review Materials}\ }\textbf
  {\bibinfo {volume} {2}},\ \bibinfo {pages} {065405} (\bibinfo {year}
  {2018})}\BibitemShut {NoStop}%
\bibitem [{\citenamefont {Kobayashi}\ \emph {et~al.}(2020)\citenamefont
  {Kobayashi}, \citenamefont {Miyaji}, \citenamefont {Nakano},\ and\
  \citenamefont {Nakayama}}]{masanobu2020cuckoo}%
  \BibitemOpen
  \bibfield  {author} {\bibinfo {author} {\bibfnamefont {R.}~\bibnamefont
  {Kobayashi}}, \bibinfo {author} {\bibfnamefont {Y.}~\bibnamefont {Miyaji}},
  \bibinfo {author} {\bibfnamefont {K.}~\bibnamefont {Nakano}}, \ and\ \bibinfo
  {author} {\bibfnamefont {M.}~\bibnamefont {Nakayama}},\ }\bibfield  {title}
  {\enquote {\bibinfo {title} {High-throughput production of force-fields for
  solid-state electrolyte materials},}\ }\href {\doibase 10.1063/5.0015373}
  {\bibfield  {journal} {\bibinfo  {journal} {APL Materials}\ }\textbf
  {\bibinfo {volume} {8}},\ \bibinfo {pages} {081111} (\bibinfo {year}
  {2020})}\BibitemShut {NoStop}%
\bibitem [{\citenamefont {Li}\ \emph {et~al.}(2020)\citenamefont {Li},
  \citenamefont {Zhou}, \citenamefont {Wang},\ and\ \citenamefont
  {Jiang}}]{icf2020jiang}%
  \BibitemOpen
  \bibfield  {author} {\bibinfo {author} {\bibfnamefont {H.-X.}\ \bibnamefont
  {Li}}, \bibinfo {author} {\bibfnamefont {X.-Y.}\ \bibnamefont {Zhou}},
  \bibinfo {author} {\bibfnamefont {Y.-C.}\ \bibnamefont {Wang}}, \ and\
  \bibinfo {author} {\bibfnamefont {H.}~\bibnamefont {Jiang}},\ }\bibfield
  {title} {\enquote {\bibinfo {title} {Theoretical study of na+ transport in
  the solid-state electrolyte na3obr based on deep potential molecular
  dynamics},}\ }\href {\doibase 10.1039/D0QI00921K} {\bibfield  {journal}
  {\bibinfo  {journal} {Inorg. Chem. Front.}\ ,\ \bibinfo {pages} {--}}
  (\bibinfo {year} {2020})}\BibitemShut {NoStop}%
\bibitem [{\citenamefont {Behler}\ and\ \citenamefont
  {Parrinello}(2007)}]{behler2007generalized}%
  \BibitemOpen
  \bibfield  {author} {\bibinfo {author} {\bibfnamefont {J.}~\bibnamefont
  {Behler}}\ and\ \bibinfo {author} {\bibfnamefont {M.}~\bibnamefont
  {Parrinello}},\ }\bibfield  {title} {\enquote {\bibinfo {title} {Generalized
  neural-network representation of high-dimensional potential-energy
  surfaces},}\ }\href@noop {} {\bibfield  {journal} {\bibinfo  {journal}
  {Physical review letters}\ }\textbf {\bibinfo {volume} {98}},\ \bibinfo
  {pages} {146401} (\bibinfo {year} {2007})}\BibitemShut {NoStop}%
\bibitem [{\citenamefont {Bart{\'o}k}\ \emph {et~al.}(2010)\citenamefont
  {Bart{\'o}k}, \citenamefont {Payne}, \citenamefont {Kondor},\ and\
  \citenamefont {Cs{\'a}nyi}}]{bartok2010gaussian}%
  \BibitemOpen
  \bibfield  {author} {\bibinfo {author} {\bibfnamefont {A.~P.}\ \bibnamefont
  {Bart{\'o}k}}, \bibinfo {author} {\bibfnamefont {M.~C.}\ \bibnamefont
  {Payne}}, \bibinfo {author} {\bibfnamefont {R.}~\bibnamefont {Kondor}}, \
  and\ \bibinfo {author} {\bibfnamefont {G.}~\bibnamefont {Cs{\'a}nyi}},\
  }\bibfield  {title} {\enquote {\bibinfo {title} {Gaussian approximation
  potentials: The accuracy of quantum mechanics, without the electrons},}\
  }\href@noop {} {\bibfield  {journal} {\bibinfo  {journal} {Physical review
  letters}\ }\textbf {\bibinfo {volume} {104}},\ \bibinfo {pages} {136403}
  (\bibinfo {year} {2010})}\BibitemShut {NoStop}%
\bibitem [{\citenamefont {Artrith}, \citenamefont {Urban},\ and\ \citenamefont
  {Ceder}(2017)}]{artrith2017atomic}%
  \BibitemOpen
  \bibfield  {author} {\bibinfo {author} {\bibfnamefont {N.}~\bibnamefont
  {Artrith}}, \bibinfo {author} {\bibfnamefont {A.}~\bibnamefont {Urban}}, \
  and\ \bibinfo {author} {\bibfnamefont {G.}~\bibnamefont {Ceder}},\ }\bibfield
   {title} {\enquote {\bibinfo {title} {Efficient and accurate machine-learning
  interpolation of atomic energies in compositions with many species},}\
  }\href@noop {} {\bibfield  {journal} {\bibinfo  {journal} {Physical Review
  B}\ }\textbf {\bibinfo {volume} {96}},\ \bibinfo {pages} {014112} (\bibinfo
  {year} {2017})}\BibitemShut {NoStop}%
\bibitem [{\citenamefont {Zhang}\ \emph
  {et~al.}(2018{\natexlab{b}})\citenamefont {Zhang}, \citenamefont {Han},
  \citenamefont {Wang}, \citenamefont {Saidi}, \citenamefont {Car},\ and\
  \citenamefont {E}}]{NIPS2018_7696}%
  \BibitemOpen
  \bibfield  {author} {\bibinfo {author} {\bibfnamefont {L.}~\bibnamefont
  {Zhang}}, \bibinfo {author} {\bibfnamefont {J.}~\bibnamefont {Han}}, \bibinfo
  {author} {\bibfnamefont {H.}~\bibnamefont {Wang}}, \bibinfo {author}
  {\bibfnamefont {W.}~\bibnamefont {Saidi}}, \bibinfo {author} {\bibfnamefont
  {R.}~\bibnamefont {Car}}, \ and\ \bibinfo {author} {\bibfnamefont
  {W.}~\bibnamefont {E}},\ }\bibfield  {title} {\enquote {\bibinfo {title}
  {End-to-end symmetry preserving inter-atomic potential energy model for
  finite and extended systems},}\ }in\ \href
  {http://papers.nips.cc/paper/7696-end-to-end-symmetry-preserving-inter-atomic-potential-energy-model-for-finite-and-extended-systems.pdf}
  {\emph {\bibinfo {booktitle} {Advances in Neural Information Processing
  Systems 31}}}\ (\bibinfo  {publisher} {Curran Associates, Inc.},\ \bibinfo
  {year} {2018})\ pp.\ \bibinfo {pages} {4436--4446}\BibitemShut {NoStop}%
\bibitem [{\citenamefont {Artrith}, \citenamefont {Urban},\ and\ \citenamefont
  {Ceder}(2018)}]{artrith2018constructing}%
  \BibitemOpen
  \bibfield  {author} {\bibinfo {author} {\bibfnamefont {N.}~\bibnamefont
  {Artrith}}, \bibinfo {author} {\bibfnamefont {A.}~\bibnamefont {Urban}}, \
  and\ \bibinfo {author} {\bibfnamefont {G.}~\bibnamefont {Ceder}},\ }\bibfield
   {title} {\enquote {\bibinfo {title} {Constructing first-principles phase
  diagrams of amorphous li x si using machine-learning-assisted sampling with
  an evolutionary algorithm},}\ }\href@noop {} {\bibfield  {journal} {\bibinfo
  {journal} {The Journal of chemical physics}\ }\textbf {\bibinfo {volume}
  {148}},\ \bibinfo {pages} {241711} (\bibinfo {year} {2018})}\BibitemShut
  {NoStop}%
\bibitem [{\citenamefont {Deringer}\ \emph
  {et~al.}(2018{\natexlab{a}})\citenamefont {Deringer}, \citenamefont {Merlet},
  \citenamefont {Hu}, \citenamefont {Lee}, \citenamefont {Kattirtzi},
  \citenamefont {Pecher}, \citenamefont {Cs{\'a}nyi}, \citenamefont {Elliott},\
  and\ \citenamefont {Grey}}]{deringer2018towards}%
  \BibitemOpen
  \bibfield  {author} {\bibinfo {author} {\bibfnamefont {V.~L.}\ \bibnamefont
  {Deringer}}, \bibinfo {author} {\bibfnamefont {C.}~\bibnamefont {Merlet}},
  \bibinfo {author} {\bibfnamefont {Y.}~\bibnamefont {Hu}}, \bibinfo {author}
  {\bibfnamefont {T.~H.}\ \bibnamefont {Lee}}, \bibinfo {author} {\bibfnamefont
  {J.~A.}\ \bibnamefont {Kattirtzi}}, \bibinfo {author} {\bibfnamefont
  {O.}~\bibnamefont {Pecher}}, \bibinfo {author} {\bibfnamefont
  {G.}~\bibnamefont {Cs{\'a}nyi}}, \bibinfo {author} {\bibfnamefont {S.~R.}\
  \bibnamefont {Elliott}}, \ and\ \bibinfo {author} {\bibfnamefont {C.~P.}\
  \bibnamefont {Grey}},\ }\bibfield  {title} {\enquote {\bibinfo {title}
  {Towards an atomistic understanding of disordered carbon electrode
  materials},}\ }\href@noop {} {\bibfield  {journal} {\bibinfo  {journal}
  {Chemical communications}\ }\textbf {\bibinfo {volume} {54}},\ \bibinfo
  {pages} {5988--5991} (\bibinfo {year} {2018}{\natexlab{a}})}\BibitemShut
  {NoStop}%
\bibitem [{\citenamefont {Fujikake}\ \emph {et~al.}(2018)\citenamefont
  {Fujikake}, \citenamefont {Deringer}, \citenamefont {Lee}, \citenamefont
  {Krynski}, \citenamefont {Elliott},\ and\ \citenamefont
  {Cs{\'a}nyi}}]{fujikake2018gaussian}%
  \BibitemOpen
  \bibfield  {author} {\bibinfo {author} {\bibfnamefont {S.}~\bibnamefont
  {Fujikake}}, \bibinfo {author} {\bibfnamefont {V.~L.}\ \bibnamefont
  {Deringer}}, \bibinfo {author} {\bibfnamefont {T.~H.}\ \bibnamefont {Lee}},
  \bibinfo {author} {\bibfnamefont {M.}~\bibnamefont {Krynski}}, \bibinfo
  {author} {\bibfnamefont {S.~R.}\ \bibnamefont {Elliott}}, \ and\ \bibinfo
  {author} {\bibfnamefont {G.}~\bibnamefont {Cs{\'a}nyi}},\ }\bibfield  {title}
  {\enquote {\bibinfo {title} {Gaussian approximation potential modeling of
  lithium intercalation in carbon nanostructures},}\ }\href@noop {} {\bibfield
  {journal} {\bibinfo  {journal} {The Journal of chemical physics}\ }\textbf
  {\bibinfo {volume} {148}},\ \bibinfo {pages} {241714} (\bibinfo {year}
  {2018})}\BibitemShut {NoStop}%
\bibitem [{\citenamefont {Lacivita}, \citenamefont {Artrith},\ and\
  \citenamefont {Ceder}(2018)}]{lacivita2018structural}%
  \BibitemOpen
  \bibfield  {author} {\bibinfo {author} {\bibfnamefont {V.}~\bibnamefont
  {Lacivita}}, \bibinfo {author} {\bibfnamefont {N.}~\bibnamefont {Artrith}}, \
  and\ \bibinfo {author} {\bibfnamefont {G.}~\bibnamefont {Ceder}},\ }\bibfield
   {title} {\enquote {\bibinfo {title} {Structural and compositional factors
  that control the li-ion conductivity in lipon electrolytes},}\ }\href@noop {}
  {\bibfield  {journal} {\bibinfo  {journal} {Chemistry of Materials}\ }\textbf
  {\bibinfo {volume} {30}},\ \bibinfo {pages} {7077--7090} (\bibinfo {year}
  {2018})}\BibitemShut {NoStop}%
\bibitem [{\citenamefont {Li}\ \emph {et~al.}(2017)\citenamefont {Li},
  \citenamefont {Ando}, \citenamefont {Minamitani},\ and\ \citenamefont
  {Watanabe}}]{li2017study}%
  \BibitemOpen
  \bibfield  {author} {\bibinfo {author} {\bibfnamefont {W.}~\bibnamefont
  {Li}}, \bibinfo {author} {\bibfnamefont {Y.}~\bibnamefont {Ando}}, \bibinfo
  {author} {\bibfnamefont {E.}~\bibnamefont {Minamitani}}, \ and\ \bibinfo
  {author} {\bibfnamefont {S.}~\bibnamefont {Watanabe}},\ }\bibfield  {title}
  {\enquote {\bibinfo {title} {Study of li atom diffusion in amorphous li3po4
  with neural network potential},}\ }\href@noop {} {\bibfield  {journal}
  {\bibinfo  {journal} {The Journal of chemical physics}\ }\textbf {\bibinfo
  {volume} {147}},\ \bibinfo {pages} {214106} (\bibinfo {year}
  {2017})}\BibitemShut {NoStop}%
\bibitem [{\citenamefont {Deng}\ \emph {et~al.}(2019)\citenamefont {Deng},
  \citenamefont {Chen}, \citenamefont {Li},\ and\ \citenamefont
  {Ong}}]{deng2019electrostatic}%
  \BibitemOpen
  \bibfield  {author} {\bibinfo {author} {\bibfnamefont {Z.}~\bibnamefont
  {Deng}}, \bibinfo {author} {\bibfnamefont {C.}~\bibnamefont {Chen}}, \bibinfo
  {author} {\bibfnamefont {X.-G.}\ \bibnamefont {Li}}, \ and\ \bibinfo {author}
  {\bibfnamefont {S.~P.}\ \bibnamefont {Ong}},\ }\bibfield  {title} {\enquote
  {\bibinfo {title} {An electrostatic spectral neighbor analysis potential for
  lithium nitride},}\ }\href@noop {} {\bibfield  {journal} {\bibinfo  {journal}
  {npj Computational Materials}\ }\textbf {\bibinfo {volume} {5}},\ \bibinfo
  {pages} {75} (\bibinfo {year} {2019})}\BibitemShut {NoStop}%
\bibitem [{\citenamefont {Marcolongo}\ \emph {et~al.}(2019)\citenamefont
  {Marcolongo}, \citenamefont {Binninger}, \citenamefont {Zipoli},\ and\
  \citenamefont {Laino}}]{Aris2019}%
  \BibitemOpen
  \bibfield  {author} {\bibinfo {author} {\bibfnamefont {A.}~\bibnamefont
  {Marcolongo}}, \bibinfo {author} {\bibfnamefont {T.}~\bibnamefont
  {Binninger}}, \bibinfo {author} {\bibfnamefont {F.}~\bibnamefont {Zipoli}}, \
  and\ \bibinfo {author} {\bibfnamefont {T.}~\bibnamefont {Laino}},\ }\bibfield
   {title} {\enquote {\bibinfo {title} {Simulating diffusion properties of
  solid-state electrolytes via a neural network potential: Performance and
  training scheme},}\ }\href {\doibase 10.1002/syst.201900031} {\bibfield
  {journal} {\bibinfo  {journal} {ChemSystemsChem}\ }\textbf {\bibinfo {volume}
  {n/a}} (\bibinfo {year} {2019}),\ 10.1002/syst.201900031},\ \Eprint
  {http://arxiv.org/abs/https://onlinelibrary.wiley.com/doi/pdf/10.1002/syst.201900031}
  {https://onlinelibrary.wiley.com/doi/pdf/10.1002/syst.201900031} \BibitemShut
  {NoStop}%
\bibitem [{\citenamefont {Bernstein}, \citenamefont {Cs{\'a}nyi},\ and\
  \citenamefont {Deringer}(2019)}]{bernstein2019novo}%
  \BibitemOpen
  \bibfield  {author} {\bibinfo {author} {\bibfnamefont {N.}~\bibnamefont
  {Bernstein}}, \bibinfo {author} {\bibfnamefont {G.}~\bibnamefont
  {Cs{\'a}nyi}}, \ and\ \bibinfo {author} {\bibfnamefont {V.~L.}\ \bibnamefont
  {Deringer}},\ }\bibfield  {title} {\enquote {\bibinfo {title} {De novo
  exploration and self-guided learning of potential-energy surfaces},}\
  }\href@noop {} {\bibfield  {journal} {\bibinfo  {journal} {arXiv preprint
  arXiv:1905.10407}\ } (\bibinfo {year} {2019})}\BibitemShut {NoStop}%
\bibitem [{\citenamefont {Nyshadham}\ \emph {et~al.}(2019)\citenamefont
  {Nyshadham}, \citenamefont {Rupp}, \citenamefont {Bekker}, \citenamefont
  {Shapeev}, \citenamefont {Mueller}, \citenamefont {Rosenbrock}, \citenamefont
  {Cs{\'{a}}nyi}, \citenamefont {Wingate},\ and\ \citenamefont
  {Hart}}]{Nyshadham2019}%
  \BibitemOpen
  \bibfield  {author} {\bibinfo {author} {\bibfnamefont {C.}~\bibnamefont
  {Nyshadham}}, \bibinfo {author} {\bibfnamefont {M.}~\bibnamefont {Rupp}},
  \bibinfo {author} {\bibfnamefont {B.}~\bibnamefont {Bekker}}, \bibinfo
  {author} {\bibfnamefont {A.~V.}\ \bibnamefont {Shapeev}}, \bibinfo {author}
  {\bibfnamefont {T.}~\bibnamefont {Mueller}}, \bibinfo {author} {\bibfnamefont
  {C.~W.}\ \bibnamefont {Rosenbrock}}, \bibinfo {author} {\bibfnamefont
  {G.}~\bibnamefont {Cs{\'{a}}nyi}}, \bibinfo {author} {\bibfnamefont {D.~W.}\
  \bibnamefont {Wingate}}, \ and\ \bibinfo {author} {\bibfnamefont {G.~L.}\
  \bibnamefont {Hart}},\ }\bibfield  {title} {\enquote {\bibinfo {title}
  {{Machine-learned multi-system surrogate models for materials prediction}},}\
  }\href {\doibase 10.1038/s41524-019-0189-9} {\bibfield  {journal} {\bibinfo
  {journal} {npj Computational Materials}\ }\textbf {\bibinfo {volume} {5}},\
  \bibinfo {pages} {1--6} (\bibinfo {year} {2019})}\BibitemShut {NoStop}%
\bibitem [{\citenamefont {Deringer}\ \emph
  {et~al.}(2018{\natexlab{b}})\citenamefont {Deringer}, \citenamefont
  {Proserpio}, \citenamefont {Csányi},\ and\ \citenamefont
  {Pickard}}]{boron2018faraday}%
  \BibitemOpen
  \bibfield  {author} {\bibinfo {author} {\bibfnamefont {V.~L.}\ \bibnamefont
  {Deringer}}, \bibinfo {author} {\bibfnamefont {D.~M.}\ \bibnamefont
  {Proserpio}}, \bibinfo {author} {\bibfnamefont {G.}~\bibnamefont {Csányi}},
  \ and\ \bibinfo {author} {\bibfnamefont {C.~J.}\ \bibnamefont {Pickard}},\
  }\bibfield  {title} {\enquote {\bibinfo {title} {Data-driven learning and
  prediction of inorganic crystal structures},}\ }\href {\doibase
  10.1039/C8FD00034D} {\bibfield  {journal} {\bibinfo  {journal} {Faraday
  Discuss.}\ }\textbf {\bibinfo {volume} {211}},\ \bibinfo {pages} {45--59}
  (\bibinfo {year} {2018}{\natexlab{b}})}\BibitemShut {NoStop}%
\bibitem [{\citenamefont {Mortazavi}\ \emph {et~al.}(2020)\citenamefont
  {Mortazavi}, \citenamefont {Podryabinkin}, \citenamefont {Novikov},
  \citenamefont {Roche}, \citenamefont {Rabczuk}, \citenamefont {Zhuang},\ and\
  \citenamefont {Shapeev}}]{mortazavi2020-conductivity}%
  \BibitemOpen
  \bibfield  {author} {\bibinfo {author} {\bibfnamefont {B.}~\bibnamefont
  {Mortazavi}}, \bibinfo {author} {\bibfnamefont {E.}~\bibnamefont
  {Podryabinkin}}, \bibinfo {author} {\bibfnamefont {I.~S.}\ \bibnamefont
  {Novikov}}, \bibinfo {author} {\bibfnamefont {S.}~\bibnamefont {Roche}},
  \bibinfo {author} {\bibfnamefont {T.}~\bibnamefont {Rabczuk}}, \bibinfo
  {author} {\bibfnamefont {X.}~\bibnamefont {Zhuang}}, \ and\ \bibinfo {author}
  {\bibfnamefont {A.}~\bibnamefont {Shapeev}},\ }\bibfield  {title} {\enquote
  {\bibinfo {title} {Efficient machine-learning based interatomic potentials
  for exploring thermal conductivity in two-dimensional materials},}\ }\href
  {\doibase 10.1088/2515-7639/ab7cbb} {\bibfield  {journal} {\bibinfo
  {journal} {Journal of Physics: Materials}\ } (\bibinfo {year} {2020}),\
  10.1088/2515-7639/ab7cbb}\BibitemShut {NoStop}%
\bibitem [{\citenamefont {Podryabinkin}\ \emph {et~al.}(2019)\citenamefont
  {Podryabinkin}, \citenamefont {Tikhonov}, \citenamefont {Shapeev},\ and\
  \citenamefont {Oganov}}]{podryabinkin2019-uspex}%
  \BibitemOpen
  \bibfield  {author} {\bibinfo {author} {\bibfnamefont {E.~V.}\ \bibnamefont
  {Podryabinkin}}, \bibinfo {author} {\bibfnamefont {E.~V.}\ \bibnamefont
  {Tikhonov}}, \bibinfo {author} {\bibfnamefont {A.~V.}\ \bibnamefont
  {Shapeev}}, \ and\ \bibinfo {author} {\bibfnamefont {A.~R.}\ \bibnamefont
  {Oganov}},\ }\bibfield  {title} {\enquote {\bibinfo {title} {Accelerating
  crystal structure prediction by machine-learning interatomic potentials with
  active learning},}\ }\href {\doibase 10.1103/PhysRevB.99.064114} {\bibfield
  {journal} {\bibinfo  {journal} {Physical Review B}\ }\textbf {\bibinfo
  {volume} {99}},\ \bibinfo {pages} {064114} (\bibinfo {year}
  {2019})}\BibitemShut {NoStop}%
\bibitem [{\citenamefont {Gubaev}, \citenamefont {Podryabinkin},\ and\
  \citenamefont {Shapeev}(2018)}]{Gubaev2018}%
  \BibitemOpen
  \bibfield  {author} {\bibinfo {author} {\bibfnamefont {K.}~\bibnamefont
  {Gubaev}}, \bibinfo {author} {\bibfnamefont {E.~V.}\ \bibnamefont
  {Podryabinkin}}, \ and\ \bibinfo {author} {\bibfnamefont {A.~V.}\
  \bibnamefont {Shapeev}},\ }\bibfield  {title} {\enquote {\bibinfo {title}
  {Machine learning of molecular properties: Locality and active learning},}\
  }\href {\doibase 10.1063/1.5005095} {\bibfield  {journal} {\bibinfo
  {journal} {The Journal of Chemical Physics}\ }\textbf {\bibinfo {volume}
  {148}},\ \bibinfo {pages} {241727} (\bibinfo {year} {2018})}\BibitemShut
  {NoStop}%
\bibitem [{\citenamefont {Stamminger}\ \emph {et~al.}(2019)\citenamefont
  {Stamminger}, \citenamefont {Ziebarth}, \citenamefont {Mrovec}, \citenamefont
  {Hammerschmidt},\ and\ \citenamefont {Drautz}}]{stamminger2019ionic}%
  \BibitemOpen
  \bibfield  {author} {\bibinfo {author} {\bibfnamefont {A.~R.}\ \bibnamefont
  {Stamminger}}, \bibinfo {author} {\bibfnamefont {B.}~\bibnamefont
  {Ziebarth}}, \bibinfo {author} {\bibfnamefont {M.}~\bibnamefont {Mrovec}},
  \bibinfo {author} {\bibfnamefont {T.}~\bibnamefont {Hammerschmidt}}, \ and\
  \bibinfo {author} {\bibfnamefont {R.}~\bibnamefont {Drautz}},\ }\bibfield
  {title} {\enquote {\bibinfo {title} {Ionic conductivity and its dependence on
  structural disorder in halogenated argyrodites li6ps5x (x= br, cl, i)},}\
  }\href@noop {} {\bibfield  {journal} {\bibinfo  {journal} {Chemistry of
  Materials}\ }\textbf {\bibinfo {volume} {31}},\ \bibinfo {pages} {8673--8678}
  (\bibinfo {year} {2019})}\BibitemShut {NoStop}%
\bibitem [{\citenamefont {Ohno}\ \emph {et~al.}(2019)\citenamefont {Ohno},
  \citenamefont {Helm}, \citenamefont {Fuchs}, \citenamefont {Dewald},
  \citenamefont {Kraft}, \citenamefont {Culver}, \citenamefont {Senyshyn},\
  and\ \citenamefont {Zeier}}]{ohno2019further}%
  \BibitemOpen
  \bibfield  {author} {\bibinfo {author} {\bibfnamefont {S.}~\bibnamefont
  {Ohno}}, \bibinfo {author} {\bibfnamefont {B.}~\bibnamefont {Helm}}, \bibinfo
  {author} {\bibfnamefont {T.}~\bibnamefont {Fuchs}}, \bibinfo {author}
  {\bibfnamefont {G.}~\bibnamefont {Dewald}}, \bibinfo {author} {\bibfnamefont
  {M.~A.}\ \bibnamefont {Kraft}}, \bibinfo {author} {\bibfnamefont {S.~P.}\
  \bibnamefont {Culver}}, \bibinfo {author} {\bibfnamefont {A.}~\bibnamefont
  {Senyshyn}}, \ and\ \bibinfo {author} {\bibfnamefont {W.~G.}\ \bibnamefont
  {Zeier}},\ }\bibfield  {title} {\enquote {\bibinfo {title} {Further evidence
  for energy landscape flattening in the superionic argyrodites li6+ x p1--x m
  x s5i (m= si, ge, sn)},}\ }\href@noop {} {\bibfield  {journal} {\bibinfo
  {journal} {Chemistry of Materials}\ }\textbf {\bibinfo {volume} {31}},\
  \bibinfo {pages} {4936--4944} (\bibinfo {year} {2019})}\BibitemShut {NoStop}%
\bibitem [{\citenamefont {Zhou}\ \emph {et~al.}(2019)\citenamefont {Zhou},
  \citenamefont {Assoud}, \citenamefont {Shyamsunder}, \citenamefont {Huq},
  \citenamefont {Zhang}, \citenamefont {Hartmann}, \citenamefont {Kulisch},\
  and\ \citenamefont {Nazar}}]{zhou2019entropically}%
  \BibitemOpen
  \bibfield  {author} {\bibinfo {author} {\bibfnamefont {L.}~\bibnamefont
  {Zhou}}, \bibinfo {author} {\bibfnamefont {A.}~\bibnamefont {Assoud}},
  \bibinfo {author} {\bibfnamefont {A.}~\bibnamefont {Shyamsunder}}, \bibinfo
  {author} {\bibfnamefont {A.}~\bibnamefont {Huq}}, \bibinfo {author}
  {\bibfnamefont {Q.}~\bibnamefont {Zhang}}, \bibinfo {author} {\bibfnamefont
  {P.}~\bibnamefont {Hartmann}}, \bibinfo {author} {\bibfnamefont
  {J.}~\bibnamefont {Kulisch}}, \ and\ \bibinfo {author} {\bibfnamefont
  {L.~F.}\ \bibnamefont {Nazar}},\ }\bibfield  {title} {\enquote {\bibinfo
  {title} {An entropically stabilized fast-ion conductor: Li3. 25 [si0. 25p0.
  75] s4},}\ }\href@noop {} {\bibfield  {journal} {\bibinfo  {journal}
  {Chemistry of Materials}\ }\textbf {\bibinfo {volume} {31}},\ \bibinfo
  {pages} {7801--7811} (\bibinfo {year} {2019})}\BibitemShut {NoStop}%
\bibitem [{\citenamefont {Hanghofer}\ \emph {et~al.}(2019)\citenamefont
  {Hanghofer}, \citenamefont {Brinek}, \citenamefont {Eisbacher}, \citenamefont
  {Bitschnau}, \citenamefont {Volck}, \citenamefont {Hennige}, \citenamefont
  {Hanzu}, \citenamefont {Rettenwander},\ and\ \citenamefont
  {Wilkening}}]{hanghofer2019substitutional}%
  \BibitemOpen
  \bibfield  {author} {\bibinfo {author} {\bibfnamefont {I.}~\bibnamefont
  {Hanghofer}}, \bibinfo {author} {\bibfnamefont {M.}~\bibnamefont {Brinek}},
  \bibinfo {author} {\bibfnamefont {S.}~\bibnamefont {Eisbacher}}, \bibinfo
  {author} {\bibfnamefont {B.}~\bibnamefont {Bitschnau}}, \bibinfo {author}
  {\bibfnamefont {M.}~\bibnamefont {Volck}}, \bibinfo {author} {\bibfnamefont
  {V.}~\bibnamefont {Hennige}}, \bibinfo {author} {\bibfnamefont
  {I.}~\bibnamefont {Hanzu}}, \bibinfo {author} {\bibfnamefont
  {D.}~\bibnamefont {Rettenwander}}, \ and\ \bibinfo {author} {\bibfnamefont
  {H.}~\bibnamefont {Wilkening}},\ }\bibfield  {title} {\enquote {\bibinfo
  {title} {Substitutional disorder: Structure and ion dynamics of the
  argyrodites li 6 ps 5 cl, li 6 ps 5 br and li 6 ps 5 i},}\ }\href@noop {}
  {\bibfield  {journal} {\bibinfo  {journal} {Physical Chemistry Chemical
  Physics}\ }\textbf {\bibinfo {volume} {21}},\ \bibinfo {pages} {8489--8507}
  (\bibinfo {year} {2019})}\BibitemShut {NoStop}%
\bibitem [{\citenamefont {Zhang}\ \emph {et~al.}(2019)\citenamefont {Zhang},
  \citenamefont {Lin}, \citenamefont {Wang}, \citenamefont {Car},\ and\
  \citenamefont {E}}]{zhang2019active}%
  \BibitemOpen
  \bibfield  {author} {\bibinfo {author} {\bibfnamefont {L.}~\bibnamefont
  {Zhang}}, \bibinfo {author} {\bibfnamefont {D.-Y.}\ \bibnamefont {Lin}},
  \bibinfo {author} {\bibfnamefont {H.}~\bibnamefont {Wang}}, \bibinfo {author}
  {\bibfnamefont {R.}~\bibnamefont {Car}}, \ and\ \bibinfo {author}
  {\bibfnamefont {W.}~\bibnamefont {E}},\ }\bibfield  {title} {\enquote
  {\bibinfo {title} {Active learning of uniformly accurate interatomic
  potentials for materials simulation},}\ }\href@noop {} {\bibfield  {journal}
  {\bibinfo  {journal} {Physical Review Materials}\ }\textbf {\bibinfo {volume}
  {3}},\ \bibinfo {pages} {023804} (\bibinfo {year} {2019})}\BibitemShut
  {NoStop}%
\bibitem [{\citenamefont {Zhang}\ \emph {et~al.}(2020)\citenamefont {Zhang},
  \citenamefont {Wang}, \citenamefont {Chen}, \citenamefont {Zeng},
  \citenamefont {Zhang}, \citenamefont {Wang},\ and\ \citenamefont
  {E}}]{zhang2020dpgen}%
  \BibitemOpen
  \bibfield  {author} {\bibinfo {author} {\bibfnamefont {Y.}~\bibnamefont
  {Zhang}}, \bibinfo {author} {\bibfnamefont {H.}~\bibnamefont {Wang}},
  \bibinfo {author} {\bibfnamefont {W.}~\bibnamefont {Chen}}, \bibinfo {author}
  {\bibfnamefont {J.}~\bibnamefont {Zeng}}, \bibinfo {author} {\bibfnamefont
  {L.}~\bibnamefont {Zhang}}, \bibinfo {author} {\bibfnamefont
  {H.}~\bibnamefont {Wang}}, \ and\ \bibinfo {author} {\bibfnamefont
  {W.}~\bibnamefont {E}},\ }\bibfield  {title} {\enquote {\bibinfo {title}
  {Dp-gen: A concurrent learning platform for the generation of reliable deep
  learning based potential energy models},}\ }\href {\doibase
  https://doi.org/10.1016/j.cpc.2020.107206} {\bibfield  {journal} {\bibinfo
  {journal} {Computer Physics Communications}\ ,\ \bibinfo {pages} {107206}}
  (\bibinfo {year} {2020})}\BibitemShut {NoStop}%
\bibitem [{\citenamefont {Jain}\ \emph {et~al.}(2013)\citenamefont {Jain},
  \citenamefont {Ong}, \citenamefont {Hautier}, \citenamefont {Chen},
  \citenamefont {Richards}, \citenamefont {Dacek}, \citenamefont {Cholia},
  \citenamefont {Gunter}, \citenamefont {Skinner}, \citenamefont {Ceder} \emph
  {et~al.}}]{jain2013commentary}%
  \BibitemOpen
  \bibfield  {author} {\bibinfo {author} {\bibfnamefont {A.}~\bibnamefont
  {Jain}}, \bibinfo {author} {\bibfnamefont {S.~P.}\ \bibnamefont {Ong}},
  \bibinfo {author} {\bibfnamefont {G.}~\bibnamefont {Hautier}}, \bibinfo
  {author} {\bibfnamefont {W.}~\bibnamefont {Chen}}, \bibinfo {author}
  {\bibfnamefont {W.~D.}\ \bibnamefont {Richards}}, \bibinfo {author}
  {\bibfnamefont {S.}~\bibnamefont {Dacek}}, \bibinfo {author} {\bibfnamefont
  {S.}~\bibnamefont {Cholia}}, \bibinfo {author} {\bibfnamefont
  {D.}~\bibnamefont {Gunter}}, \bibinfo {author} {\bibfnamefont
  {D.}~\bibnamefont {Skinner}}, \bibinfo {author} {\bibfnamefont
  {G.}~\bibnamefont {Ceder}},  \emph {et~al.},\ }\bibfield  {title} {\enquote
  {\bibinfo {title} {Commentary: The materials project: A materials genome
  approach to accelerating materials innovation},}\ }\href@noop {} {\bibfield
  {journal} {\bibinfo  {journal} {Apl Materials}\ }\textbf {\bibinfo {volume}
  {1}},\ \bibinfo {pages} {011002} (\bibinfo {year} {2013})}\BibitemShut
  {NoStop}%
\bibitem [{\citenamefont {Ong}\ \emph {et~al.}(2015)\citenamefont {Ong},
  \citenamefont {Cholia}, \citenamefont {Jain}, \citenamefont {Brafman},
  \citenamefont {Gunter}, \citenamefont {Ceder},\ and\ \citenamefont
  {Persson}}]{ong2015materials}%
  \BibitemOpen
  \bibfield  {author} {\bibinfo {author} {\bibfnamefont {S.~P.}\ \bibnamefont
  {Ong}}, \bibinfo {author} {\bibfnamefont {S.}~\bibnamefont {Cholia}},
  \bibinfo {author} {\bibfnamefont {A.}~\bibnamefont {Jain}}, \bibinfo {author}
  {\bibfnamefont {M.}~\bibnamefont {Brafman}}, \bibinfo {author} {\bibfnamefont
  {D.}~\bibnamefont {Gunter}}, \bibinfo {author} {\bibfnamefont
  {G.}~\bibnamefont {Ceder}}, \ and\ \bibinfo {author} {\bibfnamefont {K.~A.}\
  \bibnamefont {Persson}},\ }\bibfield  {title} {\enquote {\bibinfo {title}
  {The materials application programming interface (api): A simple, flexible
  and efficient api for materials data based on representational state transfer
  (rest) principles},}\ }\href@noop {} {\bibfield  {journal} {\bibinfo
  {journal} {Computational Materials Science}\ }\textbf {\bibinfo {volume}
  {97}},\ \bibinfo {pages} {209--215} (\bibinfo {year} {2015})}\BibitemShut
  {NoStop}%
\bibitem [{\citenamefont {Ong}\ \emph {et~al.}(2013{\natexlab{b}})\citenamefont
  {Ong}, \citenamefont {Richards}, \citenamefont {Jain}, \citenamefont
  {Hautier}, \citenamefont {Kocher}, \citenamefont {Cholia}, \citenamefont
  {Gunter}, \citenamefont {Chevrier}, \citenamefont {Persson},\ and\
  \citenamefont {Ceder}}]{ong2013python}%
  \BibitemOpen
  \bibfield  {author} {\bibinfo {author} {\bibfnamefont {S.~P.}\ \bibnamefont
  {Ong}}, \bibinfo {author} {\bibfnamefont {W.~D.}\ \bibnamefont {Richards}},
  \bibinfo {author} {\bibfnamefont {A.}~\bibnamefont {Jain}}, \bibinfo {author}
  {\bibfnamefont {G.}~\bibnamefont {Hautier}}, \bibinfo {author} {\bibfnamefont
  {M.}~\bibnamefont {Kocher}}, \bibinfo {author} {\bibfnamefont
  {S.}~\bibnamefont {Cholia}}, \bibinfo {author} {\bibfnamefont
  {D.}~\bibnamefont {Gunter}}, \bibinfo {author} {\bibfnamefont {V.~L.}\
  \bibnamefont {Chevrier}}, \bibinfo {author} {\bibfnamefont {K.~A.}\
  \bibnamefont {Persson}}, \ and\ \bibinfo {author} {\bibfnamefont
  {G.}~\bibnamefont {Ceder}},\ }\bibfield  {title} {\enquote {\bibinfo {title}
  {Python materials genomics (pymatgen): A robust, open-source python library
  for materials analysis},}\ }\href@noop {} {\bibfield  {journal} {\bibinfo
  {journal} {Computational Materials Science}\ }\textbf {\bibinfo {volume}
  {68}},\ \bibinfo {pages} {314--319} (\bibinfo {year}
  {2013}{\natexlab{b}})}\BibitemShut {NoStop}%
\bibitem [{\citenamefont {Wang}\ \emph {et~al.}(2018)\citenamefont {Wang},
  \citenamefont {Zhang}, \citenamefont {Han},\ and\ \citenamefont
  {E}}]{wang2018deepmd}%
  \BibitemOpen
  \bibfield  {author} {\bibinfo {author} {\bibfnamefont {H.}~\bibnamefont
  {Wang}}, \bibinfo {author} {\bibfnamefont {L.}~\bibnamefont {Zhang}},
  \bibinfo {author} {\bibfnamefont {J.}~\bibnamefont {Han}}, \ and\ \bibinfo
  {author} {\bibfnamefont {W.}~\bibnamefont {E}},\ }\bibfield  {title}
  {\enquote {\bibinfo {title} {Deepmd-kit: A deep learning package for
  many-body potential energy representation and molecular dynamics},}\
  }\href@noop {} {\bibfield  {journal} {\bibinfo  {journal} {Computer Physics
  Communications}\ }\textbf {\bibinfo {volume} {228}},\ \bibinfo {pages}
  {178--184} (\bibinfo {year} {2018})}\BibitemShut {NoStop}%
\bibitem [{\citenamefont {Bl{\"o}chl}(1994)}]{blochl1994projector}%
  \BibitemOpen
  \bibfield  {author} {\bibinfo {author} {\bibfnamefont {P.~E.}\ \bibnamefont
  {Bl{\"o}chl}},\ }\bibfield  {title} {\enquote {\bibinfo {title} {Projector
  augmented-wave method},}\ }\href@noop {} {\bibfield  {journal} {\bibinfo
  {journal} {Physical review B}\ }\textbf {\bibinfo {volume} {50}},\ \bibinfo
  {pages} {17953} (\bibinfo {year} {1994})}\BibitemShut {NoStop}%
\bibitem [{\citenamefont {Kresse}\ and\ \citenamefont
  {Furthm{\"u}ller}(1996)}]{kresse1996efficient}%
  \BibitemOpen
  \bibfield  {author} {\bibinfo {author} {\bibfnamefont {G.}~\bibnamefont
  {Kresse}}\ and\ \bibinfo {author} {\bibfnamefont {J.}~\bibnamefont
  {Furthm{\"u}ller}},\ }\bibfield  {title} {\enquote {\bibinfo {title}
  {Efficient iterative schemes for ab initio total-energy calculations using a
  plane-wave basis set},}\ }\href@noop {} {\bibfield  {journal} {\bibinfo
  {journal} {Physical review B}\ }\textbf {\bibinfo {volume} {54}},\ \bibinfo
  {pages} {11169} (\bibinfo {year} {1996})}\BibitemShut {NoStop}%
\bibitem [{\citenamefont {Kresse}\ and\ \citenamefont
  {Joubert}(1999)}]{kresse1999ultrasoft}%
  \BibitemOpen
  \bibfield  {author} {\bibinfo {author} {\bibfnamefont {G.}~\bibnamefont
  {Kresse}}\ and\ \bibinfo {author} {\bibfnamefont {D.}~\bibnamefont
  {Joubert}},\ }\bibfield  {title} {\enquote {\bibinfo {title} {From ultrasoft
  pseudopotentials to the projector augmented-wave method},}\ }\href@noop {}
  {\bibfield  {journal} {\bibinfo  {journal} {Physical Review B}\ }\textbf
  {\bibinfo {volume} {59}},\ \bibinfo {pages} {1758} (\bibinfo {year}
  {1999})}\BibitemShut {NoStop}%
\bibitem [{\citenamefont {Kohn}\ and\ \citenamefont {Sham}(1965)}]{lda}%
  \BibitemOpen
  \bibfield  {author} {\bibinfo {author} {\bibfnamefont {W.}~\bibnamefont
  {Kohn}}\ and\ \bibinfo {author} {\bibfnamefont {L.~J.}\ \bibnamefont
  {Sham}},\ }\bibfield  {title} {\enquote {\bibinfo {title} {Self-consistent
  equations including exchange and correlation effects},}\ }\href {\doibase
  10.1103/PhysRev.140.A1133} {\bibfield  {journal} {\bibinfo  {journal} {Phys.
  Rev.}\ }\textbf {\bibinfo {volume} {140}},\ \bibinfo {pages} {A1133--A1138}
  (\bibinfo {year} {1965})}\BibitemShut {NoStop}%
\bibitem [{\citenamefont {Perdew}, \citenamefont {Burke},\ and\ \citenamefont
  {Ernzerhof}(1996{\natexlab{a}})}]{gga-pbe}%
  \BibitemOpen
  \bibfield  {author} {\bibinfo {author} {\bibfnamefont {J.~P.}\ \bibnamefont
  {Perdew}}, \bibinfo {author} {\bibfnamefont {K.}~\bibnamefont {Burke}}, \
  and\ \bibinfo {author} {\bibfnamefont {M.}~\bibnamefont {Ernzerhof}},\
  }\bibfield  {title} {\enquote {\bibinfo {title} {Generalized gradient
  approximation made simple},}\ }\href {\doibase 10.1103/PhysRevLett.77.3865}
  {\bibfield  {journal} {\bibinfo  {journal} {Phys. Rev. Lett.}\ }\textbf
  {\bibinfo {volume} {77}},\ \bibinfo {pages} {3865--3868} (\bibinfo {year}
  {1996}{\natexlab{a}})}\BibitemShut {NoStop}%
\bibitem [{\citenamefont {Perdew}\ \emph {et~al.}(2008)\citenamefont {Perdew},
  \citenamefont {Ruzsinszky}, \citenamefont {Csonka}, \citenamefont {Vydrov},
  \citenamefont {Scuseria}, \citenamefont {Constantin}, \citenamefont {Zhou},\
  and\ \citenamefont {Burke}}]{pbesol}%
  \BibitemOpen
  \bibfield  {author} {\bibinfo {author} {\bibfnamefont {J.~P.}\ \bibnamefont
  {Perdew}}, \bibinfo {author} {\bibfnamefont {A.}~\bibnamefont {Ruzsinszky}},
  \bibinfo {author} {\bibfnamefont {G.~I.}\ \bibnamefont {Csonka}}, \bibinfo
  {author} {\bibfnamefont {O.~A.}\ \bibnamefont {Vydrov}}, \bibinfo {author}
  {\bibfnamefont {G.~E.}\ \bibnamefont {Scuseria}}, \bibinfo {author}
  {\bibfnamefont {L.~A.}\ \bibnamefont {Constantin}}, \bibinfo {author}
  {\bibfnamefont {X.}~\bibnamefont {Zhou}}, \ and\ \bibinfo {author}
  {\bibfnamefont {K.}~\bibnamefont {Burke}},\ }\bibfield  {title} {\enquote
  {\bibinfo {title} {Restoring the density-gradient expansion for exchange in
  solids and surfaces},}\ }\href {\doibase 10.1103/PhysRevLett.100.136406}
  {\bibfield  {journal} {\bibinfo  {journal} {Phys. Rev. Lett.}\ }\textbf
  {\bibinfo {volume} {100}},\ \bibinfo {pages} {136406} (\bibinfo {year}
  {2008})}\BibitemShut {NoStop}%
\bibitem [{\citenamefont {Sun}\ \emph {et~al.}(2015)\citenamefont {Sun},
  \citenamefont {Remsing}, \citenamefont {Zhang}, \citenamefont {Sun},
  \citenamefont {Ruzsinszky}, \citenamefont {Peng}, \citenamefont {Yang},
  \citenamefont {Paul}, \citenamefont {Waghmare}, \citenamefont {Wu} \emph
  {et~al.}}]{sun2015scan}%
  \BibitemOpen
  \bibfield  {author} {\bibinfo {author} {\bibfnamefont {J.}~\bibnamefont
  {Sun}}, \bibinfo {author} {\bibfnamefont {R.~C.}\ \bibnamefont {Remsing}},
  \bibinfo {author} {\bibfnamefont {Y.}~\bibnamefont {Zhang}}, \bibinfo
  {author} {\bibfnamefont {Z.}~\bibnamefont {Sun}}, \bibinfo {author}
  {\bibfnamefont {A.}~\bibnamefont {Ruzsinszky}}, \bibinfo {author}
  {\bibfnamefont {H.}~\bibnamefont {Peng}}, \bibinfo {author} {\bibfnamefont
  {Z.}~\bibnamefont {Yang}}, \bibinfo {author} {\bibfnamefont {A.}~\bibnamefont
  {Paul}}, \bibinfo {author} {\bibfnamefont {U.}~\bibnamefont {Waghmare}},
  \bibinfo {author} {\bibfnamefont {X.}~\bibnamefont {Wu}},  \emph {et~al.},\
  }\bibfield  {title} {\enquote {\bibinfo {title} {Scan: An efficient density
  functional yielding accurate structures and energies of diversely-bonded
  materials},}\ }\href@noop {} {\bibfield  {journal} {\bibinfo  {journal}
  {arXiv preprint arXiv:1511.01089}\ } (\bibinfo {year} {2015})}\BibitemShut
  {NoStop}%
\bibitem [{\citenamefont {Perdew}, \citenamefont {Ernzerhof},\ and\
  \citenamefont {Burke}(1996)}]{pbe0-1}%
  \BibitemOpen
  \bibfield  {author} {\bibinfo {author} {\bibfnamefont {J.~P.}\ \bibnamefont
  {Perdew}}, \bibinfo {author} {\bibfnamefont {M.}~\bibnamefont {Ernzerhof}}, \
  and\ \bibinfo {author} {\bibfnamefont {K.}~\bibnamefont {Burke}},\ }\bibfield
   {title} {\enquote {\bibinfo {title} {Rationale for mixing exact exchange
  with density functional approximations},}\ }\href {\doibase 10.1063/1.472933}
  {\bibfield  {journal} {\bibinfo  {journal} {The Journal of Chemical Physics}\
  }\textbf {\bibinfo {volume} {105}},\ \bibinfo {pages} {9982--9985} (\bibinfo
  {year} {1996})}\BibitemShut {NoStop}%
\bibitem [{\citenamefont {Adamo}\ and\ \citenamefont {Barone}(1999)}]{pbe0-2}%
  \BibitemOpen
  \bibfield  {author} {\bibinfo {author} {\bibfnamefont {C.}~\bibnamefont
  {Adamo}}\ and\ \bibinfo {author} {\bibfnamefont {V.}~\bibnamefont {Barone}},\
  }\bibfield  {title} {\enquote {\bibinfo {title} {Toward reliable density
  functional methods without adjustable parameters: The pbe0 model},}\ }\href
  {\doibase 10.1063/1.478522} {\bibfield  {journal} {\bibinfo  {journal} {The
  Journal of Chemical Physics}\ }\textbf {\bibinfo {volume} {110}},\ \bibinfo
  {pages} {6158--6170} (\bibinfo {year} {1999})}\BibitemShut {NoStop}%
\bibitem [{\citenamefont {Kuhn}, \citenamefont {Duppel},\ and\ \citenamefont
  {Lotsch}(2013)}]{C3EE41728J}%
  \BibitemOpen
  \bibfield  {author} {\bibinfo {author} {\bibfnamefont {A.}~\bibnamefont
  {Kuhn}}, \bibinfo {author} {\bibfnamefont {V.}~\bibnamefont {Duppel}}, \ and\
  \bibinfo {author} {\bibfnamefont {B.~V.}\ \bibnamefont {Lotsch}},\ }\bibfield
   {title} {\enquote {\bibinfo {title} {Tetragonal li10gep2s12 and li7geps8 –
  exploring the li ion dynamics in lgps li electrolytes},}\ }\href {\doibase
  10.1039/C3EE41728J} {\bibfield  {journal} {\bibinfo  {journal} {Energy
  Environ. Sci.}\ }\textbf {\bibinfo {volume} {6}},\ \bibinfo {pages}
  {3548--3552} (\bibinfo {year} {2013})}\BibitemShut {NoStop}%
\bibitem [{\citenamefont {Whiteley}\ \emph {et~al.}(2014)\citenamefont
  {Whiteley}, \citenamefont {Woo}, \citenamefont {Hu}, \citenamefont {Nam},\
  and\ \citenamefont {Lee}}]{whiteley2014empowering}%
  \BibitemOpen
  \bibfield  {author} {\bibinfo {author} {\bibfnamefont {J.~M.}\ \bibnamefont
  {Whiteley}}, \bibinfo {author} {\bibfnamefont {J.~H.}\ \bibnamefont {Woo}},
  \bibinfo {author} {\bibfnamefont {E.}~\bibnamefont {Hu}}, \bibinfo {author}
  {\bibfnamefont {K.-W.}\ \bibnamefont {Nam}}, \ and\ \bibinfo {author}
  {\bibfnamefont {S.-H.}\ \bibnamefont {Lee}},\ }\bibfield  {title} {\enquote
  {\bibinfo {title} {Empowering the lithium metal battery through a
  silicon-based superionic conductor},}\ }\href@noop {} {\bibfield  {journal}
  {\bibinfo  {journal} {Journal of the Electrochemical Society}\ }\textbf
  {\bibinfo {volume} {161}},\ \bibinfo {pages} {A1812--A1817} (\bibinfo {year}
  {2014})}\BibitemShut {NoStop}%
\bibitem [{\citenamefont {Plimpton}(1995)}]{plimpton1995fast}%
  \BibitemOpen
  \bibfield  {author} {\bibinfo {author} {\bibfnamefont {S.}~\bibnamefont
  {Plimpton}},\ }\bibfield  {title} {\enquote {\bibinfo {title} {Fast parallel
  algorithms for short-range molecular dynamics},}\ }\href@noop {} {\bibfield
  {journal} {\bibinfo  {journal} {Journal of computational physics}\ }\textbf
  {\bibinfo {volume} {117}},\ \bibinfo {pages} {1--19} (\bibinfo {year}
  {1995})}\BibitemShut {NoStop}%
\bibitem [{\citenamefont {Weber}\ \emph {et~al.}(2016)\citenamefont {Weber},
  \citenamefont {Senyshyn}, \citenamefont {Weldert}, \citenamefont {Wenzel},
  \citenamefont {Zhang}, \citenamefont {Kaiser}, \citenamefont {Berendts},
  \citenamefont {Janek},\ and\ \citenamefont {Zeier}}]{Weber2016}%
  \BibitemOpen
  \bibfield  {author} {\bibinfo {author} {\bibfnamefont {D.~A.}\ \bibnamefont
  {Weber}}, \bibinfo {author} {\bibfnamefont {A.}~\bibnamefont {Senyshyn}},
  \bibinfo {author} {\bibfnamefont {K.~S.}\ \bibnamefont {Weldert}}, \bibinfo
  {author} {\bibfnamefont {S.}~\bibnamefont {Wenzel}}, \bibinfo {author}
  {\bibfnamefont {W.}~\bibnamefont {Zhang}}, \bibinfo {author} {\bibfnamefont
  {R.}~\bibnamefont {Kaiser}}, \bibinfo {author} {\bibfnamefont
  {S.}~\bibnamefont {Berendts}}, \bibinfo {author} {\bibfnamefont
  {J.}~\bibnamefont {Janek}}, \ and\ \bibinfo {author} {\bibfnamefont {W.~G.}\
  \bibnamefont {Zeier}},\ }\bibfield  {title} {\enquote {\bibinfo {title}
  {{Structural Insights and 3D Diffusion Pathways within the Lithium Superionic
  Conductor Li 10 GeP 2 S 12}},}\ }\href {\doibase
  10.1021/acs.chemmater.6b02424} {\bibfield  {journal} {\bibinfo  {journal}
  {Chemistry of Materials}\ }\textbf {\bibinfo {volume} {28}},\ \bibinfo
  {pages} {5905--5915} (\bibinfo {year} {2016})}\BibitemShut {NoStop}%
\bibitem [{\citenamefont {Yeh}\ and\ \citenamefont
  {Hummer}(2004)}]{yeh2004system}%
  \BibitemOpen
  \bibfield  {author} {\bibinfo {author} {\bibfnamefont {I.-C.}\ \bibnamefont
  {Yeh}}\ and\ \bibinfo {author} {\bibfnamefont {G.}~\bibnamefont {Hummer}},\
  }\bibfield  {title} {\enquote {\bibinfo {title} {System-size dependence of
  diffusion coefficients and viscosities from molecular dynamics simulations
  with periodic boundary conditions},}\ }\href {\doibase 10.1021/jp0477147}
  {\bibfield  {journal} {\bibinfo  {journal} {The Journal of Physical Chemistry
  B}\ }\textbf {\bibinfo {volume} {108}},\ \bibinfo {pages} {15873--15879}
  (\bibinfo {year} {2004})}\BibitemShut {NoStop}%
\bibitem [{\citenamefont {Ong}\ \emph {et~al.}(2013{\natexlab{c}})\citenamefont
  {Ong}, \citenamefont {Mo}, \citenamefont {Richards}, \citenamefont {Miara},
  \citenamefont {Lee},\ and\ \citenamefont {Ceder}}]{Ong2013}%
  \BibitemOpen
  \bibfield  {author} {\bibinfo {author} {\bibfnamefont {S.~P.}\ \bibnamefont
  {Ong}}, \bibinfo {author} {\bibfnamefont {Y.}~\bibnamefont {Mo}}, \bibinfo
  {author} {\bibfnamefont {W.~D.}\ \bibnamefont {Richards}}, \bibinfo {author}
  {\bibfnamefont {L.}~\bibnamefont {Miara}}, \bibinfo {author} {\bibfnamefont
  {H.~S.}\ \bibnamefont {Lee}}, \ and\ \bibinfo {author} {\bibfnamefont
  {G.}~\bibnamefont {Ceder}},\ }\bibfield  {title} {\enquote {\bibinfo {title}
  {{Phase stability, electrochemical stability and ionic conductivity of the
  Li10+-1MP2X12 (M = Ge, Si, Sn, Al or P, and X = O, S or Se) family of
  superionic conductors}},}\ }\href {\doibase 10.1039/c2ee23355j} {\bibfield
  {journal} {\bibinfo  {journal} {Energy and Environmental Science}\ }\textbf
  {\bibinfo {volume} {6}},\ \bibinfo {pages} {148--156} (\bibinfo {year}
  {2013}{\natexlab{c}})}\BibitemShut {NoStop}%
\bibitem [{\citenamefont {Bachman}\ \emph {et~al.}(2016)\citenamefont
  {Bachman}, \citenamefont {Muy}, \citenamefont {Grimaud}, \citenamefont
  {Chang}, \citenamefont {Pour}, \citenamefont {Lux}, \citenamefont {Paschos},
  \citenamefont {Maglia}, \citenamefont {Lupart}, \citenamefont {Lamp},
  \citenamefont {Giordano},\ and\ \citenamefont {Shao-Horn}}]{shao2016CR}%
  \BibitemOpen
  \bibfield  {author} {\bibinfo {author} {\bibfnamefont {J.~C.}\ \bibnamefont
  {Bachman}}, \bibinfo {author} {\bibfnamefont {S.}~\bibnamefont {Muy}},
  \bibinfo {author} {\bibfnamefont {A.}~\bibnamefont {Grimaud}}, \bibinfo
  {author} {\bibfnamefont {H.-H.}\ \bibnamefont {Chang}}, \bibinfo {author}
  {\bibfnamefont {N.}~\bibnamefont {Pour}}, \bibinfo {author} {\bibfnamefont
  {S.~F.}\ \bibnamefont {Lux}}, \bibinfo {author} {\bibfnamefont
  {O.}~\bibnamefont {Paschos}}, \bibinfo {author} {\bibfnamefont
  {F.}~\bibnamefont {Maglia}}, \bibinfo {author} {\bibfnamefont
  {S.}~\bibnamefont {Lupart}}, \bibinfo {author} {\bibfnamefont
  {P.}~\bibnamefont {Lamp}}, \bibinfo {author} {\bibfnamefont {L.}~\bibnamefont
  {Giordano}}, \ and\ \bibinfo {author} {\bibfnamefont {Y.}~\bibnamefont
  {Shao-Horn}},\ }\bibfield  {title} {\enquote {\bibinfo {title} {Inorganic
  solid-state electrolytes for lithium batteries: Mechanisms and properties
  governing ion conduction},}\ }\href {\doibase 10.1021/acs.chemrev.5b00563}
  {\bibfield  {journal} {\bibinfo  {journal} {Chemical Reviews}\ }\textbf
  {\bibinfo {volume} {116}},\ \bibinfo {pages} {140--162} (\bibinfo {year}
  {2016})},\ \bibinfo {note} {pMID: 26713396},\ \Eprint
  {http://arxiv.org/abs/https://doi.org/10.1021/acs.chemrev.5b00563}
  {https://doi.org/10.1021/acs.chemrev.5b00563} \BibitemShut {NoStop}%
\bibitem [{\citenamefont {Fitzhugh}, \citenamefont {Ye},\ and\ \citenamefont
  {Li}(2019)}]{C9TA05248H}%
  \BibitemOpen
  \bibfield  {author} {\bibinfo {author} {\bibfnamefont {W.}~\bibnamefont
  {Fitzhugh}}, \bibinfo {author} {\bibfnamefont {L.}~\bibnamefont {Ye}}, \ and\
  \bibinfo {author} {\bibfnamefont {X.}~\bibnamefont {Li}},\ }\bibfield
  {title} {\enquote {\bibinfo {title} {The effects of mechanical constriction
  on the operation of sulfide based solid-state batteries},}\ }\href {\doibase
  10.1039/C9TA05248H} {\bibfield  {journal} {\bibinfo  {journal} {J. Mater.
  Chem. A}\ }\textbf {\bibinfo {volume} {7}},\ \bibinfo {pages} {23604--23627}
  (\bibinfo {year} {2019})}\BibitemShut {NoStop}%
\bibitem [{\citenamefont {Ohno}\ \emph {et~al.}(2020)\citenamefont {Ohno},
  \citenamefont {Bernges}, \citenamefont {Buchheim}, \citenamefont {Duchardt},
  \citenamefont {Hatz}, \citenamefont {Kraft}, \citenamefont {Kwak},
  \citenamefont {Santhosha}, \citenamefont {Liu}, \citenamefont {Minafra},
  \citenamefont {Tsuji}, \citenamefont {Sakuda}, \citenamefont {Schlem},
  \citenamefont {Xiong}, \citenamefont {Zhang}, \citenamefont {Adelhelm},
  \citenamefont {Chen}, \citenamefont {Hayashi}, \citenamefont {Jung},
  \citenamefont {Lotsch}, \citenamefont {Roling}, \citenamefont
  {Vargas-Barbosa},\ and\ \citenamefont {Zeier}}]{zeier2020centainty}%
  \BibitemOpen
  \bibfield  {author} {\bibinfo {author} {\bibfnamefont {S.}~\bibnamefont
  {Ohno}}, \bibinfo {author} {\bibfnamefont {T.}~\bibnamefont {Bernges}},
  \bibinfo {author} {\bibfnamefont {J.}~\bibnamefont {Buchheim}}, \bibinfo
  {author} {\bibfnamefont {M.}~\bibnamefont {Duchardt}}, \bibinfo {author}
  {\bibfnamefont {A.-K.}\ \bibnamefont {Hatz}}, \bibinfo {author}
  {\bibfnamefont {M.~A.}\ \bibnamefont {Kraft}}, \bibinfo {author}
  {\bibfnamefont {H.}~\bibnamefont {Kwak}}, \bibinfo {author} {\bibfnamefont
  {A.~L.}\ \bibnamefont {Santhosha}}, \bibinfo {author} {\bibfnamefont
  {Z.}~\bibnamefont {Liu}}, \bibinfo {author} {\bibfnamefont {N.}~\bibnamefont
  {Minafra}}, \bibinfo {author} {\bibfnamefont {F.}~\bibnamefont {Tsuji}},
  \bibinfo {author} {\bibfnamefont {A.}~\bibnamefont {Sakuda}}, \bibinfo
  {author} {\bibfnamefont {R.}~\bibnamefont {Schlem}}, \bibinfo {author}
  {\bibfnamefont {S.}~\bibnamefont {Xiong}}, \bibinfo {author} {\bibfnamefont
  {Z.}~\bibnamefont {Zhang}}, \bibinfo {author} {\bibfnamefont
  {P.}~\bibnamefont {Adelhelm}}, \bibinfo {author} {\bibfnamefont
  {H.}~\bibnamefont {Chen}}, \bibinfo {author} {\bibfnamefont {A.}~\bibnamefont
  {Hayashi}}, \bibinfo {author} {\bibfnamefont {Y.~S.}\ \bibnamefont {Jung}},
  \bibinfo {author} {\bibfnamefont {B.~V.}\ \bibnamefont {Lotsch}}, \bibinfo
  {author} {\bibfnamefont {B.}~\bibnamefont {Roling}}, \bibinfo {author}
  {\bibfnamefont {N.~M.}\ \bibnamefont {Vargas-Barbosa}}, \ and\ \bibinfo
  {author} {\bibfnamefont {W.~G.}\ \bibnamefont {Zeier}},\ }\bibfield  {title}
  {\enquote {\bibinfo {title} {How certain are the reported ionic
  conductivities of thiophosphate-based solid electrolytes? an interlaboratory
  study},}\ }\href {\doibase 10.1021/acsenergylett.9b02764} {\bibfield
  {journal} {\bibinfo  {journal} {ACS Energy Letters}\ }\textbf {\bibinfo
  {volume} {5}},\ \bibinfo {pages} {910--915} (\bibinfo {year}
  {2020})}\BibitemShut {NoStop}%
\bibitem [{\citenamefont {Murch}(1982)}]{murch1982haven}%
  \BibitemOpen
  \bibfield  {author} {\bibinfo {author} {\bibfnamefont {G.}~\bibnamefont
  {Murch}},\ }\bibfield  {title} {\enquote {\bibinfo {title} {The haven ratio
  in fast ionic conductors},}\ }\href@noop {} {\bibfield  {journal} {\bibinfo
  {journal} {Solid State Ionics}\ }\textbf {\bibinfo {volume} {7}},\ \bibinfo
  {pages} {177--198} (\bibinfo {year} {1982})}\BibitemShut {NoStop}%
\bibitem [{\citenamefont {Bart{\'o}k}, \citenamefont {Kondor},\ and\
  \citenamefont {Cs{\'a}nyi}(2013)}]{bartok2013representing}%
  \BibitemOpen
  \bibfield  {author} {\bibinfo {author} {\bibfnamefont {A.~P.}\ \bibnamefont
  {Bart{\'o}k}}, \bibinfo {author} {\bibfnamefont {R.}~\bibnamefont {Kondor}},
  \ and\ \bibinfo {author} {\bibfnamefont {G.}~\bibnamefont {Cs{\'a}nyi}},\
  }\bibfield  {title} {\enquote {\bibinfo {title} {On representing chemical
  environments},}\ }\href@noop {} {\bibfield  {journal} {\bibinfo  {journal}
  {Physical Review B}\ }\textbf {\bibinfo {volume} {87}},\ \bibinfo {pages}
  {184115} (\bibinfo {year} {2013})}\BibitemShut {NoStop}%
\bibitem [{\citenamefont {Bernstein}\ \emph {et~al.}(2019)\citenamefont
  {Bernstein}, \citenamefont {Bhattarai}, \citenamefont {Cs{\'a}nyi},
  \citenamefont {Drabold}, \citenamefont {Elliott},\ and\ \citenamefont
  {Deringer}}]{bernstein2019quantifying}%
  \BibitemOpen
  \bibfield  {author} {\bibinfo {author} {\bibfnamefont {N.}~\bibnamefont
  {Bernstein}}, \bibinfo {author} {\bibfnamefont {B.}~\bibnamefont
  {Bhattarai}}, \bibinfo {author} {\bibfnamefont {G.}~\bibnamefont
  {Cs{\'a}nyi}}, \bibinfo {author} {\bibfnamefont {D.~A.}\ \bibnamefont
  {Drabold}}, \bibinfo {author} {\bibfnamefont {S.~R.}\ \bibnamefont
  {Elliott}}, \ and\ \bibinfo {author} {\bibfnamefont {V.~L.}\ \bibnamefont
  {Deringer}},\ }\bibfield  {title} {\enquote {\bibinfo {title} {Quantifying
  chemical structure and machine-learned atomic energies in amorphous and
  liquid silicon},}\ }\href@noop {} {\bibfield  {journal} {\bibinfo  {journal}
  {Angewandte Chemie}\ }\textbf {\bibinfo {volume} {131}},\ \bibinfo {pages}
  {7131--7135} (\bibinfo {year} {2019})}\BibitemShut {NoStop}%
\bibitem [{\citenamefont {Kato}\ \emph {et~al.}(2016)\citenamefont {Kato},
  \citenamefont {Hori}, \citenamefont {Saito}, \citenamefont {Suzuki},
  \citenamefont {Hirayama}, \citenamefont {Mitsui}, \citenamefont {Yonemura},
  \citenamefont {Iba},\ and\ \citenamefont {Kanno}}]{kato2016high}%
  \BibitemOpen
  \bibfield  {author} {\bibinfo {author} {\bibfnamefont {Y.}~\bibnamefont
  {Kato}}, \bibinfo {author} {\bibfnamefont {S.}~\bibnamefont {Hori}}, \bibinfo
  {author} {\bibfnamefont {T.}~\bibnamefont {Saito}}, \bibinfo {author}
  {\bibfnamefont {K.}~\bibnamefont {Suzuki}}, \bibinfo {author} {\bibfnamefont
  {M.}~\bibnamefont {Hirayama}}, \bibinfo {author} {\bibfnamefont
  {A.}~\bibnamefont {Mitsui}}, \bibinfo {author} {\bibfnamefont
  {M.}~\bibnamefont {Yonemura}}, \bibinfo {author} {\bibfnamefont
  {H.}~\bibnamefont {Iba}}, \ and\ \bibinfo {author} {\bibfnamefont
  {R.}~\bibnamefont {Kanno}},\ }\bibfield  {title} {\enquote {\bibinfo {title}
  {High-power all-solid-state batteries using sulfide superionic conductors},}\
  }\href@noop {} {\bibfield  {journal} {\bibinfo  {journal} {Nature Energy}\
  }\textbf {\bibinfo {volume} {1}},\ \bibinfo {pages} {16030} (\bibinfo {year}
  {2016})}\BibitemShut {NoStop}%
\bibitem [{\citenamefont {Bachman}\ \emph {et~al.}(2015)\citenamefont
  {Bachman}, \citenamefont {Muy}, \citenamefont {Grimaud}, \citenamefont
  {Chang}, \citenamefont {Pour}, \citenamefont {Lux}, \citenamefont {Paschos},
  \citenamefont {Maglia}, \citenamefont {Lupart}, \citenamefont {Lamp} \emph
  {et~al.}}]{bachman2015inorganic}%
  \BibitemOpen
  \bibfield  {author} {\bibinfo {author} {\bibfnamefont {J.~C.}\ \bibnamefont
  {Bachman}}, \bibinfo {author} {\bibfnamefont {S.}~\bibnamefont {Muy}},
  \bibinfo {author} {\bibfnamefont {A.}~\bibnamefont {Grimaud}}, \bibinfo
  {author} {\bibfnamefont {H.-H.}\ \bibnamefont {Chang}}, \bibinfo {author}
  {\bibfnamefont {N.}~\bibnamefont {Pour}}, \bibinfo {author} {\bibfnamefont
  {S.~F.}\ \bibnamefont {Lux}}, \bibinfo {author} {\bibfnamefont
  {O.}~\bibnamefont {Paschos}}, \bibinfo {author} {\bibfnamefont
  {F.}~\bibnamefont {Maglia}}, \bibinfo {author} {\bibfnamefont
  {S.}~\bibnamefont {Lupart}}, \bibinfo {author} {\bibfnamefont
  {P.}~\bibnamefont {Lamp}},  \emph {et~al.},\ }\bibfield  {title} {\enquote
  {\bibinfo {title} {Inorganic solid-state electrolytes for lithium batteries:
  mechanisms and properties governing ion conduction},}\ }\href@noop {}
  {\bibfield  {journal} {\bibinfo  {journal} {Chemical reviews}\ }\textbf
  {\bibinfo {volume} {116}},\ \bibinfo {pages} {140--162} (\bibinfo {year}
  {2015})}\BibitemShut {NoStop}%
\bibitem [{\citenamefont {Xie}\ \emph {et~al.}(2019)\citenamefont {Xie},
  \citenamefont {France-Lanord}, \citenamefont {Wang}, \citenamefont
  {Shao-Horn},\ and\ \citenamefont {Grossman}}]{xie2019graph}%
  \BibitemOpen
  \bibfield  {author} {\bibinfo {author} {\bibfnamefont {T.}~\bibnamefont
  {Xie}}, \bibinfo {author} {\bibfnamefont {A.}~\bibnamefont {France-Lanord}},
  \bibinfo {author} {\bibfnamefont {Y.}~\bibnamefont {Wang}}, \bibinfo {author}
  {\bibfnamefont {Y.}~\bibnamefont {Shao-Horn}}, \ and\ \bibinfo {author}
  {\bibfnamefont {J.~C.}\ \bibnamefont {Grossman}},\ }\bibfield  {title}
  {\enquote {\bibinfo {title} {Graph dynamical networks for unsupervised
  learning of atomic scale dynamics in materials},}\ }\href@noop {} {\bibfield
  {journal} {\bibinfo  {journal} {Nature communications}\ }\textbf {\bibinfo
  {volume} {10}},\ \bibinfo {pages} {2667} (\bibinfo {year}
  {2019})}\BibitemShut {NoStop}%
\bibitem [{\citenamefont {Deng}\ \emph {et~al.}(2015)\citenamefont {Deng},
  \citenamefont {Eames}, \citenamefont {Chotard}, \citenamefont {Lalère},
  \citenamefont {Seznec}, \citenamefont {Emge}, \citenamefont {Pecher},
  \citenamefont {Grey}, \citenamefont {Masquelier},\ and\ \citenamefont
  {Islam}}]{deng2015structural}%
  \BibitemOpen
  \bibfield  {author} {\bibinfo {author} {\bibfnamefont {Y.}~\bibnamefont
  {Deng}}, \bibinfo {author} {\bibfnamefont {C.}~\bibnamefont {Eames}},
  \bibinfo {author} {\bibfnamefont {J.-N.}\ \bibnamefont {Chotard}}, \bibinfo
  {author} {\bibfnamefont {F.}~\bibnamefont {Lalère}}, \bibinfo {author}
  {\bibfnamefont {V.}~\bibnamefont {Seznec}}, \bibinfo {author} {\bibfnamefont
  {S.}~\bibnamefont {Emge}}, \bibinfo {author} {\bibfnamefont {O.}~\bibnamefont
  {Pecher}}, \bibinfo {author} {\bibfnamefont {C.~P.}\ \bibnamefont {Grey}},
  \bibinfo {author} {\bibfnamefont {C.}~\bibnamefont {Masquelier}}, \ and\
  \bibinfo {author} {\bibfnamefont {M.~S.}\ \bibnamefont {Islam}},\ }\bibfield
  {title} {\enquote {\bibinfo {title} {Structural and mechanistic insights into
  fast lithium-ion conduction in li4sio4--li3po4 solid electrolytes},}\
  }\href@noop {} {\bibfield  {journal} {\bibinfo  {journal} {Journal of the
  American Chemical Society}\ }\textbf {\bibinfo {volume} {137}},\ \bibinfo
  {pages} {9136--9145} (\bibinfo {year} {2015})}\BibitemShut {NoStop}%
\bibitem [{\citenamefont {Kahle}\ \emph {et~al.}(2019)\citenamefont {Kahle},
  \citenamefont {Musaelian}, \citenamefont {Marzari},\ and\ \citenamefont
  {Kozinsky}}]{kahle2019unsupervised}%
  \BibitemOpen
  \bibfield  {author} {\bibinfo {author} {\bibfnamefont {L.}~\bibnamefont
  {Kahle}}, \bibinfo {author} {\bibfnamefont {A.}~\bibnamefont {Musaelian}},
  \bibinfo {author} {\bibfnamefont {N.}~\bibnamefont {Marzari}}, \ and\
  \bibinfo {author} {\bibfnamefont {B.}~\bibnamefont {Kozinsky}},\ }\bibfield
  {title} {\enquote {\bibinfo {title} {Unsupervised landmark analysis for jump
  detection in molecular dynamics simulations},}\ }\href@noop {} {\bibfield
  {journal} {\bibinfo  {journal} {Physical Review Materials}\ }\textbf
  {\bibinfo {volume} {3}},\ \bibinfo {pages} {055404} (\bibinfo {year}
  {2019})}\BibitemShut {NoStop}%
\bibitem [{\citenamefont {Muy}\ \emph {et~al.}(2018)\citenamefont {Muy},
  \citenamefont {Bachman}, \citenamefont {Chang}, \citenamefont {Giordano},
  \citenamefont {Maglia}, \citenamefont {Lupart}, \citenamefont {Lamp},
  \citenamefont {Zeier},\ and\ \citenamefont {Shao-Horn}}]{muy2018lithium}%
  \BibitemOpen
  \bibfield  {author} {\bibinfo {author} {\bibfnamefont {S.}~\bibnamefont
  {Muy}}, \bibinfo {author} {\bibfnamefont {J.~C.}\ \bibnamefont {Bachman}},
  \bibinfo {author} {\bibfnamefont {H.-H.}\ \bibnamefont {Chang}}, \bibinfo
  {author} {\bibfnamefont {L.}~\bibnamefont {Giordano}}, \bibinfo {author}
  {\bibfnamefont {F.}~\bibnamefont {Maglia}}, \bibinfo {author} {\bibfnamefont
  {S.}~\bibnamefont {Lupart}}, \bibinfo {author} {\bibfnamefont
  {P.}~\bibnamefont {Lamp}}, \bibinfo {author} {\bibfnamefont {W.~G.}\
  \bibnamefont {Zeier}}, \ and\ \bibinfo {author} {\bibfnamefont
  {Y.}~\bibnamefont {Shao-Horn}},\ }\bibfield  {title} {\enquote {\bibinfo
  {title} {Lithium conductivity and meyer-neldel rule in
  li3po4--li3vo4--li4geo4 lithium superionic conductors},}\ }\href@noop {}
  {\bibfield  {journal} {\bibinfo  {journal} {Chemistry of Materials}\ }\textbf
  {\bibinfo {volume} {30}},\ \bibinfo {pages} {5573--5582} (\bibinfo {year}
  {2018})}\BibitemShut {NoStop}%
\bibitem [{\citenamefont {Rao}\ \emph {et~al.}(2013)\citenamefont {Rao},
  \citenamefont {Sharma}, \citenamefont {Peterson},\ and\ \citenamefont
  {Adams}}]{rao2013formation}%
  \BibitemOpen
  \bibfield  {author} {\bibinfo {author} {\bibfnamefont {R.~P.}\ \bibnamefont
  {Rao}}, \bibinfo {author} {\bibfnamefont {N.}~\bibnamefont {Sharma}},
  \bibinfo {author} {\bibfnamefont {V.}~\bibnamefont {Peterson}}, \ and\
  \bibinfo {author} {\bibfnamefont {S.}~\bibnamefont {Adams}},\ }\bibfield
  {title} {\enquote {\bibinfo {title} {Formation and conductivity studies of
  lithium argyrodite solid electrolytes using in-situ neutron diffraction},}\
  }\href@noop {} {\bibfield  {journal} {\bibinfo  {journal} {Solid State
  Ionics}\ }\textbf {\bibinfo {volume} {230}},\ \bibinfo {pages} {72--76}
  (\bibinfo {year} {2013})}\BibitemShut {NoStop}%
\bibitem [{\citenamefont {Kingma}\ and\ \citenamefont
  {Ba}(2014)}]{kingma2014adam}%
  \BibitemOpen
  \bibfield  {author} {\bibinfo {author} {\bibfnamefont {D.~P.}\ \bibnamefont
  {Kingma}}\ and\ \bibinfo {author} {\bibfnamefont {J.}~\bibnamefont {Ba}},\
  }\bibfield  {title} {\enquote {\bibinfo {title} {Adam: A method for
  stochastic optimization},}\ }\href@noop {} {\bibfield  {journal} {\bibinfo
  {journal} {arXiv preprint arXiv:1412.6980}\ } (\bibinfo {year}
  {2014})}\BibitemShut {NoStop}%
\bibitem [{\citenamefont {Zhang}\ \emph
  {et~al.}(2018{\natexlab{c}})\citenamefont {Zhang}, \citenamefont {Han},
  \citenamefont {Wang}, \citenamefont {Car},\ and\ \citenamefont
  {E}}]{zhang2018deep}%
  \BibitemOpen
  \bibfield  {author} {\bibinfo {author} {\bibfnamefont {L.}~\bibnamefont
  {Zhang}}, \bibinfo {author} {\bibfnamefont {J.}~\bibnamefont {Han}}, \bibinfo
  {author} {\bibfnamefont {H.}~\bibnamefont {Wang}}, \bibinfo {author}
  {\bibfnamefont {R.}~\bibnamefont {Car}}, \ and\ \bibinfo {author}
  {\bibfnamefont {W.}~\bibnamefont {E}},\ }\bibfield  {title} {\enquote
  {\bibinfo {title} {Deep potential molecular dynamics: a scalable model with
  the accuracy of quantum mechanics},}\ }\href@noop {} {\bibfield  {journal}
  {\bibinfo  {journal} {Physical review letters}\ }\textbf {\bibinfo {volume}
  {120}},\ \bibinfo {pages} {143001} (\bibinfo {year}
  {2018}{\natexlab{c}})}\BibitemShut {NoStop}%
\bibitem [{\citenamefont {Henkelman}, \citenamefont {Uberuaga},\ and\
  \citenamefont {J{\'o}nsson}(2000)}]{henkelman2000climbing}%
  \BibitemOpen
  \bibfield  {author} {\bibinfo {author} {\bibfnamefont {G.}~\bibnamefont
  {Henkelman}}, \bibinfo {author} {\bibfnamefont {B.~P.}\ \bibnamefont
  {Uberuaga}}, \ and\ \bibinfo {author} {\bibfnamefont {H.}~\bibnamefont
  {J{\'o}nsson}},\ }\bibfield  {title} {\enquote {\bibinfo {title} {A climbing
  image nudged elastic band method for finding saddle points and minimum energy
  paths},}\ }\href@noop {} {\bibfield  {journal} {\bibinfo  {journal} {The
  Journal of chemical physics}\ }\textbf {\bibinfo {volume} {113}},\ \bibinfo
  {pages} {9901--9904} (\bibinfo {year} {2000})}\BibitemShut {NoStop}%
\bibitem [{\citenamefont {Perdew}, \citenamefont {Burke},\ and\ \citenamefont
  {Ernzerhof}(1996{\natexlab{b}})}]{perdew1996generalized}%
  \BibitemOpen
  \bibfield  {author} {\bibinfo {author} {\bibfnamefont {J.~P.}\ \bibnamefont
  {Perdew}}, \bibinfo {author} {\bibfnamefont {K.}~\bibnamefont {Burke}}, \
  and\ \bibinfo {author} {\bibfnamefont {M.}~\bibnamefont {Ernzerhof}},\
  }\bibfield  {title} {\enquote {\bibinfo {title} {Generalized gradient
  approximation made simple},}\ }\href@noop {} {\bibfield  {journal} {\bibinfo
  {journal} {Physical review letters}\ }\textbf {\bibinfo {volume} {77}},\
  \bibinfo {pages} {3865} (\bibinfo {year} {1996}{\natexlab{b}})}\BibitemShut
  {NoStop}%
\bibitem [{\citenamefont {Hori}\ \emph
  {et~al.}(2015{\natexlab{a}})\citenamefont {Hori}, \citenamefont {Taminato},
  \citenamefont {Suzuki}, \citenamefont {Hirayama}, \citenamefont {Kato},\ and\
  \citenamefont {Kanno}}]{Hori2015}%
  \BibitemOpen
  \bibfield  {author} {\bibinfo {author} {\bibfnamefont {S.}~\bibnamefont
  {Hori}}, \bibinfo {author} {\bibfnamefont {S.}~\bibnamefont {Taminato}},
  \bibinfo {author} {\bibfnamefont {K.}~\bibnamefont {Suzuki}}, \bibinfo
  {author} {\bibfnamefont {M.}~\bibnamefont {Hirayama}}, \bibinfo {author}
  {\bibfnamefont {Y.}~\bibnamefont {Kato}}, \ and\ \bibinfo {author}
  {\bibfnamefont {R.}~\bibnamefont {Kanno}},\ }\bibfield  {title} {\enquote
  {\bibinfo {title} {{Structure–property relationships in lithium superionic
  conductors having a Li 10 GeP 2 S 12 -type structure}},}\ }\href {\doibase
  10.1107/S2052520615022283} {\bibfield  {journal} {\bibinfo  {journal} {Acta
  Crystallographica Section B Structural Science, Crystal Engineering and
  Materials}\ }\textbf {\bibinfo {volume} {71}},\ \bibinfo {pages} {727--736}
  (\bibinfo {year} {2015}{\natexlab{a}})}\BibitemShut {NoStop}%
\bibitem [{\citenamefont {Hori}\ \emph
  {et~al.}(2015{\natexlab{b}})\citenamefont {Hori}, \citenamefont {Kato},
  \citenamefont {Suzuki}, \citenamefont {Hirayama}, \citenamefont {Kato},\ and\
  \citenamefont {Kanno}}]{Hori2015a}%
  \BibitemOpen
  \bibfield  {author} {\bibinfo {author} {\bibfnamefont {S.}~\bibnamefont
  {Hori}}, \bibinfo {author} {\bibfnamefont {M.}~\bibnamefont {Kato}}, \bibinfo
  {author} {\bibfnamefont {K.}~\bibnamefont {Suzuki}}, \bibinfo {author}
  {\bibfnamefont {M.}~\bibnamefont {Hirayama}}, \bibinfo {author}
  {\bibfnamefont {Y.}~\bibnamefont {Kato}}, \ and\ \bibinfo {author}
  {\bibfnamefont {R.}~\bibnamefont {Kanno}},\ }\bibfield  {title} {\enquote
  {\bibinfo {title} {{Phase Diagram of the Li 4 GeS 4 -Li 3 PS 4 Quasi-Binary
  System Containing the Superionic Conductor Li 10 GeP 2 S 12}},}\ }\href
  {\doibase 10.1111/jace.13694} {\bibfield  {journal} {\bibinfo  {journal}
  {Journal of the American Ceramic Society}\ }\textbf {\bibinfo {volume}
  {98}},\ \bibinfo {pages} {3352--3360} (\bibinfo {year}
  {2015}{\natexlab{b}})}\BibitemShut {NoStop}%
\bibitem [{\citenamefont {Hu}\ \emph {et~al.}(2014)\citenamefont {Hu},
  \citenamefont {Wang}, \citenamefont {Sun},\ and\ \citenamefont
  {Ouyang}}]{Hu2014}%
  \BibitemOpen
  \bibfield  {author} {\bibinfo {author} {\bibfnamefont {C.~H.}\ \bibnamefont
  {Hu}}, \bibinfo {author} {\bibfnamefont {Z.~Q.}\ \bibnamefont {Wang}},
  \bibinfo {author} {\bibfnamefont {Z.~Y.}\ \bibnamefont {Sun}}, \ and\
  \bibinfo {author} {\bibfnamefont {C.~Y.}\ \bibnamefont {Ouyang}},\ }\bibfield
   {title} {\enquote {\bibinfo {title} {{Insights into structural stability and
  Li superionic conductivity of Li 10 GeP 2 S 12 from first-principles
  calculations}},}\ }\href {\doibase 10.1016/j.cplett.2013.11.003} {\bibfield
  {journal} {\bibinfo  {journal} {Chemical Physics Letters}\ }\textbf {\bibinfo
  {volume} {591}},\ \bibinfo {pages} {16--20} (\bibinfo {year}
  {2014})}\BibitemShut {NoStop}%
\bibitem [{\citenamefont {Liang}\ \emph {et~al.}(2015)\citenamefont {Liang},
  \citenamefont {Wang}, \citenamefont {Jiang}, \citenamefont {Wang},
  \citenamefont {Luo}, \citenamefont {Liu},\ and\ \citenamefont
  {Feng}}]{Liang2015}%
  \BibitemOpen
  \bibfield  {author} {\bibinfo {author} {\bibfnamefont {X.}~\bibnamefont
  {Liang}}, \bibinfo {author} {\bibfnamefont {L.}~\bibnamefont {Wang}},
  \bibinfo {author} {\bibfnamefont {Y.}~\bibnamefont {Jiang}}, \bibinfo
  {author} {\bibfnamefont {J.}~\bibnamefont {Wang}}, \bibinfo {author}
  {\bibfnamefont {H.}~\bibnamefont {Luo}}, \bibinfo {author} {\bibfnamefont
  {C.}~\bibnamefont {Liu}}, \ and\ \bibinfo {author} {\bibfnamefont
  {J.}~\bibnamefont {Feng}},\ }\bibfield  {title} {\enquote {\bibinfo {title}
  {{In-Channel and In-Plane Li Ion Diffusions in the Superionic Conductor Li 10
  GeP 2 S 12 Probed by Solid-State NMR}},}\ }\href {\doibase
  10.1021/acs.chemmater.5b01384} {\bibfield  {journal} {\bibinfo  {journal}
  {Chemistry of Materials}\ }\textbf {\bibinfo {volume} {27}},\ \bibinfo
  {pages} {5503--5510} (\bibinfo {year} {2015})}\BibitemShut {NoStop}%
\bibitem [{\citenamefont {Perdew}\ \emph {et~al.}(1999)\citenamefont {Perdew},
  \citenamefont {Kurth}, \citenamefont {Zupan},\ and\ \citenamefont
  {Blaha}}]{PhysRevLett.82.2544}%
  \BibitemOpen
  \bibfield  {author} {\bibinfo {author} {\bibfnamefont {J.~P.}\ \bibnamefont
  {Perdew}}, \bibinfo {author} {\bibfnamefont {S.}~\bibnamefont {Kurth}},
  \bibinfo {author} {\bibfnamefont {A.~c.~v.}\ \bibnamefont {Zupan}}, \ and\
  \bibinfo {author} {\bibfnamefont {P.}~\bibnamefont {Blaha}},\ }\bibfield
  {title} {\enquote {\bibinfo {title} {Accurate density functional with correct
  formal properties: A step beyond the generalized gradient approximation},}\
  }\href {\doibase 10.1103/PhysRevLett.82.2544} {\bibfield  {journal} {\bibinfo
   {journal} {Phys. Rev. Lett.}\ }\textbf {\bibinfo {volume} {82}},\ \bibinfo
  {pages} {2544--2547} (\bibinfo {year} {1999})}\BibitemShut {NoStop}%
\bibitem [{\citenamefont {Kuhn}\ \emph {et~al.}(2014)\citenamefont {Kuhn},
  \citenamefont {Gerbig}, \citenamefont {Zhu}, \citenamefont {Falkenberg},
  \citenamefont {Maier},\ and\ \citenamefont {Lotsch}}]{C4CP02046D}%
  \BibitemOpen
  \bibfield  {author} {\bibinfo {author} {\bibfnamefont {A.}~\bibnamefont
  {Kuhn}}, \bibinfo {author} {\bibfnamefont {O.}~\bibnamefont {Gerbig}},
  \bibinfo {author} {\bibfnamefont {C.}~\bibnamefont {Zhu}}, \bibinfo {author}
  {\bibfnamefont {F.}~\bibnamefont {Falkenberg}}, \bibinfo {author}
  {\bibfnamefont {J.}~\bibnamefont {Maier}}, \ and\ \bibinfo {author}
  {\bibfnamefont {B.~V.}\ \bibnamefont {Lotsch}},\ }\bibfield  {title}
  {\enquote {\bibinfo {title} {A new ultrafast superionic li-conductor: ion
  dynamics in li11si2ps12 and comparison with other tetragonal lgps-type
  electrolytes},}\ }\href {\doibase 10.1039/C4CP02046D} {\bibfield  {journal}
  {\bibinfo  {journal} {Phys. Chem. Chem. Phys.}\ }\textbf {\bibinfo {volume}
  {16}},\ \bibinfo {pages} {14669--14674} (\bibinfo {year} {2014})}\BibitemShut
  {NoStop}%
\bibitem [{\citenamefont {Wang}\ \emph {et~al.}(2014)\citenamefont {Wang},
  \citenamefont {Wu}, \citenamefont {Liu}, \citenamefont {Lei}, \citenamefont
  {Xu},\ and\ \citenamefont {Ouyang}}]{Wang2014}%
  \BibitemOpen
  \bibfield  {author} {\bibinfo {author} {\bibfnamefont {Z.~Q.}\ \bibnamefont
  {Wang}}, \bibinfo {author} {\bibfnamefont {M.~S.}\ \bibnamefont {Wu}},
  \bibinfo {author} {\bibfnamefont {G.}~\bibnamefont {Liu}}, \bibinfo {author}
  {\bibfnamefont {X.~L.}\ \bibnamefont {Lei}}, \bibinfo {author} {\bibfnamefont
  {B.}~\bibnamefont {Xu}}, \ and\ \bibinfo {author} {\bibfnamefont {C.~Y.}\
  \bibnamefont {Ouyang}},\ }\bibfield  {title} {\enquote {\bibinfo {title}
  {{Elastic properties of new solid state electrolyte material Li10GeP2S12: A
  study from first-principles calculations}},}\ }\href@noop {} {\bibfield
  {journal} {\bibinfo  {journal} {International Journal of Electrochemical
  Science}\ }\textbf {\bibinfo {volume} {9}},\ \bibinfo {pages} {562--568}
  (\bibinfo {year} {2014})}\BibitemShut {NoStop}%
\bibitem [{\citenamefont {Klime{\v{s}}}, \citenamefont {Bowler},\ and\
  \citenamefont {Michaelides}(2009{\natexlab{a}})}]{Klime2009}%
  \BibitemOpen
  \bibfield  {author} {\bibinfo {author} {\bibfnamefont {J.}~\bibnamefont
  {Klime{\v{s}}}}, \bibinfo {author} {\bibfnamefont {D.~R.}\ \bibnamefont
  {Bowler}}, \ and\ \bibinfo {author} {\bibfnamefont {A.}~\bibnamefont
  {Michaelides}},\ }\bibfield  {title} {\enquote {\bibinfo {title} {Chemical
  accuracy for the van der waals density functional},}\ }\href {\doibase
  10.1088/0953-8984/22/2/022201} {\bibfield  {journal} {\bibinfo  {journal}
  {Journal of Physics: Condensed Matter}\ }\textbf {\bibinfo {volume} {22}},\
  \bibinfo {pages} {022201} (\bibinfo {year} {2009}{\natexlab{a}})}\BibitemShut
  {NoStop}%
\bibitem [{\citenamefont {Klime{\v{s}}}, \citenamefont {Bowler},\ and\
  \citenamefont {Michaelides}(2009{\natexlab{b}})}]{vdw-optb88}%
  \BibitemOpen
  \bibfield  {author} {\bibinfo {author} {\bibfnamefont {J.}~\bibnamefont
  {Klime{\v{s}}}}, \bibinfo {author} {\bibfnamefont {D.~R.}\ \bibnamefont
  {Bowler}}, \ and\ \bibinfo {author} {\bibfnamefont {A.}~\bibnamefont
  {Michaelides}},\ }\bibfield  {title} {\enquote {\bibinfo {title} {Chemical
  accuracy for the van der waals density functional},}\ }\href {\doibase
  10.1088/0953-8984/22/2/022201} {\bibfield  {journal} {\bibinfo  {journal}
  {Journal of Physics: Condensed Matter}\ }\textbf {\bibinfo {volume} {22}},\
  \bibinfo {pages} {022201} (\bibinfo {year} {2009}{\natexlab{b}})}\BibitemShut
  {NoStop}%
\bibitem [{\citenamefont {Bart\'ok}\ \emph {et~al.}(2010)\citenamefont
  {Bart\'ok}, \citenamefont {Payne}, \citenamefont {Kondor},\ and\
  \citenamefont {Cs\'anyi}}]{GAP2010PRL}%
  \BibitemOpen
  \bibfield  {author} {\bibinfo {author} {\bibfnamefont {A.~P.}\ \bibnamefont
  {Bart\'ok}}, \bibinfo {author} {\bibfnamefont {M.~C.}\ \bibnamefont {Payne}},
  \bibinfo {author} {\bibfnamefont {R.}~\bibnamefont {Kondor}}, \ and\ \bibinfo
  {author} {\bibfnamefont {G.}~\bibnamefont {Cs\'anyi}},\ }\bibfield  {title}
  {\enquote {\bibinfo {title} {Gaussian approximation potentials: The accuracy
  of quantum mechanics, without the electrons},}\ }\href {\doibase
  10.1103/PhysRevLett.104.136403} {\bibfield  {journal} {\bibinfo  {journal}
  {Phys. Rev. Lett.}\ }\textbf {\bibinfo {volume} {104}},\ \bibinfo {pages}
  {136403} (\bibinfo {year} {2010})}\BibitemShut {NoStop}%
\bibitem [{\citenamefont {Hori}\ \emph
  {et~al.}(2015{\natexlab{c}})\citenamefont {Hori}, \citenamefont {Suzuki},
  \citenamefont {Hirayama}, \citenamefont {Kato}, \citenamefont {Saito},
  \citenamefont {Yonemura},\ and\ \citenamefont {Kanno}}]{hori2015synthesis}%
  \BibitemOpen
  \bibfield  {author} {\bibinfo {author} {\bibfnamefont {S.}~\bibnamefont
  {Hori}}, \bibinfo {author} {\bibfnamefont {K.}~\bibnamefont {Suzuki}},
  \bibinfo {author} {\bibfnamefont {M.}~\bibnamefont {Hirayama}}, \bibinfo
  {author} {\bibfnamefont {Y.}~\bibnamefont {Kato}}, \bibinfo {author}
  {\bibfnamefont {T.}~\bibnamefont {Saito}}, \bibinfo {author} {\bibfnamefont
  {M.}~\bibnamefont {Yonemura}}, \ and\ \bibinfo {author} {\bibfnamefont
  {R.}~\bibnamefont {Kanno}},\ }\bibfield  {title} {\enquote {\bibinfo {title}
  {Synthesis, structure, and ionic conductivity of solid solution, li10+
  $\delta$m1+ $\delta$p2- $\delta$s12 (m= si, sn)},}\ }\href@noop {} {\bibfield
   {journal} {\bibinfo  {journal} {Faraday Discussions}\ }\textbf {\bibinfo
  {volume} {176}},\ \bibinfo {pages} {83--94} (\bibinfo {year}
  {2015}{\natexlab{c}})}\BibitemShut {NoStop}%
\end{thebibliography}%


%

\end{document}